\documentclass[twoside,11pt]{article}

\usepackage{blindtext}

\usepackage{thmtools}

\usepackage[preprint]{jmlr2e}

\usepackage[T1]{fontenc}
\usepackage{newtxtext}
\usepackage{newtxmath}
\usepackage[bb=boondox,cal=boondox]{mathalpha}

\usepackage{amsmath}
\usepackage{amsfonts}
\usepackage{thmtools}
\usepackage{mathtools}

\usepackage[usenames,dvipsnames]{xcolor}

\usepackage{graphicx}
\usepackage[labelfont=bf]{caption}
\usepackage[format=hang]{subcaption}

\usepackage{algpseudocode}
\usepackage{algorithm}

\usepackage{enumitem}

\usepackage{tikz}
\usetikzlibrary{bayesnet}

\usepackage[nameinlink]{cleveref}

\newcommand{\Sc}{\mathcal{S}}
\newcommand{\Uc}{\mathcal{U}}

\newcommand{\Qc}{\mathcal{Q}}
\newcommand{\E}[1]{\mathbb{E}\left[ #1 \right]}
\newcommand{\EE}[2]{\mathbb{E}_{#1}\left[ #2 \right]}
\newcommand{\rmdo}{\mathrm{do}}
\newcommand{\s}{\, ; \,}
\newcommand{\g}{\,\vert\,}
\newcommand{\di}{\mathrm{d}}
\newcommand{\pa}{\mathrm{pa}}
\newcommand{\dd}{\mathrm{dd}}
\newcommand{\da}{\mathrm{da}}

\newcommand{\pr}{\mathrm{p}}
\newcommand{\Prob}{\mathrm{P}}
\newcommand{\f}{\mathrm{f}}
\newcommand{\obs}{\mathrm{obs}}
\newcommand{\col}{\mathrm{col}}
\newcommand{\aug}{\mathrm{aug}}
\newcommand{\mar}{\mathrm{mar}}
\newcommand{\ep}{\mathrm{ep}}
\newcommand{\Was}{\mathrm{W}}
\newcommand{\m}{\mathrm{m}}

\newcommand{\Mc}{\mathcal{M}}
\newcommand{\cgm}{\mathrm{cgm}}
\newcommand{\scm}{\mathrm{scm}}

\newcommand{\QLR}{Q^{\mathcal{L}|\mathcal{R}}}
\newcommand{\QLp}{Q^{\mathcal{L}|\pa_\Sc(\mathcal{L})}}
\newcommand{\qLR}{q^{\mathcal{L}|\mathcal{R}}}
\newcommand{\qLp}{q^{\mathcal{L}|\pa_\Sc(\mathcal{L})}}
\newcommand{\fLR}{\f^{\mathcal{L}|\mathcal{R}}}
\newcommand{\qsLR}{q_\star^{\mathcal{L}|\mathcal{R}}}
\newcommand{\an}{\mathrm{an}}
\newcommand{\ch}{\mathrm{ch}}

\newtheorem{assumption}{Assumption}

\newcommand{\parhead}[1]{\vspace{0.1in} \noindent \textbf{#1}. \,\,}
\newcommand{\parheadsmall}[1]{\vspace{0.1in} \noindent \textbf{#1} \,\,}

\usepackage{lastpage}
\jmlrheading{27}{2026}{1-\pageref{LastPage}}{4/25; Revised
10/25}{1/26}{25-0899}{Eli N. Weinstein and David M. Blei}

\ShortHeadings{Hierarchical Causal Models}{Weinstein and Blei}
\firstpageno{1}

\begin{document}

\title{Hierarchical Causal Models}

\author{\name Eli N. Weinstein \email enawe@dtu.dk \\
       \addr Department of Chemistry\\
       Technical University of Denmark\\
       Kgs. Lyngby, DK 2800, Denmark
       \AND
       \name David M. Blei \email david.blei@columbia.edu \\
       \addr Department of Computer Science and Department of Statistics\\
       Columbia University\\
       New York, NY 10027, USA}

\editor{Kun Zhang}

\maketitle

\begin{abstract}

Causal questions often arise in settings where data are hierarchical:
subunits are nested within units. Consider students in schools, cells
in patients, or cities in states. In these settings, unit-level
variables (e.g., a school's budget) may affect subunit-level outcomes
(e.g., student test scores), and subunit-level characteristics may
aggregate to influence unit-level outcomes. In this paper, we show how
to analyze hierarchical data for causal inference. We introduce
hierarchical causal models, which extend structural causal models and
graphical models by incorporating inner plates to represent nested
data structures. We develop a graphical identification technique for
these models that generalizes do-calculus. We show that hierarchical
data can enable causal identification even when it would be impossible
with non-hierarchical data--for example, when only unit-level
summaries are available. We develop estimation strategies, including
using hierarchical Bayesian models. We illustrate our results in
simulation and through a reanalysis of the classic ``eight schools''
study.

 \end{abstract}

\begin{keywords}
  causality, graphical models, multiple environments, Bayesian models
\end{keywords}

\section{Introduction}
\label{sec:intro}

We propose \textit{hierarchical causal models} (HCMs), a graphical
modeling framework for causal inference from nested or hierarchical
data. Hierarchical graphical models are a mainstay of modern Bayesian
statistics, but causal graphical models have largely remained
``flat''. In contrast, HCMs describe causal relationships among nested
variables. We develop the semantics and theory behind HCMs, and we
derive identification and estimation tools for using them to answer
causal questions from hierarchical data. We show how HCMs can help us
exploit the nested structure of the data to correct for confounders
and other sources of causal bias.

Why study hierarchical causal models? Many phenomena across the
natural and social sciences can be framed in terms of nested data.
Consider the following domains where hierarchical causal modeling
could be useful.
\begin{enumerate}[leftmargin=*]
\item \textit{Political science.} We observe citizens (subunits)
  within states (units). How do individual citizens' political
  preferences (a subunit variable) determine which political party
  governs (a unit variable)? How do states' economic policies (a unit
  variable) affect citizens' incomes (a subunit variable)?

\item \textit{Biology.} We observe cells (subunits) within patients
  (units). How do individual cells' genetic mutations (a subunit
  variable) determine whether the patient develops cancer (a unit
  variable)? How do patients' chemotherapy treatments (a unit
  variable) determine cells' survival (a subunit variable)?

\item \textit{Physical chemistry.} We observe molecules (subunits)
  within a gas (unit). How does altering molecules' motion (a subunit
  variable) alter the pressure exerted by the gas (a unit variable)?
  How does increasing the temperature of the gas's container (a unit
  variable) alter individual molecules' motion (a subunit variable)?

\end{enumerate}
In all these settings, we may encounter complex causal graphs, with
many different unit and subunit-level variables affecting one another.
This paper shows how to estimate causal effects from data following
given graph.

\begin{figure}[t!]
\centering
\begin{subfigure}[t]{0.24\textwidth}
\caption{} \label{fig:hcm_confounder}
\centering
\begin{tikzpicture}

\node[obs]                               (y) {$Y_{ij}$};
  \node[obs, left=1cm of y] (a) {$A_{ij}$};
  \node[latent, above=0.4cm of a, xshift=0.8cm]  (u) {$U_i$};

\edge {a,u} {y} ; \edge {u} {a} ;

\plate[dashed] {in} {(a)(y)} {$m$} ;
  \plate {out} {(in)(u)} {$n$} ;

\end{tikzpicture}
\end{subfigure}
\begin{subfigure}[t]{0.24\textwidth}
\caption{}
\label{fig:hcm_interfere}
\centering
\begin{tikzpicture}

\node[obs]                               (y) {$Y_{ij}$};
  \node[obs, left=1cm of y] (a) {$A_{ij}$};
  \node[latent, above=0.4cm of a, xshift=0.8cm]  (u) {$U_i$};
  \node[obs, below=.6cm of a] (z) {$Z_i$};

\edge {a,u} {y} ; \edge {u} {a} ;
  \edge {a} {z} ;
  \edge {z} {y} ;

\plate[dashed] {in} {(a)(y)} {$m$} ;
  \plate {out} {(in)(u)(z)} {$n$} ;

\end{tikzpicture}
\end{subfigure}
\begin{subfigure}[t]{0.24\textwidth}
\caption{} \label{fig:hcm_instrument}
\centering
\begin{tikzpicture}

  \node[obs]                               (y) {$Y_{i}$};
  \node[obs, left=0.4cm of y] (a) {$A_{ij}$};
  \node[obs, left=0.4cm of a] (z) {$Z_{ij}$};
  \node[latent, above=0.4cm of a, xshift=0.6cm]  (u) {$U_i$};

  \edge {a,u} {y} ;
  \edge {u} {a} ;
  \edge {z} {a} ;

  \plate[dashed] {in} {(a)(z)} {$m$} ;
  \plate {out} {(in)(u)(y)} {$n$} ;

\end{tikzpicture}
\end{subfigure}
\caption{\textbf{Example hierarchical causal models.} \Cref{fig:hcm_confounder}: the \textsc{unit confounder} graph. \Cref{fig:hcm_interfere}: the \textsc{unit confounder \& unit interference} graph. \Cref{fig:hcm_instrument}: the \textsc{subunit instrument} graph.} \label{fig:hcm_motifs}
\end{figure}

As a simplest first example, consider \Cref{fig:hcm_confounder}, where we observe a treatment $a$ and an outcome $y$, and our goal is estimate the causal effect of $a$ on $y$. For example, consider data from students in different schools, where $a$ is a test preparation course and $y$ is test score.
The challenge is that this graph also contains an unobserved confounder $u$, which affects both $a$ and $y$.
In a typical flat causal graph, $u$ would prevent us from learning the causal effect. 
However, in \Cref{fig:hcm_confounder}, the data is nested: we observe $a_{ij}$ and $y_{ij}$ for different subunits ($j$) within each unit ($i$), e.g. different students ($j$) within each school ($i$). 
The confounder $u_i$, meanwhile, is at the unit level, so it is fixed across subunits.
For example, a school's budget is a per-school quantity that affects individual students' test preparation and test scores.
Using our HCM methodology, we can show that the causal effect of $a$ on $y$ is identifiable, thanks to the nested structure of the data. Indeed, identification recovers classical results on fixed-effect models~\citep{Wooldridge2005-bk,Angrist2009-ah,Witty2020-et}.

Besides \Cref{fig:hcm_confounder}, there are many other causal graphs a scientist might consider, and causal questions they might ask (\Cref{fig:hcm_motifs}; \Cref{def:hscm}).
Besides confounding, there may be interference, as tutoring one student influences the test scores of others.
Besides student outcomes, there may be outcomes at the school level, such as whether the school is ranked highly on a statewide list.
For many graphs, and many causal questions, fixed-effect estimators do not provide a reliable estimate, or cannot be applied to the data at all.
Our identification and estimation tools let us reason about different hierarchical causal graphs, and derive appropriate methods for each.

We begin by formally defining hierarchical causal models (HCMs), a
causal generalization of hierarchical probabilistic models. Following
Pearl's hierarchy, we propose a definition of HCMs in terms of structural
equations with deterministic causal
mechanisms~\citep{Pearl2009-fh,Bareinboim2021-ro}. Next we analyze
causal identification in HCMs, to determine when and how we can
estimate causal effects from data. We consider arbitrary graphs and
nonparametric causal mechanisms, and we develop techniques for causal
identification in HCMs that build on the do calculus. Finally, we
craft new causal estimation techniques, by extending methods for
inference in hierarchical probabilistic models.

Overall, this paper presents a broad toolkit for
accomplishing causal inference with nested data. The approach
systematically extends hierarchical probabilistic and Bayesian models,
to encompass interventions and causality. This framework encompasses
many previous causal methods as special cases, such as fixed-effect
models and clustered interference
models~\citep{Wooldridge2005-bk,Angrist2009-ah,Goetgeluk2008-tm,Neuhaus1998-eq,Greenland2000-cq,Jensen2020-ds} (\Cref{apx:previous}).
More importantly, however, it enables causal inference across
novel problem settings and datasets, with different causal graphs.

 \paragraph{Related work.}

\label{sec:related_work}

There has been substantial research into causal inference from
hierarchical data. Many of these proposed models can be understood as
instances of HCMs with particular graphs and particular parametric
assumptions (\Cref{apx:previous}).

Fixed-effects models are widely used for correcting for unit-level
confounding~\citep[Chapters
10-11]{Wooldridge2005-bk,Wooldridge2010-dv}. One way to interpret
these methods is as HCMs that follow the \textsc{unit confounder}
graph, with a particular linear parameterization for the mechanism
generating $Y$ and sometimes with additional observed subunit-level
and unit-level confounders; see \Cref{apx:fixed_effect}. Closely
related are difference-in-difference and synthetic control methods,
which can be understood as following the same graph, but use alternate
model parameterizations~\citep[Chapter 5][]{Angrist1999-hm,
  Abadie2010-nn}. Hierarchical Bayesian methods are often used for
inference in fixed-effects
models~\citep{Gelman2006-ft,Hill2013-jq,Feller2015-eb}. More recently,
researchers have considered extensions of fixed-effects models that
allow for nonparametric causal mechanisms. Specifically,
\citet{Witty2020-et} propose an HCM following the \textsc{unit
  confounder} graph with a nonparametric parameterization for the
mechanism generating $Y$, along with a Gaussian process-based
inference method. We build on this work, studying arbitrary graphs
with nonparametric mechanisms.

This paper also relates to clustered interference, spillover effects,
and peer effects~\citep[Chapter
11]{Hudgens2008-hn,Tchetgen_Tchetgen2012-lm,Wooldridge2005-bk,Wooldridge2010-dv}.
Many of these models can be understood as HCMs with an unobserved
unit-level variable between the subunit treatment and outcome, and a linear
parameterization of the causal mechanisms; see \Cref{apx:interference}. 
The unobserved unit-level variable gives rise to interference among the subunits within each unit.
Other models, however, go beyond by allowing for general
interaction between
subunits, not just those mediated by a unit-level variable~\citep{Hudgens2008-hn,Ogburn2014-rp,Savje2021-zy}.

Another line of related work considers instrumental variable models in
multi-site trials, that is, repeated across multiple
units~\citep{Raudenbush2012-es,Reardon2014-fy}. These models
correspond to HCMs with an IV graph entirely at the subunit level,
together with an unobserved unit confounder, and using linear causal mechanisms; see
\Cref{apx:multisite_iv}.

At a high level, our contribution to this long literature is to
formalize and study HCMs under broad assumptions and conditions.
Specifically we study (1) arbitrary causal graphs and (2) arbitrary
(nonparametric) causal mechanisms. Our key finding is that there are
many novel scenarios where causal inference is possible.

Finally, an important thread of research has studied grouped data from
multiple ``environments'', focusing especially on problems where the causal graph is not fully known.
In some work the principal aim is to discover the causal graph~\citep[Chapters 2,4][]{Peters2017-ww,Tian2013-kn,Peters2015-uw,Perry2022-ah,Guo2022-bl}.
In other work, the aim is to develop prediction or estimation methods that are robust to the unknown graph~\citep{Arjovsky2019-bx,Yin2021-ma,Shi2020-cm,Krueger2020-ik}.
In either case, we can interpret multi-environment models as HCMs in which each unit corresponds to an environment, and the graph of the observed subunit variables is not entirely known.
In this paper, however, we focus on problems where the causal graph is known.
One consequence is that we do not require invariance assumptions or independent causal mechanism assumptions, as is standard in the multi-environment literature~\citep[Chapters 2,4][]{Peters2017-ww,Peters2015-uw,Guo2025-rd}; see \Cref{apx:multi-enviro}.

 \section{A first look at hierarchical causal models} \label{sec:first_look}
\parhead{Three examples of hierarchical causal models.} Causal models describe the world in
terms of variables and their impact on one another, mapped out in a
graph. A hierarchical causal model (HCM) contains a plate,
which denotes a systematic replication of variables within each unit.
In an HCM, variables that fall inside the plate are called
\textit{subunit-level variables}; variables that fall outside the
plate are called \textit{unit-level variables}.

\Cref{fig:hcm_confounder} shows a hierarchical causal model we call \textsc{unit confounder}. The confounder variable $U_i$ is a hidden
unit-level variable, while the treatment $A_{ij}$ and outcome $Y_{ij}$ are
observed subunit-level variables. For example, consider a scenario where the
units are schools and the subunits are students within the schools.
The treatment $A_{ij}$ is the amount of tutoring each student receives, and the outcome $Y_{ij}$ is their test score.
Both tutoring and test scores can be impacted by unobserved school-level variables such 
as resources, budget, or teaching philosophy. These quantities are captured by the hidden unit-level confounder $U_i$.

\begin{figure}[t]
\centering
\begin{subfigure}[t]{0.33\textwidth}
\caption{\textsc{unit confounder}} \label{fig:hcm_confounder_intervene}
\centering
\begin{tikzpicture}

\node[obs]                               (y) {$Y_{ij}$};
  \node[obs, left=.8cm of y] (a) {$A_{ij}$};
  \node[latent, above=.8cm of a, xshift=1cm]  (u) {$U_i$};
  \node[factor, above=.5cm of a, fill=black!0] (qa) {$q^{a}_\star$};

\edge {a,u} {y} ; \edge {qa} {a} ;

\plate[dashed] {in} {(a)(y)} {$m$} ;
  \plate {out} {(in)(u)(qa)} {$n$} ;

\end{tikzpicture}
\end{subfigure}
\begin{subfigure}[t]{0.3\textwidth}
\caption{\textsc{unit confounder \& unit interference}} \label{fig:hcm_interfere_intervene}
\centering
\begin{tikzpicture}

\node[obs]                               (y) {$Y_{ij}$};
  \node[obs, left=.8cm of y] (a) {$A_{ij}$};
  \node[latent, above=.8cm of a, xshift=0.8cm]  (u) {$U_i$};
  \node[obs, below=.6cm of a] (z) {$Z_i$};
  \node[factor, above=.5cm of a, fill=black!0] (qa) {$q^{a}_\star$};

\edge {a,u} {y} ; \edge {qa} {a} ;
  \edge {a} {z} ;
  \edge {z} {y} ;

\plate[dashed] {in} {(a)(y)} {$m$} ;
  \plate {out} {(in)(u)(z)} {$n$} ;

\end{tikzpicture}
\end{subfigure}
\begin{subfigure}[t]{0.3\textwidth}
\caption{\textsc{subunit instrument}} \label{fig:hcm_instrument_intervene}
\centering
\begin{tikzpicture}

  \node[obs]                               (y) {$Y_{i}$};
  \node[obs, left=0.8cm of y] (a) {$A_{ij}$};
  \node[obs, left=0.6cm of a] (z) {$Z_{ij}$};
  \node[latent, above=.8cm of a, xshift=.8cm]  (u) {$U_i$};
  \node[factor, above=.5cm of a, fill=black!0] (qa) {$q^{a}_\star$};

  \edge {a,u} {y} ;
  \edge {qa} {a} ;

  \plate[dashed] {in} {(a)(z)} {$m$} ;
  \plate {out} {(in)(u)(y)} {$n$} ;

\end{tikzpicture}
\end{subfigure}
\caption{\textbf{Examples of unconditional soft interventions in HCMs.} In these examples, each of the HCMs of \Cref{fig:hcm_motifs} is intervened on by drawing $A_{ij} \sim q^a_\star(a)$. So, arrows into $A_{ij}$ from other causal variables are removed. } \label{fig:hcm_examples_interventions}
\end{figure}

The \textsc{unit confounder} model in \Cref{fig:hcm_confounder} captures the setup of many existing methods for causal inference from nested data. 
For example, we will see that fixed-effect models can be re-derived from \Cref{fig:hcm_confounder}, through our identification and estimation procedure.
\Cref{fig:hcm_interfere,fig:hcm_instrument} show two other motivating HCM graphs that represent novel scenarios.
Here existing methods for causal inference from nested data do not apply.
Our approach will enable nonparametric identification and inference in these new graphs, as well as many others.

\Cref{fig:hcm_interfere} is called \textsc{unit confounder \& unit interference}. It includes unobserved unit-level confounding. But it further captures unit-level interference: the subunit-level treatment variables ($A_{ij}$) might affect an observable unit-level variable ($Z_i$) that, in turn, affects the outcome of all subunits ($Y_{ij}$). For example, if more students are tutored then there may be more school-wide discussion about academic subjects. This discussion might then lead to better academic performance for everyone in the school.

\Cref{fig:hcm_instrument} is called \textsc{subunit instrument}. Unlike the other two graphs, the outcome variable $Y_i$ is now at the unit level. Further, there is a subunit-level instrument $Z_{ij}$, which affects the treatment $A_{ij}$. For example, $Y_i$ might be a school-level outcome variable, such as whether school $i$ is published in a list of best schools. The instrument $Z_{ij}$ might be a randomly administered incentive for students to enroll in extra tutoring.

We will use these three examples as illustrations in our discussion of HCMs, but they are just examples;
we are interested in HCMs with any graph. See \Cref{fig:ID_exs} and \Cref{fig:nonID_exs} for many more examples.

\parhead{Interventions in hierarchical causal models.} The goal of causal modeling is to study the effect of an intervention. What types of interventions can we consider in an HCM?  One possibility is to ask about the effect of assigning every student to receive $a_\star$ hours of tutoring, setting $a_{ij} = a_\star$ for all $i$ and $j$.  This is a deterministic or \textit{hard} intervention. Another possibility is to ask about the effect of drawing $a_{ij}$ stochastically from a distribution $q_\star^a(a)$, of randomly distributing tutoring hours to the students.  This is an example of a stochastic policy or \textit{soft} intervention~\citep[Chap. 4][]{Pearl2009-fh,Dawid2002-yq,Correa2020-df}.

Graphically, an intervention can disconnect the treatment from its parents. \Cref{fig:hcm_examples_interventions} illustrates stochastic interventions on the three motifs.  Note that these stochastic interventions generalize the deterministic intervention; we write a deterministic intervention as a point-mass $q_\star^a(a) = \delta_{a_\star}(a)$.

\begin{figure}
\centering
	\begin{subfigure}[c]{0.3\textwidth}
\caption{Hierarchical causal model.} \label{fig:hcm_obs_confound}
\centering
\begin{tikzpicture}

  \node[obs]                               (y) {$Y_{ij}$};
  \node[obs, left=1.5cm of y] (a) {$A_{ij}$};
  \node[obs, above=.01cm of a, xshift=1cm] (x) {$X_{ij}$};
  \node[latent, above=1.2cm of a, xshift=1cm]  (u) {$U_i$};
  \node[obs, below=.8cm of a, xshift=1cm] (z) {$Z_i$};

  \edge {a,u} {y} ;
  \edge {u} {a} ;
  \edge {a} {z} ;
  \edge {z} {y} ;
  \edge {u} {x} ;
  \edge {x} {a} ;
  \edge {x} {y} ;
  \edge {x} {z} ;

  \plate[dashed] {in} {(a)(y)(x)} {$m$} ;
  \plate {out} {(in)(u)(z)(x)} {$n$} ;

\end{tikzpicture}
\end{subfigure}
\begin{subfigure}[c]{0.3\textwidth}
\caption{Intervened model.} \label{fig:hcm_obs_confound_intervene}
\centering
\begin{tikzpicture}

  \node[obs]                               (y) {$Y_{ij}$};
  \node[obs, above=.01cm of a, xshift=1cm] (x) {$X_{ij}$};
  \node[obs, left=1.5cm of y] (a) {$A_{ij}$};
  \node[latent, above=1.2cm of a, xshift=1cm]  (u) {$U_i$};
  \node[obs, below=.8cm of a, xshift=1cm] (z) {$Z_i$};
  \node[factor, left=of a, fill=black!0] (qax) {$q^{a\mid x}_\star$};

  \edge {a,u} {y} ;
  \edge {a} {z} ;
  \edge {z} {y} ;
  \edge {u} {x} ;
  \edge {x} {a} ;
  \edge {x} {y} ;
  \edge {x} {z} ;
  \edge {qax} {a} ;

  \plate[dashed] {in} {(a)(y)(x)} {$m$} ;
  \plate {out} {(in)(u)(z)(x)(qax)} {$n$} ;

\end{tikzpicture}
\end{subfigure}
\caption{\textbf{Example of a conditional soft intervention.} The HCM (\Cref{fig:hcm_obs_confound}) is intervened on (\Cref{fig:hcm_obs_confound_intervene}) by drawing $A_{ij} \sim q_\star^{a|x}(a \mid x_{ij})$. In this intervention, the arrow from the parent $X_{ij}$ to $A_{ij}$ remains, but the unobserved confounder $U_i$ no longer affects $A_{ij}$.}
\end{figure}

Finally, we consider more targeted interventions. Suppose the HCM contains an additional subunit variable $X_{ij}$, such as the student's previous grades. The past performance might naturally affect a student's seeking out tutoring ($A_{ij}$) and their test performance ($Y_{ij}$); see \Cref{fig:hcm_obs_confound}. With this model, we can ask about the effect of providing more tutoring to students with low grades, where the tutoring hours are drawn from a fixed conditional distribution $q_\star^{a|x}(a \g x_{ij})$; see \Cref{fig:hcm_obs_confound_intervene}.
This is a \textit{conditional soft} intervention~\citep[Chap. 4][]{Pearl2009-fh,Dawid2002-yq,Didelez2006-sf,Correa2020-df}.

\parhead{A roadmap for formalizing hierarchical causal models.} We
study hierarchical causal models with arbitrary graphs.
\citet{Pearl2009-fh}'s theory of causality involves models at two
levels of descriptive detail: structural causal models and causal
graphical models (or causal Bayesian networks). We will use and build
on these ideas. We first define hierarchical structural causal models,
which describe how each variable is generated according to a
deterministic causal mechanism (\Cref{sec:hscm}). We then derive
hierarchical causal graphical models, which involve stochastic causal
mechanisms (\Cref{sec:hcgm}).

With these formalisms in place, we will turn to identification and
estimation in HCMs (\Cref{sec:identification_problem}). In the HCM
identification problem, we consider infinite data from both units and
subunits. In this setting, we will develop a systematic procedure to
identify the effects of interventions. We will find, for example, that
for each of the HCMs in \Cref{fig:hcm_motifs} and interventions in
\Cref{fig:hcm_examples_interventions}, we can identify the effect of
the treatment $A$ on the outcome $Y$. We provide a detailed theory in
\Cref{sec:theory}. We define our notation where it is first used, but
\Cref{table:notation} also provides a central reference.

 \section{Hierarchical Structural Causal Models}
\label{sec:hscm}

We begin by defining the structural equations of hierarchical causal models, extending classical ``flat'' structural causal models by introducing subunit-level variables that lie inside an inner plate.  In flat structural causal models, we have unit-level variables, affected by unit-level causes.  In hierarchical structural causal models, we have both unit-level and subunit-level variables. Each type of variable can affect and be affected by unit-level and subunit-level causes.

\subsection{Structural causal models}
\label{sec:intro-scm}

We first review flat structural causal models~\citep{Pearl2009-fh,Peters2017-ww}.  In a \textit{structural causal model} (SCM), each causal variable is generated through a deterministic function of other causal variables and independent noise.

\Cref{fig:flat_subunit_gen} shows a causal graph. The corresponding structural causal model is
\begin{align}
  \label{eqn:ex-scm}
  \begin{split}
    \gamma_i^{x} \sim \pr(\gamma^{x}) \quad & \,\,\,\,\,\,\, x_i = \f^x(\gamma^{x}_i)\\
    \gamma_i^{\bar{a}} \sim \pr(\gamma^{\bar{a}}) \,\,\,\,\,\,&\,\,\,\,\,\,\, \bar{a}_i = \f^{\bar{a}}(\gamma^{\bar{a}}_i)\\
    \gamma_i^{\bar{y}} \sim \pr(\gamma^{\bar{y}}) \,\,\,\,\,\,&\,\,\,\,\,\,\, \bar{y}_i = \f^{\bar{y}}(\bar{a}_i, x_i, \gamma_i^{\bar{y}}),
  \end{split}
\end{align}
for all $i \in \{1, \ldots, n\}$. The random variables $X_i$, $\bar{A}_i$ and $\bar{Y}_i$ are called \textit{endogenous} causal variables; the variables $\gamma^{x}_i, \gamma^{a}_i, \gamma^{\bar{y}}_i$ are called \textit{exogenous} noise variables; the deterministic functions $\f^{x}, \f^{\bar{a}}, \f^{\bar{y}}$ are called \textit{mechanisms}.

Each endogenous variable depends on a set of direct causes through its mechanism. Some direct causes are other causal variables in the model---the parents in the causal graph.  The other direct cause is the noise variable, which accounts for the remaining unobserved factors that influence the endogenous variable. In \Cref{fig:flat_subunit_gen}, suppose $\bar{Y}_i$ represents average test scores at a school, $\bar{A}_i$ represents average studying hours, and $X_i$ represents teacher quality. The value of the score $\bar{Y}_i=\bar{y}_i$ is determined through the mechanism $\f^{\bar{y}}(\bar{a}_i, x_i, \gamma_i^{\bar{y}})$. It is a function of teacher quality, tutoring hours, and its noise variable $\gamma_i^{\bar{y}}$. Here the ``noise'' might account for textbook choice, school budget, cafeteria menu, and other unobserved factors which affect the test score.

A requirement of an SCM is that the noise variables are i.i.d. across units and independent of one another, and that each appears as an argument in a single mechanism.  If an unobserved cause affects more than one causal variable, i.e., it is a \textit{confounder}, then we must include it in the model as an unobserved causal variable.

Given an SCM, we can describe the effects of a hypothetical intervention by modifying its structural equations.  For example, consider an unconditional soft intervention on $\bar{A}$, where it is drawn from $q^{\bar{a}}_\star$. This intervention corresponds to the following modified SCM:
\begin{align}
  \label{eqn:ex-scm-intervene}
  \begin{split}
    \gamma_i^{x} \sim \pr(\gamma^{x}) \,\,\,\,\,\,
    &\,\,\,\,\,\,\, x_i = \f^x(\gamma^{x}_i)\\
    &\,\,\,\,\,\,\,\bar{A}_i \sim q_\star^{\bar{a}}(\bar{a})\\
    \gamma_i^{\bar{y}} \sim \pr(\gamma^{\bar{y}}) \,\,\,\,\,\,
    &\,\,\,\,\,\,\, \bar{y}_i = \f^{\bar{y}}(\bar{a}_i, x_i, \gamma_i^{\bar{y}}).
  \end{split}
\end{align}
We can form a hard intervention on $\bar{a}$ with a point-mass
$q_\star^{\bar{a}} = \delta_{\bar{a}_\star}$.

\subsection{Hierarchical structural causal models} \label{sec:hscm-intro}

\begin{figure}[t]
\centering
\begin{subfigure}[t]{0.4\textwidth}
\caption{Flat causal model.} \label{fig:flat_subunit_gen}
\centering
\begin{tikzpicture}

\node[obs]                               (y) {$\bar{Y}_{i}$};
  \node[obs, left=.8cm of y] (z) {$\bar{A}_{i}$};
  \node[obs, below=.4cm of z, xshift=.5cm] (x) {$X_i$};

\edge {x,z} {y} ;

\plate {out} {(x)(y)(z)} {$n$} ;

\end{tikzpicture}
\end{subfigure}
\begin{subfigure}[t]{0.4\textwidth}
\caption{Detailed view of the SCM for $\bar{Y}_i$.} \label{fig:flat_subunit_gen_detailed}
\centering
\begin{tikzpicture}

\node[obs] (y) {$\bar{Y}_{i}$};
  \node[latent, below=.6cm of y, color=white, text=black] (gy) {$ \gamma^{\bar{y}}_i$};
  \node[obs, left=.8cm of y] (z) {$\bar{A}_{i}$};
  \node[obs, below=.4cm of z, xshift=.5cm] (x) {$X_i$};

  \path (x) edge [very thick, ->]  (y) ;
  \path (x) edge [color=white, thick] (y) ;
  \path (z) edge [very thick, ->]  (y) ;
  \path (z) edge [color=white, thick] (y) ;
  \path (gy) edge [very thick, ->]  (y) ;
  \path (gy) edge [color=white, thick] (y) ;

\plate {out} {(x)(y)(z)(gy)} {$n$} ;

\end{tikzpicture}
\end{subfigure}	
\caption{\textbf{Generating variables in structural causal models.} \Cref{fig:flat_subunit_gen} shows a flat causal model. In the corresponding SCM, $\bar{Y}_i$ is generated from its parents $\bar{A}_i,X_i$ and an exogenous noise variable $\gamma^{\bar{y}}_i$ via a deterministic mechanism $\f^{\bar{y}}$ (\Cref{fig:flat_subunit_gen_detailed}). We depict deterministic mechanisms with double arrows.} \label{fig:flat_scm}
\end{figure}

\begin{figure}[t]
\centering
\begin{subfigure}[t]{0.4\textwidth}
\caption{HCM with subunit-level $Y_{ij}$.} \label{fig:hcm_subunit_gen}
\centering
\begin{tikzpicture}

\node[obs]                               (y) {$Y_{ij}$};
  \node[obs, left=1cm of y] (z) {$A_{ij}$};
  \node[obs, below=.6cm of z, xshift=.5cm] (x) {$X_i$};

\edge {x,z} {y} ;

\plate[dashed] {in} {(y)(z)} {$m$};
  \plate {out} {(x)(in)} {$n$} ;

\end{tikzpicture}

\end{subfigure}
\begin{subfigure}[t]{0.4\textwidth}
\caption{Detailed view of the HSCM for $Y_{ij}$.}
\label{fig:hcm_subunit_gen_detailed}
\centering
\begin{tikzpicture}

\node[obs] (y) {$Y_{ij}$};
  \node[latent, below=.8cm of y, color=white, text=black] (gy) {$\gamma^{y}_i$};
  \node[latent, above=.3cm of y, xshift=-1cm, color=white, text=black] (ey) {$\epsilon^{y}_{ij}$};
  \node[obs, left=1cm of y] (z) {$A_{ij}$};
  \node[obs, below=.6cm of z, xshift=.5cm] (x) {$X_i$};
  \node[factor, right=.03cm of y, fill=black!0] (d) {\,};

    \path (x) edge [very thick, ->]  (y) ;
  \path (x) edge [color=white, thick] (y) ;
  \path (z) edge [very thick, ->]  (y) ;
  \path (z) edge [color=white, thick] (y) ;
  \path (gy) edge [very thick, ->]  (y) ;
  \path (gy) edge [color=white, thick] (y) ;
  \path (ey) edge [very thick, ->]  (y) ;
  \path (ey) edge [color=white, thick] (y) ;

\plate[dashed] {in} {(y)(z)(ey)(d)} {$m$};
  \plate {out} {(x)(in)(gy)} {$n$} ;

\end{tikzpicture}

\end{subfigure}

\begin{subfigure}[t]{0.4\textwidth}
\caption{HCM with unit-level $Y_{i}$.} \label{fig:hcm_unit_gen}
\centering
\begin{tikzpicture}

\node[obs]                               (y) {$Y_{i}$};
  \node[obs, left=1cm of y] (z) {$A_{ij}$};
  \node[obs, below=.6cm of z, xshift=.5cm] (x) {$X_i$};

\edge {x,z} {y} ;

\plate[dashed] {in} {(z)} {$m$};
  \plate {out} {(x)(in)(y)} {$n$} ;

\end{tikzpicture}
\end{subfigure}
\begin{subfigure}[t]{0.4\textwidth}
\caption{Detailed view of the HSCM for $Y_{i}$.} \label{fig:hcm_unit_gen_detailed}
\centering
\begin{tikzpicture}

\node[obs] (y) {$Y_{i}$};
  \node[latent, below=.8cm of y, color=white, text=black] (gy) {$\gamma^{y}_i$};
  \node[latent, above=.3cm of y, xshift=-1cm, color=white, text=black] (ey) {$\epsilon^{y}_{ij}$};
  \node[obs, left=1cm of y] (z) {$A_{ij}$};
  \node[obs, below=.8cm of z, xshift=.6cm] (x) {$X_i$};

  \path (x) edge [very thick, ->]  (y) ;
  \path (x) edge [color=white, thick] (y) ;
  \path (z) edge [very thick, ->]  (y) ;
  \path (z) edge [color=white, thick] (y) ;
  \path (gy) edge [very thick, ->]  (y) ;
  \path (gy) edge [color=white, thick] (y) ;
  \path (ey) edge [very thick, ->]  (y) ;
  \path (ey) edge [color=white, thick] (y) ;

\plate[dashed] {in} {(z)(ey)} {$m$};
  \plate {out} {(x)(in)(gy)(y)} {$n$} ;

\end{tikzpicture}
\end{subfigure}

\caption{\textbf{Generating subunit and unit variables in hierarchical structural causal models.} \Cref{fig:hcm_subunit_gen} and \Cref{fig:hcm_unit_gen} show HCMs. In the corresponding HSCMs, the subunit variable $Y_{ij}$ (\Cref{fig:hcm_subunit_gen_detailed}) and unit variable $Y_i$ (\Cref{fig:hcm_unit_gen_detailed}) are generated from their parents $A_{ij},X_i$ and from exogenous unit and subunit noise variables $\gamma_i^{y},\epsilon^y_{ij}$ via a deterministic mechanism $\f^{y}$.}

\label{fig:hcm_examples}

\end{figure}
 
While \Cref{fig:flat_subunit_gen} shows a flat causal graph, \Cref{fig:hcm_subunit_gen} shows a hierarchical causal graph, a model of $n$ units where each one contains $m$ subunits.  In this graph, the variable $X_i$ is an endogenous unit-level variable; the variables $A_{ij}$ and $Y_{ij}$ are endogenous subunit-level variables. For example, suppose $\bar{A}_i$ and $\bar{Y}_i$ are per-group averages in the flat model of \Cref{fig:flat_subunit_gen}.  Then $A_{ij}$ and $Y_{ij}$ from \Cref{fig:hcm_subunit_gen} might be disaggregated variables, the individual values that formed the averages.

Given a hierarchical causal graph, how do we write its \textit{hierarchical structural causal model} (HSCM)?  Flat structural causal models only contain unit variables.  Now we must account for unit variables and subunit variables.

\parhead{Subunit-level variables}  We first describe how an HSCM generates endogenous subunit variables.  In a flat structural causal model, endogenous unit variables are affected by other unit-level variables, including their endogenous parents and exogenous noise.  In an HSCM, endogenous subunit variables can be affected by both unit-level and subunit-level variables.  These can include both endogenous parents and exogenous noise.

Consider the graph in \Cref{fig:hcm_subunit_gen}. The HSCM posits that $Y_{ij}$ is generated as (\Cref{fig:hcm_subunit_gen_detailed}),
\begin{align} \label{eqn:hscm_subunit_gen}
  \gamma_i^y \sim \pr(\gamma^y)
  \quad \quad
  \epsilon_{ij}^y \sim \pr(\epsilon^y)
  \quad \quad
  y_{ij} = \f^y(x_i, \gamma_i^y, a_{ij}, \epsilon^y_{ij}).
\end{align}
In this equation, the subunit-level variable $Y_{ij}$ depends on its endogenous unit-level parent $X_i$ and exogenous unit-level noise $\gamma_i^y$.  It also depends on its endogenous subunit-level parent $A_{ij}$ and exogenous subunit-level noise $\epsilon_{ij}^y$.  Since $Y_{ij}$ is specific to subunit $j$, its subunit-level parents must be in the same subunit, that is, $A_{ij'}$ and $\epsilon_{ij'}^y$ for $j' \neq j$ do not enter into the mechanism for $Y_{ij}$.

Continuing the running example, suppose $Y_{ij}$ is the test performance of student $j$ in school $i$.  Then, $X_i$ might describe the budget of their school (an endogenous unit-level cause) and $A_{ij}$ might describe the number of hours of tutoring they received (an endogenous subunit-level cause).  The unit level noise variable $\gamma_i^y$ accounts for unobserved school-level causes, such as the school's learning environment, the textbooks on their syllabi, or the quality of their teachers.  The subunit level noise variable $\epsilon_{ij}^y$ accounts for unobserved student-level causes, such as the student's academic interests or what they had for breakfast the morning of the test.

Notice that for subunit variables, unlike in a flat model, there are two sources of noise. The unit noise $\gamma_i^{y}$ is shared across all values of $Y_{ij}$ in unit $i$. The subunit noise $\epsilon_{ij}$ is involved only in determining $Y_{ij}$. More generally, all the unit-$i$ subunit variables $Y_{i1}, \ldots, Y_{im}$ are affected by the same unit level variables, the observed $X_i$ and the unobserved unit-level noise $\gamma_i^y$.  This pattern of dependence helps capture the intuition that subunits within a unit are more similar to each other than they are to subunits in other units.

Also note the subunit noise $\epsilon_{ij}^y$ is independent across subunits.
This ensures that the subunits within each unit are exchangeable, just like in hierarchical probabilistic models with inner plates.

\parhead{Unit-level variables}  We now describe how the HSCM generates unit-level endogenous variables.  In a flat SCM, endogenous variables are only affected by other unit-level variables.  In an HSCM, unit variables can also be affected by subunit variables.

\Cref{fig:hcm_unit_gen} shows a unit-level variable $Y_i$ that depends on subunit variables $A_{ij}$. Its HSCM posits that $Y_i$ is generated as (\Cref{fig:hcm_unit_gen_detailed}),
\begin{align} \label{eqn:hscm_unit_gen} \gamma_i^y \sim \pr(\gamma^y)
  \quad \quad
    \epsilon_{ij}^y \sim \pr(\epsilon^y)
  \quad \quad
    y_{i} = \f^y(x_i, \gamma_i^y, \{(a_{ij}, \epsilon^y_{ij})\}_{j=1}^m).
\end{align}
In this equation, the unit-level variable $Y_i$ depends on its unit-level parent $X_i$ and exogenous unit-level noise $\gamma_i^y$.  It also depends on its subunit-level parents $A_{ij}$ and exogenous subunit-level noise $\epsilon_{ij}^y$, for $j=1, \ldots, m$.  Note $Y_i$ can depend on all the subunits within the unit.

For example, suppose $Y_i$ indicates whether school $i$ is published in a list of best schools. Then $X_i$ might describe school budget and $A_{ij}$ might describe student tutoring hours.  The noise $\gamma_i^y$ accounts for unobserved unit-level causes of inclusion in the list (e.g., teacher quality, school location) while $\epsilon^y_{ij}$ accounts for unobserved subunit-level causes (e.g., student extracurricular activities, student achievements).

In an HSCM, we require that the dependence of unit-level variables $Y_i$ on subunit variables $A_{ij}$ and $\epsilon^y_{ij}$ is expressed as a dependence on a set of $m$ items $\{(a_{ij}, \epsilon^y_{ij})\}_{j=1}^m$, so that the value of $\f^y$ does not depend on the order of the elements of this set.\footnote{More precisely, we use $\{x_j\}_{j=1}^m$ to denote a \textit{multiset} or \textit{bag}, as identical elements are allowed to appear more than once, e.g. if $x_j = x_{j'}$ for $j' \neq j$.}
This assumption reflects the idea that subunits are exchangeable within each unit, as in hierarchical probabilistic models with inner plates.
Intuitively, the requirement ensures there are no \textit{a priori} privileged subunits.  For example, the school's appearance on the ``best schools'' list $Y_i$ cannot just depend on how much the first student in the data is tutored, $A_{i1}$.

\subsection{Theory}
\label{sec:general-hscm}

In this section, we propose a formal definition of hierarchical structural causal models. We clarify what is required to specify them, and discuss further their relationship to flat structural causal models.
To see concrete examples of HSCMs, skip to \Cref{sec:ex-hscm}.

An HSCM has endogenous variables $X^1, \ldots, X^V$, each generated according to its mechanism $\f^1,\ldots,\f^V$. The variables are ordered causally, such that $X^v$ can only depend on $X^1, \ldots, X^{v-1}$.  Let $\mathcal{V} = \{1, \ldots, V\}$. Some variables $X_{ij}^v$ are subunit-level, and they fall inside an inner plate. We denote their coordinate $\mathcal{S} \subseteq \mathcal{V}$ and they are indexed by both the unit $i$ and subunit $j$.  The rest of the variables $X_i^v$ are unit-level, and fall outside the inner plate. We denote their coordinates $\mathcal{U}$, where $\mathcal{S} \cup \mathcal{U} = \mathcal{V}$; they are indexed only by their unit $i$. Let $\pa(v) \subseteq \{1, \ldots, v-1\}$ denote the indices of the parents of $X^v$ in the graph.  The parents that are subunit-level are denoted $\pa_{\mathcal{S}}(v)$; the parents that are unit-level are denoted $\pa_{\,\mathcal{U}}(v)$.

Each endogenous variable $X^v$ depends on parents, unit-level noise, and subunit-level noise. For the subunit-level variables $X^v_{ij}$, they depend on unit-level variables $X_{i}^{\pa_\mathcal{U}(v)}$, unit-level noise $\gamma_i^v$, and the subunit variables from subunit $j$, specifically, subunit-level noise $\epsilon^v_{ij}$ and subunit-level parents $X^{\pa_\mathcal{S}(v)}_{ij}$. For the unit-level variables $X^v_{i}$, they depend on other unit-level variables $X_{i}^{\pa_\mathcal{U}(v)}$ and $\gamma_i^v$, and \textit{all} the subunit variables $\epsilon^v_{ij}$ and parents $X^{\pa_\mathcal{S}(v)}_{ij}$ for $j \in \{1, \ldots, m\}$.

Thus we define an HSCM as follows.

\begin{definition}[Hierarchical structural causal model] \label{def:hscm} A hierarchical structural causal model (HSCM) $\mathcal{M}^{\scm}$ is defined by (1) a directed acyclic graph $\mathcal{G}$, (2) a set of endogenous variables $X^{\mathcal{V}}$, of which $X^{\mathcal{S}}$ are subunit-level, (3) probability distributions $\pr(\gamma^v)$ and $\pr(\epsilon^v)$ over unit-level and subunit-level noise variables for all $v \in \mathcal{V}$, and (4) a set of mechanisms $\f^{\mathcal{V}}$.  Each endogenous variable $X^v$ for $v \in \mathcal{V}$ is generated as,
\begin{equation} \label{eqn:hscm}
\begin{split}
\gamma_i^v &\sim \pr(\gamma^v)\\
\epsilon_{ij}^v &\sim \pr(\epsilon^v)\\
x^v_{ij} &= \f^v(x^{\pa_\mathcal{U}(v)}_i, \gamma_i^v, x^{\pa_\mathcal{S}(v)}_{ij}, \epsilon^v_{ij}) \,\,\,\,\,\,\,\,\,\,\,\,\,\,\,\,\,\,\,\,\,\,\,\,\,\, \text{ if } v \in \mathcal{S}\\
x^v_i &= \f^v(x^{\pa_\mathcal{U}(v)}_i, \gamma_i^v, \{(x^{\pa_\mathcal{S}(v)}_{ij}, \epsilon^v_{ij})\}_{j=1}^m)  \,\,\,\,\,\, \text{ if } v \in \mathcal{U},
\end{split}
\end{equation}
for $j \in \{1, \ldots, m\}$ and $i \in \{1, \ldots, n\}$.
\end{definition}
This gives an axiomatic definition for hierarchical causal models \citep[Chap. 7]{Pearl2009-fh}.

With this class of models in place, we now describe the interventions we will consider.  Our analysis focuses on hard interventions on unit variables, and soft interventions on subunit variables.  In particular, we study soft interventions that may condition on parent subunit variables, though not unit variables.
It is possible to define other types of interventions in HSCMs, such as soft interventions on subunit variables. But these are outside the scope of the identification and estimation methods we develop in this paper, and are left for future work.
\begin{definition}[HSCM after intervention] \label{def:intervention} An
  intervention $\Delta$ on a hierarchical structural causal model is
  defined by (1) a set of variables
  $\mathcal{I} \subseteq \mathcal{V}$ that are intervened on, (2) the
  values $\{x_\star^{v}: v \in \mathcal{I} \cap \mathcal{U}\}$ that the
  unit-level variables are set to, and (3) the distributions
  $\{q^{v \mid \pa_\mathcal{S}(v)}_\star(x^v \mid x^{\pa_\mathcal{S}(v)}): v \in \mathcal{I} \cap \mathcal{S}\}$ that the subunit-level
  variables are drawn from.  
Post-intervention, the mechanisms generating the variables $X^v: v \in \mathcal{I}$ are,
  \begin{equation}
  \begin{split}	
    X^v_{ij} &\sim  q^{v \mid \pa_\mathcal{S}(v)}_\star(x^v \mid x^{\pa_\mathcal{S}(v)}_{ij}) \quad \text{ if } v \in  \mathcal{S} \cap \mathcal{I} \\
    x^v_i & = x_\star^v \quad \text{ if } v \in \mathcal{U} \cap \mathcal{I},
  \end{split}
  \end{equation}
  for $j \in \{1, \ldots, m\}$ and $i \in \{1, \ldots, n\}$.
\end{definition}
\noindent Note this class of interventions includes hard interventions on subunit variables, which correspond to the special case of using a fixed delta mass as a soft intervention. 

Which variables should be included in an HSCM? In a flat SCM, any variable that has more than one child -- i.e. a confounder -- must be included as an endogenous variable. Variables with one or no children, by contrast, can be marginalized out of the model, soaked into their noise variables and mechanisms~\citep[see e.g.][Def. 5,6]{Spirtes2010-cu,Richardson2002-sp,Janzing2022-rn}. 
In an HSCM, unlike an SCM, we cannot marginalize out any variable with only one child. Rather, we must include in an HSCM those single-child unit-level variables for which both their child and a parent are subunit-level; see \Cref{appx:marginalization}.
The variable $Z$ in the ``interference'' graph of \Cref{fig:hcm_interfere} is an example of such an ``interferer'' variable. Just as we must include confounders, whether observed or not, in a flat SCM, we must include interferers (as well as confounders) in an HSCM.

One might ask why the distribution of subunit noise $\pr(\epsilon^v)$ does not vary across units.  In fact, \Cref{def:hscm} can cover such situations without loss of generality; variation in $\pr(\epsilon^v)$ can be absorbed into the unit noise $\gamma_i^v$ and mechanism $\f^v$. One may also ask why subunit noise is included in the equation for generating unit variables. It is true the subunit noise does not increase the model's expressivity in describing observational and interventional distributions (see \Cref{sec:hcgm_ex_derivation} below). But it does increase the model's expressivity in describing counterfactual distributions (\Cref{apx:counterfactuals}).

\paragraph{Relationship to classical SCMs} How does the HSCM formalism relate to classical SCMs? 

First, HSCMs add subunit-level variables to SCMs. We recover a classical SCM if there are no subunit-level variables, i.e. if $\mathcal{S} = \emptyset$. (We can absorb the subunit noise $\{\epsilon_{ij}^v\}_{j=1:m}$ into the unit noise $\gamma_i^v$ in \Cref{eqn:hscm}.)
Second, we recover a classical SCM if there is no variability among subunits. That is, if $\epsilon^v$ is constant for all $i$ and $j$. (Absorb $\epsilon^v$ into $f^v$ in \Cref{eqn:hscm}.)
The resulting model has the same graph as the HSCM, but with the inner plate erased.
Third, we recover a classical SCM if there is just one subunit. That is, if $m = 1$. (Absorb $\epsilon_{i1}^v$ into $\gamma_{i1}^v$ in \Cref{eqn:hscm}.)
Fourth, we recover a flat SCM if there are no unit-level variables and no unit-level variability.
That is, if $\mathcal{U} = \emptyset$ and $p(\gamma^{\mathcal{V}}) = \delta_{\tilde \gamma}$. (Absorb $\gamma_i^v = \tilde{\gamma}^v$ into $f^v$ in \Cref{eqn:hscm}.)
From this perspective, HSCMs allow for variation in causal mechanisms across different units.

Finally, one might ask also how an HSCM model relates to an SCM with the same
graph, but where the inner plate is expanded
(\Cref{fig:expanded_inner_plate_ex}). The difference is that the HSCM
places the restriction that each subunit's variables are generated by
the same mechanism, i.e. the mechanism generating $a_{i1}$ from $z_{i1}$ is the same as that generating $a_{i2}$ from $z_{i2}$. Further, the mechanism for each unit variable must be invariant to the ordering of its parent subunit variables and subunit noise.

\subsection{Examples of hierarchical structural causal models} \label{sec:ex-hscm}

We now return to the three examples of hierarchical causal models in \Cref{fig:hcm_motifs}. For each, we detail the corresponding hierarchical structural causal model.

The \textsc{unit confounder} graph in \Cref{fig:hcm_confounder} has a unit-level variable $U_i$ and subunit-level variables $A_{ij}$ and $Y_{ij}$. It corresponds to the following HSCM:
\begin{align}
  \label{eqn:ex-hscm}
\begin{split}
\gamma_i^{u} \sim \pr(\gamma^{u}) \,\,\,\,\,\,\,\,\,\,\,\epsilon_{ij}^{u} &\sim \pr(\epsilon^{u})\,\,\,\,\,\,\,\,\,\,\,\,u_i = \f^u(\gamma_i^{u}, \{\epsilon_{ij}^u\}_{j=1}^m)\\
\gamma_i^{a} \sim \pr(\gamma^{a}) \,\,\,\,\,\,\,\,\,\,\,\epsilon_{ij}^{a} &\sim \pr(\epsilon^{a}) \,\,\,\,\,\,\,\,\,\,\, a_{ij} = \f^a(u_i, \gamma^{a}_i, \epsilon_{ij}^{a})\\
\gamma_i^{y} \sim \pr(\gamma^{y}) \,\,\,\,\,\,\,\,\,\,\, \epsilon_{ij}^{y} &\sim \pr(\epsilon^{y}) \,\,\,\,\,\,\,\,\,\,\, y_{ij} = \f^y(u_i, \gamma_i^{y}, a_{ij}, \epsilon_{ij}^{y}),
\end{split}
\end{align}
for all $i \in \{1, \ldots, n\}$ and $j \in \{1, \ldots, m\}$.  Note in some scenarios, subunit noise is unlikely to contribute to $U_i$; for example, if $U_i$ consists only of factors that are determined before the start of the school year, they are unlikely to depend on student-level variables.  In this special case, $\f^u$ will not depend on any subunit noise $\epsilon_{ij}^u$, and so $u_i = \f^u(\gamma_i^u)$.

The \textsc{unit confounder \& unit interference} graph in \Cref{fig:hcm_interfere} includes a unit-level variable, $Z_{i}$, with an observed subunit-level parent, $A_{ij}$. It corresponds to the following HSCM:
\begin{align} \label{eqn:ex-hscm-interfere}
\begin{split}
\gamma_i^{u} \sim \pr(\gamma^{u}) \,\,\,\,\,\,\,\,\,\,\,\epsilon_{ij}^{u} &\sim \pr(\epsilon^{u})\,\,\,\,\,\,\,\,\,\,\,\,u_i = \f^u(\gamma_i^{u}, \{\epsilon_{ij}^u\}_{j=1}^m)\\
\gamma_i^{a} \sim \pr(\gamma^{a}) \,\,\,\,\,\,\,\,\,\,\,\epsilon_{ij}^{a} &\sim \pr(\epsilon^{a}) \,\,\,\,\,\,\,\,\,\,\, a_{ij} = \f^a(u_i, \gamma^{a}_i, \epsilon_{ij}^{a})\\
\gamma_i^{z} \sim \pr(\gamma^{z}) \,\,\,\,\,\,\,\,\,\,\,\epsilon_{ij}^{z} &\sim \pr(\epsilon^{z}) \,\,\,\,\,\,\,\,\,\,\,\, z_i = \f^z(\gamma_i^{z}, \{(a_{ij}, \epsilon_{ij}^{z})\}_{j=1}^m)\\
\gamma_i^{y} \sim \pr(\gamma^{y}) \,\,\,\,\,\,\,\,\,\,\, \epsilon_{ij}^{y} &\sim \pr(\epsilon^{y}) \,\,\,\,\,\,\,\,\,\,\, y_{ij} = \f^y(z_i, u_i, \gamma_i^{y}, a_{ij}, \epsilon_{ij}^{y}),
\end{split}
\end{align}
for all $i \in \{1, \ldots, n\}$ and $j \in \{1, \ldots, m\}$.  In this model, different subunits can indirectly impact one another: for $j' \neq j$, $A_{ij}$ does not directly impact $Y_{ij'}$ via $\f^y$, but $A_{ij}$ does impact $Z_i$, which in turn impacts $Y_{ij'}$.  Hierarchical structural causal models can thus describe a special case of interference or spillover between subunits, namely interference that goes through a unit-level variable~\citep{Hudgens2008-hn,Tchetgen_Tchetgen2012-lm}.
(See \Cref{apx:interference} for more detailed connections to notions of ``clustered" interference.)

Finally, the \textsc{subunit instrument} graph in \Cref{fig:hcm_instrument} corresponds to the following HSCM:
\begin{equation} \label{eqn:ex-hscm-instrument}
\begin{split}
\gamma_i^{u} \sim \pr(\gamma^{u}) \,\,\,\,\,\,\,\,\,\,\,\epsilon_{ij}^{u} &\sim \pr(\epsilon^{u})\,\,\,\,\,\,\,\,\,\,\,\,u_i = \f^u(\gamma_i^{u}, \{\epsilon_{ij}^u\}_{j=1}^m)\\
\gamma_i^{z} \sim \pr(\gamma^{z}) \,\,\,\,\,\,\,\,\,\,\,\epsilon_{ij}^{z} &\sim \pr(\epsilon^{z}) \,\,\,\,\,\,\,\,\,\,\,\, z_{ij} = \f^z(\gamma^{z}_i, \epsilon_{ij}^{z})\\
\gamma_i^{a} \sim \pr(\gamma^{a}) \,\,\,\,\,\,\,\,\,\,\,\epsilon_{ij}^{a} &\sim \pr(\epsilon^{a}) \,\,\,\,\,\,\,\,\,\,\,  a_{ij} = \f^a(u_i, \gamma^{a}_i, z_{ij}, \epsilon_{ij}^{a})\\
\gamma_i^{y} \sim \pr(\gamma^{y}) \,\,\,\,\,\,\,\,\,\,\, \epsilon_{ij}^{y} &\sim \pr(\epsilon^{y}) \,\,\,\,\,\,\,\,\,\,\, y_i = \f^y(u_i, \gamma_i^{y}, \{(a_{ij}, \epsilon_{ij})\}_{j=1}^m),
\end{split}
\end{equation}
for all $i \in \{1, \ldots, n\}$ and $j \in \{1, \ldots, m\}$.

 \section{Hierarchical Causal Graphical Models}
\label{sec:hcgm}

While structural causal models (SCMs) describe causal processes using
deterministic mechanisms and exogenous noise, \textit{causal graphical models}
(CGMs), also known as \textit{causal Bayesian networks} or \textit{agnostic causal DAGs}, describe causal processes with stochastic mechanisms~\citep{shalizi2013advanced,Lauritzen2001-vi,Bareinboim2021-ro,Richardson2013-yy}. Any classical flat SCM can be
written as a CGM by integrating out the noise. Here we apply the same
idea to develop \textit{hierarchical causal graphical models} (HCGMs), deriving them from hierarchical structural causal models (\Cref{def:hscm}).

\subsection{Causal graphical models} We first review how to derive a classical causal graphical model from a classical SCM.  Consider again the flat causal model in \Cref{fig:flat_subunit_gen} and recall the structural equation for $\bar{Y}$ in \Cref{eqn:ex-scm}. We can integrate out the noise $\gamma^{\bar{y}}$ to form a stochastic mechanism describing how $\bar{Y}$ is generated given $X$ and $\bar{A}$,
\begin{align*}
  \bar{Y}_i \sim \pr(\bar{y} \mid x_i, \bar{a}_i) \quad \text{ where } \Prob(\bar{Y} \in \Xi \mid X = x, \bar{A} = \bar{a})
  =
  \int \mathbb{I}\Big(\f^{\bar{y}}(x, \bar{a}, \gamma^{\bar{y}}) \in \Xi \Big) \pr(\gamma^{\bar{y}})\di\gamma^{\bar{y}}.
\end{align*}
Note that this derivation is made possible by the assumption in SCMs that the generative process is \textit{stable} across units, in the sense that the mechanism $\f^{\bar{y}}$ is fixed and the noise $\gamma^{\bar{y}}_i$ is i.i.d. across units.

If we repeat this derivation for all the endogenous variables in an SCM, we find an alternative characterization of the model as a cascade of random variables. The distribution of each, conditional on its parents, is formed by integrating out the random noise from its deterministic mechanism. For example, the model of \Cref{fig:flat_subunit_gen} and \Cref{eqn:ex-scm} corresponds to the following CGM,
\begin{align}
  \label{eqn:cgm}
  \begin{split}
    X_i &\sim \pr(x) \\
    \bar{A}_i &\sim \pr(\bar{a}) \\
    \bar{Y}_i  &\sim \pr(\bar{y} \mid \bar{a}_i, x_i).
  \end{split}
\end{align}

A probabilistic graphical model represents the family of joint distributions that respect the factorization implied by the graph. A causal graphical model additionally implies a distribution under intervention on the underlying SCM. Interventions in a CGM work the same way as in an SCM, where we replace the mechanism of a variable with an intervened mechanism. Consider a soft unconditional intervention on $\bar{A}_i$, where it is drawn from an intervention distribution $q_\star^{\bar{a}}$. This intervention results in the following CGM, \begin{align}
  \begin{split}
    X_i &\sim \pr(x) \\
    \bar{A}_i &\sim q_\star^{\bar{a}}(\bar{a}) \\
    \bar{Y}_i  &\sim \pr(\bar{y} \mid \bar{a}_i, x_i).
  \end{split}
\end{align}
The intervened CGM derives directly from the intervened SCM, again by integrating out the noise. Note while CGMs can describe interventions, they differ from SCMs in that they cannot describe counterfactuals~\citep[Chap. 1]{Pearl2009-fh}. This paper focuses on interventions.

\subsection{Hierarchical causal graphical models} \label{sec:hcgm_ex_derivation}

We described how flat CGMs are derived from flat SCMs; now we apply the same reasoning to derive hierarchical causal graphical models from hierarchical structural causal models.

\parhead{Subunit variables} We first consider the subunit variables. Recall the example graph in \Cref{fig:hcm_subunit_gen} and the structural equation for subunit variable $Y_{ij}$ in \Cref{eqn:hscm_subunit_gen}. We will form the distribution of $Y_{ij}$ in two stages. First, we consider it as a random variable within unit $i$. Then we consider the variation in its conditional distribution across units.

What is the distribution of $Y_{ij}$ \textit{within} its unit $i$? To derive this distribution, we hold the unit variables $\gamma_i^{y}$ and $X_i$ fixed, and marginalize out the subunit noise $\epsilon_{ij}^{y}$. The result is the conditional distribution of $Y_{ij} \g A_{ij}$ for subunits of unit $i$, which we denote $Q_i^{y \mid a}$.  In the running example, $Q_i^{y \mid a}(Y=y \mid A=a)$ describes the probability that a student at school $i$ will receive a test score $y$ after studying for $a$ hours.

Now consider the variation of $Q_i^{y \mid a}$ \textit{across} units. This conditional distribution depends deterministically on the unit noise $\gamma_i^{y}$ as well as the unit-level parents $x_i$.  If we marginalize out $\gamma_i^{y}$, we produce a two-stage generative process for $Y_{ij}$:
\begin{align}
  \label{eqn:hcgm_subunit} \begin{split}
    Q_i^{y \mid a} &\sim \pr\big (q^{y \mid a}\, \big|\, x_i\big) \\
    Y_{ij} &\sim q_i^{y \mid a}(y \mid a_{ij}).
  \end{split}
\end{align}
Notice that $\pr\big (q^{y \mid a}\, \big|\, x_i\big)$ is a
\textit{distribution over distributions}.  It describes how the
conditional distribution of $Y$ given $A$ varies across units, due to
unit-level noise, i.e., unobserved unit-level causes of $Y$.
Continuing the running example,
$\pr\big (q^{y \mid a}\, \big|\, x_i\big)$ tells us how the effectiveness
of the tutoring program at producing good test scores changes across
schools.

More formally, we can derive the conditional distribution of $Q^{y\mid a}$ in two stages. First we define the distribution that marginalizes out the subunit-level noise from a mechanism, holding the unit-level noise fixed. Then we marginalize out the unit-level noise. The result is,
\begin{align*}
  \begin{split}
    [g^{y \mid a}_q(x, \gamma^{y})](Y \in \Xi \mid A= a) &\triangleq \int \mathbb{I}\Big(\f^{y}(x, \gamma^y, a, \epsilon^{y}) \in \Xi \Big) \pr(\epsilon^{y}) \di\epsilon^{y}\\
    \Prob\big(Q^{y \mid a} \in \Pi \, \big|\, X = x\big) &\triangleq \int \mathbb{I}\Big([g^{y \mid a}_q(x, \gamma^{y})] \in \Pi \Big) \pr(\gamma^{y}) \di\gamma^{y}.
  \end{split}
\end{align*}
Here $g^{y \mid a}_q(x, \gamma^{y})$ is a function which takes as input $x$ and $\gamma^{y}$ and returns a distribution $q^{y \mid a}$.  Thus $q^{y \mid a}_i(y \mid a) = [g^{y \mid a}_q(x_i, \gamma^{y}_i)](y \mid a)$.  The second line defines $\pr\big(q^{y \mid a}\, \big|\, x_i\big)$.  Note this derivation is made possible by the HSCM assumption that the generative process is stable across subunits (as well as units), in the sense that the mechanism $\f^y$ is fixed and the noise $\epsilon^y_{ij}$ is i.i.d. across subunits.

Setting aside for a moment the possibility of interventions, \Cref{eqn:hcgm_subunit} takes the form of a hierarchical \textit{probabilistic} model.  First, for each unit $i$, we draw a conditional distribution over subunit variables $Q_i^{y \mid a}$. Second, for each subunit $j$ within each unit $i$, we draw the subunit variable $Y_{ij}$ from $Q_i^{y \mid a}(\cdot \mid a_{ij})$.  The subunit variables $Y_{i1}, \ldots, Y_{im}$ within each unit $i$ are thus similar, as they are drawn from the same distribution $Q_i^{y \mid a}$. Such hierarchical models of grouped data are a mainstay of Bayesian statistics~\citep[e.g.][]{Gelman2006-ft}, and the idea of building hierarchical models by drawing random distributions is at the foundations of Bayesian nonparametric statistics~\citep[e.g.][]{Ghosh2003-yv}.

\parhead{Unit variables} We now turn to the unit-level variables.  Consider the variable $Y$ in the example graph in \Cref{fig:hcm_unit_gen}. Recall from \Cref{eqn:hscm_unit_gen} that $Y$ is generated from $X$ and $A$ as,
\begin{align*}
  \gamma_i^y \sim \pr(\gamma^y) \,\,\,\,\,\,\,\,\,\,\, \epsilon_{ij}^y \sim \pr(\epsilon^y) \,\,\,\,\,\,\,\,\,\,\, y_{i} = \f^y(x_i, \gamma_i^y, \{(a_{ij}, \epsilon_{ij}^y)\}_{j=1}^m).
\end{align*}
Since $Y_i$ is unit-level, we can form the random variable by simultaneously marginalizing out both unit noise $\gamma_i^{y}$ and subunit noise $\epsilon_{i1}^{y}, \ldots, \epsilon_{im}^{y}$. The result is,
\begin{align} \label{eqn:hcgm_unit}
  Y_i \sim \pr(y \mid x_i, \{a_{ij}\}_{j=1}^m).
\end{align}
This form of the conditional stems from the fact that the deterministic mechanism $\f^{y}$ in the HSCM is invariant to permutations of $(a_{i1}, \epsilon_{i1}^y), \ldots, (a_{im}, \epsilon_{im}^y)$ and because $\epsilon_{i1}^{y}, \ldots, \epsilon_{im}^{y}$ are i.i.d.. Thus the stochastic mechanism $\pr(y \mid x_i, \{a_{ij}\}_{j=1}^m)$ must also be permutation invariant, depending only on the set of values $\{a_{ij}\}_{j=1}^m$. \Cref{eqn:hcgm_unit} defines how a unit endogenous variable is generated in a hierarchical causal graphical model.

\parhead{General case} We now define hierarchical causal graphical models in general.
\begin{definition}[Hierarchical causal graphical model] \label{def:hcgm}
Consider a hierarchical structural causal model $\mathcal{M}^{\scm}$ (\Cref{def:hscm}). The corresponding hierarchical causal graphical model $\mathcal{M}^{\cgm}$ has the same graph and endogenous variables, with stochastic mechanisms,
\begin{equation} \label{eqn:hcgm}
\begin{split}
        Q^{v\mid \pa_\mathcal{S}(v)}_i &\sim \pr\big(q^{v\mid \pa_\mathcal{S}(v)} \, \big| \, x^{\pa_\mathcal{U}(v)}_{i}\big) \,\,\,\,\,\,\,\,\,\,\,\,\,\,\,\,\,\,\,\,\,\,\,\,\,\,\,\,\,\,\,\,\,\,\,\,\,\,\,\,\,\,\,\,\,\,\,\, \text{ for } v \in \mathcal{S}\\
        X^{v}_{i j} &\sim q^{v\mid \pa_\mathcal{S}(v)}_i\big(x^{v} \,\big|\, x^{\pa_\mathcal{S}(v)}_{i j} \big) \,\,\,\,\,\,\,\,\,\,\,\,\,\,\,\,\,\,\,\,\,\,\,\,\,\,\,\,\,\,\,\,\,\,\,\,\,\,\,\,\,\,\,\,\, \text{ for } v \in \mathcal{S}\\
X^{v}_i &\sim \pr\big(x^{v} \, \big| \, x^{\pa_\mathcal{U}(v)}_{i}, \{x^{\pa_\mathcal{S}(v)}_{i j} \big\}_{j=1}^m\big) \,\,\,\,\,\,\,\,\,\,\,\,\,\,\,\,\,\,\,\,\,\,\,\,\,\,\text{ for } v \in \mathcal{U},
\end{split}
\end{equation}
for $i \in \{1, \ldots, n\}$ and $j \in \{1, \ldots, m\}$.
\end{definition}
\noindent A derivation of hierarchical causal graphical models from hierarchical structural causal models is given in \Cref{apx:hcgm}.
Note the distribution $Q^{v \mid \pa_\mathcal{S}(v)}$ over a subunit level variable $X^{v}$ describes the subunit-level variable's dependence on its subunit-level ancestors, while the mechanism $\pr\big(q^{v\mid \pa_\mathcal{S}(v)} \, \big| \, x^{\pa_\mathcal{U}(v)}_{i}\big)$ generating $Q^{v \mid \pa_\mathcal{S}(v)}$ only depends on the unit-level ancestors, $X^{\pa_\mathcal{U}(v)}_{i}$.
Interventions in hierarchical causal graphical models work just as in hierarchical structural causal models, with the mechanism for the intervened variable replaced by its intervened value (\Cref{def:intervention}).

\subsection{Examples} \label{sec:hcgm_examples}

We illustrate hierarchical causal graphical models through the three examples in \Cref{fig:hcm_motifs}. Consider the \textsc{unit confounder} graph in \Cref{fig:hcm_confounder}.  It corresponds to the HSCM in \Cref{eqn:ex-hscm}, which becomes the following HCGM:
\begin{align}
  \label{eqn:hcgm_confound}
  \begin{split}
    U_i &\sim \pr(u)\\
    Q^{a}_i &\sim \pr\big(q^{a} \, \big|\, u_i \big) \,\,\,\,\,\,\,\,\,\,\,\,\,\,\,\,\,\,\,\,\,\,\,\,\,\,\,\,\, A_{ij} \sim q_i^{a}(a)\\
    Q^{y \mid a}_i &\sim \pr\big (q^{y\mid a}\, \big|\, u_i\big)\,\,\,\,\,\,\,\,\,\,\,\,\,\,\,\,\,\,\,\,\,\,\,\,\, Y_{ij} \sim q_i^{y\mid a}(y \mid a_{ij}),
  \end{split}
\end{align}
for all $i \in \{1, \ldots, n\}$ and $j \in \{1, \ldots, m\}$. \Cref{fig:hcm_intro_hpgm} depicts this model with the $Q$ variables shown explicitly.\footnote{Notationally, in HCM graphs such as \Cref{fig:hcm_confounder}, we use a dashed rather than a solid line for the inner plate because subunit variables are not conditionally independent given their parent endogenous variables. They are, however, conditionally independent given their $Q$ variable and parent subunit variables. We therefore use a solid line for the inner plate in \Cref{fig:hcm_intro_hpgm}.}

Now turn to the \textsc{unit confounder \& unit interference} graph in \Cref{fig:hcm_interfere}. Its HSCM is in \Cref{eqn:ex-hscm-interfere}. It yields the following HCGM:
\begin{align} \label{eqn:hcgm_interfere}
  \begin{split}
    U_i &\sim \pr(u)\\
    Q^{a}_i &\sim \pr\big(q^{a} \, \big|\, u_i \big) \,\,\,\,\,\,\,\,\,\,\,\,\,\,\,\,\,\,\,\,\,\,\,\,\,\,\,\,\,\,\,\,\,\,\,\,\,\,\,\, A_{ij} \sim q_i^{a}(a)\\
    Z_i &\sim \pr(z \mid \{a_{ij}\}_{j=1}^m)\\
    Q^{y\mid a}_i &\sim \pr\big (q^{y\mid a}\, \big|\, z_i, u_i\big)\,\,\,\,\,\,\,\,\,\,\,\,\,\,\,\,\,\,\,\,\,\,\,\,\,\,\,\, Y_{ij} \sim q_i^{y \mid a}(y \mid a_{ij}),
  \end{split}
\end{align}
for all $i \in \{1, \ldots, n\}$ and $j \in \{1, \ldots, m\}$ (\Cref{fig:hcm_interfere_hpgm}).

Finally, consider the \textsc{subunit instrument} graph of \Cref{fig:hcm_instrument}. Its HSCM is in ~\Cref{eqn:ex-hscm-instrument}. It yields the following HCGM:
\begin{align}
  \label{eqn:ex-hcgm-instrument} \begin{split}
    U_i &\sim \pr(u)\\
    Q^{z}_i &\sim \pr\big(q^{z} \big) \,\,\,\,\,\,\,\,\,\,\,\,\,\,\,\,\,\,\,\,\,\,\,\,\,\,\,\,\,\,\,\,\,\,\,\,\,\,\,\,\,\, Z_{ij} \sim q_i^{z}(z)\\
    Q^{a\mid z}_i &\sim \pr\big (q^{a\mid z}\, \big|\, u_i\big)\,\,\,\,\,\,\,\,\,\,\,\,\,\,\,\,\,\,\,\,\,\,\,\,\,\,\, A_{ij} \sim q_i^{a\mid z}(a \mid z_{ij})\\
    Y_i &\sim \pr(y \mid \{a_{ij}\}_{j=1}^m),\\
  \end{split}
\end{align}
for all $i \in \{1, \ldots, n\}$ and $j \in \{1, \ldots, m\}$ (\Cref{fig:hcm_instrument_hpgm}).

Hierarchical causal graphical models produce the same post-intervention distribution as the hierarchical structural causal model from which they are derived, since the process generating each variable from its parents is unchanged.  A hard intervention on $A$ in the \textsc{subunit instrument} model corresponds to replacing the third line of \Cref{eqn:ex-hcgm-instrument} with $A_{ij} = a_\star$.  A conditional soft intervention given $Z$ uses instead $A_{ij} \sim q_\star^{a \mid z}(a \mid z_{ij})$.
Equivalently, however, we can model either intervention on $A$ as a hard intervention on $Q^{a \mid z}$, which leaves unchanged the expression $A_{ij} \sim q^{a\mid z}(a \mid z_{ij})$. For a hard intervention on $A$, we replace $Q_i^{a \mid z} \sim \pr(q^{a \mid z} \mid u_i)$ with $q_i^{a \mid z} = \delta_{a_\star}$. For a conditional soft intervention, we set $q_i^{a \mid z} = q_\star^{a \mid z}$.
In short, we can describe a soft intervention on a subunit variable as a hard intervention on their underlying $Q$ distribution.

 \section{Identification and Estimation}
\label{sec:identification_problem}

The purpose of causal modeling is to understand the effect of a
hypothetical intervention on the system. A crucial step in this
process is to solve the problem of \textit{causal identification}
(causal ID).

In causal ID we posit a causal model and consider a hypothetical
intervention together with a chosen causal quantity. We assume that some
variables are observed while others are not. We ask: if given an
infinite number of data points from the pre-intervention distribution,
can we calculate the post-intervention distribution over the causal quantity? If we can then
the quantity is identified. If we cannot, for example because of which
variables are unobserved, then the quantity is not identified.

In a hierarchical causal ID problem, we again consider a hypothetical
intervention and assume some variables are observed and some are
hidden. But we now consider infinite data at both the subunit and unit
level. With the data from infinite subunits, we effectively observe
the subunit joint $q_i(x^{\mathcal{S}_\obs})$, which is the joint
distribution of the observable subunit variables within each unit $i$.
Notice that it is a \textit{random distribution}, its randomness
governed by the population distribution over units. With the data from infinite
units, we effectively observe the distribution
$\pr(x^{\mathcal{U}_\obs}, q(x^{\mathcal{S}_\obs}))$, which is the
joint distribution of the observable unit variables and the observable
subunit distributions. 
(There are technical details to these claims;
see \Cref{apx:hier_empirical_convergence}.)

Increasing the number of subunits can change how causal variables are generated in the model.
We will assume the model converges to a stable limit as the number of subunits increases (\Cref{sec:collapse_theory}).
Note this assumption restricts what models can be identified.

We study hierarchical causal ID for the class of interventions in
\Cref{def:intervention}.  In particular, we focus on (a) hard
interventions on unit variables and (b) soft interventions on subunit
variables, which may condition on other subunit variables.  Note, however, these are in some sense both hard interventions on
unit-level variables: \Cref{sec:hcgm_examples} showed that a soft
intervention on a subunit variable is equivalent to a hard
intervention on the $Q$ distribution for that variable, which is at the
unit level.

\begin{figure}[t!]
\centering
\begin{subfigure}[t]{0.24\textwidth}
\caption{} \label{fig:hcm_confounder_transforms}
\centering
\begin{tikzpicture}

\node[obs]                               (y) {$Y_{ij}$};
  \node[obs, left=1cm of y] (a) {$A_{ij}$};
  \node[latent, above=0.4cm of a, xshift=0.8cm]  (u) {$U_i$};

\edge {a,u} {y} ; \edge {u} {a} ;

\plate[dashed] {in} {(a)(y)} {$m$} ;
  \plate {out} {(in)(u)} {$n$} ;

\node[rotate=90] at (-2.7,0.2) {\textsc{unit confounder}};
\node at (-0.8,3) {\textit{Hierarchical causal}};
\node at (-0.8,2.6) {\textit{model (HCM)}};
\draw[white, opacity=0] (current bounding box.north west) rectangle (current bounding box.south east);
\end{tikzpicture}
\end{subfigure}
\begin{subfigure}[t]{0.24\textwidth}
\caption{} \label{fig:hcm_intro_hpgm}
\centering
\begin{tikzpicture}
\node[obs] (y) {$Y_{ij}$};
  \node[obs, left=1cm of y] (a) {$A_{ij}$};
  \node[latent, above=1 of a, xshift=0.8cm]  (u) {$U_i$};
  \node[latent, above=.3 of a, color=blue, fill=blue!0] (qa) {$\color{blue} \scriptstyle Q^{a}_i$};
  \node[latent, above=.3 of y, color=blue, fill=blue!0] (qya) {$\color{blue} \scriptstyle Q^{y \mid a}_i$};

\edge {u} {qya} ;
  \edge {qya} {y} ;
  \edge {u} {qa} ;
  \edge {qa} {a};
  \edge {a} {y};

\plate {ay} {(a)(y)} {$m$} ;
  \plate {ayu} {(ay)(u)} {$n$} ;

\node at (-0.8,3) {\textit{HCGM}}; 
\node at (-0.8,2.6) {\small \textit{w/ $Q$ variables shown}};
\draw[white, opacity=0] (current bounding box.north west) rectangle (current bounding box.south east);

\end{tikzpicture}
\end{subfigure}
\begin{subfigure}[t]{0.24\textwidth}
\caption{} \label{fig:collapse_intro}
\centering
\begin{tikzpicture}

\node[obs]                               (y) {$\scriptstyle Q_i^{y \mid a}$};
  \node[obs, left=1cm of y] (a) {$\scriptstyle Q_{i}^{a}$};
  \node[latent, above=0.4cm of a, xshift=0.8cm]  (u) {$U_i$};

\edge {u} {y} ; \edge {u} {a} ;

\plate {ayu} {(a)(y)(u)} {$n$} ;

\node at (-0.8,2.9) {\textit{Collapsed}};
\draw[white, opacity=0] (current bounding box.north west) rectangle (current bounding box.south east);

\end{tikzpicture}
\end{subfigure}
\begin{subfigure}[t]{0.24\textwidth}
\caption{} \label{fig:augment_intro}
\centering
\begin{tikzpicture}

\node[obs]                               (y) {$\scriptstyle Q_i^{y \mid a}$};
  \node[obs, left=1cm of y] (a) {$\scriptstyle Q_{i}^{a}$};
  \node[latent, above=0.4cm of a, xshift=0.8cm]  (u) {$U_i$};
  \node[obs, below=.2cm of a, xshift=0.8cm]  (qy) {$\scriptstyle Q_i^{y}$};

\edge {u} {y} ; \edge {u} {a} ;
  \path (a) edge [very thick, ->]  (qy) ;
  \path (a) edge [color=white, thick] (qy) ;
  \path (y) edge [very thick, ->]  (qy) ;
  \path (y) edge [color=white, thick] (qy) ;

\plate {ayu} {(a)(y)(u)(qy)} {$n$} ;

\node at (-0.8,2.5) {\textit{Augmented/}};
\node at (-0.8,2.05) {\textit{Marginalized}};
\draw[white, opacity=0] (current bounding box.north west) rectangle (current bounding box.south east);

\end{tikzpicture}
\end{subfigure}

\begin{subfigure}[t]{0.24\textwidth}
\caption{}
\label{fig:hcm_interfere_transforms}
\centering
\begin{tikzpicture}

\node[obs]                               (y) {$Y_{ij}$};
  \node[obs, left=1cm of y] (a) {$A_{ij}$};
  \node[latent, above=0.4cm of a, xshift=0.8cm]  (u) {$U_i$};
  \node[obs, below=.6cm of a] (z) {$Z_i$};

\edge {a,u} {y} ; \edge {u} {a} ;
  \edge {a} {z} ;
  \edge {z} {y} ;

\plate[dashed] {in} {(a)(y)} {$m$} ;
  \plate {out} {(in)(u)(z)} {$n$} ;

\node[rotate=90] at (-3,0) {\textsc{unit confounder \&}};
\node[rotate=90] at (-2.6,0) {\textsc{unit interference}};
\draw[white, opacity=0] (current bounding box.north west) rectangle (current bounding box.south east);
\end{tikzpicture}
\end{subfigure}
\begin{subfigure}[t]{0.24\textwidth}
\caption{} \label{fig:hcm_interfere_hpgm}
\centering
\begin{tikzpicture}
\node[obs] (y) {$Y_{ij}$};
  \node[obs, left=1cm of y] (a) {$A_{ij}$};
  \node[latent, above=1cm of a, xshift=0.8cm]  (u) {$U_i$};
  \node[latent, above=.4 of a, color=blue, fill=blue!0] (qa) {$\color{blue} \scriptstyle Q^{a}_i$};
  \node[obs, below=.6cm of a] (z) {$Z_i$};
  \node[latent, above=.3 of y, color=blue, fill=blue!0] (qya) {$\color{blue} \scriptstyle Q^{y \mid a}_i$};

\edge {u} {qa} ; \edge {u} {qya} ;
  \edge {qa} {a} ;
  \edge {a} {z} ;
  \edge {z} {qya} ;
  \edge {a} {y} ;
  \edge {qya} {y} ;

\plate {ay} {(a)(y)} {$m$} ;
  \plate {ayu} {(ay)(u)(z)(qa)(qya)} {$n$} ;

\draw[white, opacity=0] (current bounding box.north west) rectangle (current bounding box.south east);
\end{tikzpicture}
\end{subfigure}
\begin{subfigure}[t]{0.24\textwidth}
\caption{} \label{fig:collapsed_interfere}
\centering
\begin{tikzpicture}

\node[obs]                               (qya) {$\scriptstyle Q^{y \mid a}_i$};
  \node[obs, left=1.5cm of qya] (qa) {$\scriptstyle Q^{a}_i$};
  \node[obs, left=.4cm of qya] (z) {$Z_i$};
  \node[latent, above=0.4cm of z]  (u) {$U_i$};

\edge {u} {qa, qya} ; \edge {qa} {z} ;
  \edge {z} {qya} ;

\plate {ayu} {(qa)(qya)(u)(z)} {$n$} ;

\draw[white, opacity=0] (current bounding box.north west) rectangle (current bounding box.south east);
\end{tikzpicture}
\end{subfigure}
\begin{subfigure}[t]{0.24\textwidth}
\caption{} \label{fig:augment_interfere}
\centering
\begin{tikzpicture}

\node[obs]                               (qya) {$\scriptstyle Q^{y \mid a}_i$};
  \node[obs, left=1.5cm of qya] (qa) {$\scriptstyle Q^{a}_i$};
  \node[obs, left=.4cm of qya] (z) {$Z_i$};
  \node[latent, above=.4cm of z]  (u) {$U_i$};
  \node[obs, below=.2cm of z] (qy) {$\scriptstyle Q^{y}_i$};

\edge {u} {qa, qya} ; \edge {qa} {z} ;
  \edge {z} {qya} ;
  \path (qa) edge [very thick, ->]  (qy) ;
  \path (qa) edge [color=white, thick] (qy) ;
  \path (qya) edge [very thick, ->]  (qy) ;
  \path (qya) edge [color=white, thick] (qy) ;

\plate {ayu} {(qa)(qy)(qya)(u)(z)} {$n$} ;

\draw[white, opacity=0] (current bounding box.north west) rectangle (current bounding box.south east);
\end{tikzpicture}
\end{subfigure}

\begin{subfigure}[t]{0.24\textwidth}
\caption{} \label{fig:hcm_instrument_transforms}
\centering
\begin{tikzpicture}

  \node[obs]                               (y) {$Y_{i}$};
  \node[obs, left=0.4cm of y] (a) {$A_{ij}$};
  \node[obs, left=0.4cm of a] (z) {$Z_{ij}$};
  \node[latent, above=0.4cm of a, xshift=0.6cm]  (u) {$U_i$};

  \edge {a,u} {y} ;
  \edge {u} {a} ;
  \edge {z} {a} ;

  \plate[dashed] {in} {(a)(z)} {$m$} ;
  \plate {out} {(in)(u)(y)} {$n$} ;

\node[rotate=90] at (-3.1,0.2) {\textsc{subunit instrument}};
\draw[white, opacity=0] (current bounding box.north west) rectangle (current bounding box.south east);

\end{tikzpicture}
\end{subfigure}
\begin{subfigure}[t]{0.24\textwidth}
\caption{} \label{fig:hcm_instrument_hpgm}
\centering
\begin{tikzpicture}

  \node[obs] (y) {$Y_{i}$};
  \node[obs, left=0.3cm of y] (a) {$A_{ij}$};
  \node[obs, left=.4cm of a] (z) {$Z_{ij}$};
  \node[latent, above=1cm of a, xshift=1cm]  (u) {$U_i$};
  \node[latent, above=.35 of z, color=blue, fill=blue!0] (qz) {$\color{blue} \scriptstyle Q^{z}_i$};
  \node[latent, above=.3 of a, color=blue, fill=blue!0] (qaz) {$\color{blue} \scriptstyle Q^{a \mid z}_i$};

  \edge {u} {y} ;
  \edge{u} {qaz} ;
  \edge {qaz} {a} ;
  \edge {qz} {z} ;
  \edge {z} {a} ;
  \edge {a} {y} ;

  \plate {in} {(a)(z)} {$m$} ;
  \plate {out} {(in)(u)(y)(qaz)(qz)} {$n$} ;
\draw[white, opacity=0] (current bounding box.north west) rectangle (current bounding box.south east);
\end{tikzpicture}
\end{subfigure}
\begin{subfigure}[t]{0.24\textwidth}
\caption{} \label{fig:collapse_instrument}
\centering
\begin{tikzpicture}

  \node[obs]                               (y) {$Y_{i}$};
  \node[obs, left=0.8cm of y] (qaz) {$\scriptstyle Q^{a\mid z}_i$};
  \node[obs, below=.2cm of qaz, xshift=0.8cm] (qz) {$\scriptstyle Q^{z}_i$};
  \node[latent, above=1cm of qz]  (u) {$U_i$};

  \edge {u} {qaz, y} ;
  \edge {qaz,qz} {y} ;

  \plate {out} {(qz)(qaz)(u)(y)} {$n$} ;
\draw[white, opacity=0] (current bounding box.north west) rectangle (current bounding box.south east);
\end{tikzpicture}
\end{subfigure}
\begin{subfigure}[t]{0.24\textwidth}
\caption{} \label{fig:marginalize_instrument}
\centering
\begin{tikzpicture}

  \node[obs]                               (y) {$Y_{i}$};
  \node[obs, left=1.5cm of qya] (qaz) {$\scriptstyle Q^{a\mid z}_i$};
  \node[obs, left=.4cm of y] (qa) {$\scriptstyle Q^{a}_i$};
  \node[latent, above=.4cm of qa]  (u) {$U_i$};

  \edge {u} {qaz, y} ;
  \edge {qa} {y} ;
  \edge {qaz} {qa} ;

  \plate {out} {(qaz)(u)(y)} {$n$} ;
\draw[white, opacity=0] (current bounding box.north west) rectangle (current bounding box.south east);
\end{tikzpicture}
\end{subfigure}
\caption{\textbf{Transformations of example hierarchical causal models.} The first column shows three example HCMs, introduced in~\Cref{sec:first_look} (repeated from \Cref{fig:hcm_motifs}). The second column shows explicitly each HCM's latent $Q$ variables, as described in~\Cref{sec:hcgm}. The third column shows each HCM's matching collapsed model, a flat causal model (\Cref{sec:identification_problem}). The final column shows the augmented and/or marginalized collapsed model (\Cref{sec:identification_problem}). We apply do-calculus to this final model to perform identification in the original HCM.} \label{fig:hcm_transforms}
\end{figure}

The idea behind our strategy for hierarchical causal ID is to form a
flat causal model from the hierarchical causal model such that causal
identification in the flat model is equivalent to identification in
the hierarchical model. To transform the hierarchical model, we
develop three types of graphical steps: \textit{collapsing},
\textit{augmenting}, and \textit{marginalizing} (\Cref{fig:hcm_transforms}). When we collapse, we promote
the $Q$ variables to endogenous variables, and remove the subunit
endogenous variables, forming a flat causal model; when we augment, we add new unit-level
endogenous variables; and when we marginalize, we remove some of the
unit-level endogenous variables. Finally, with the flat model in hand,
we apply the do-calculus to determine whether and how identification
is possible. The next sections demonstrate these steps with the three
motifs of \Cref{fig:hcm_motifs}, and then show how the resulting ID formulae can be translated into practical estimators in concrete settings.  \Cref{sec:theory} presents the
theory that justifies these steps, and algorithms for performing them
on arbitrary graphs.

\subsection{The unit confounder graph}
\label{sec:confounder_id}

We first study identification for the \textsc{unit confounder} graph. The
hierarchical causal model is in \Cref{fig:hcm_confounder_transforms} (repeated from \Cref{fig:hcm_confounder}); the
$Q$ variables are shown explicitly in
\Cref{fig:hcm_intro_hpgm}. In the example application, the units are
schools and the subunits are students within them. The variable $A$ is
the number of tutoring hours for a student, $Y$ is their score on a
standardized test, and there are school-level confounders $U$ that
affect both how tutoring is dispersed and the performance of the
students. The target intervention is one where we provide $a_\star$
tutoring hours to all students in all schools, $\rmdo(a =
a_\star)$. At the unit level, this intervention is equivalent to
setting the subunit distribution of tutoring hours equal to a point
mass, $q^{a}_\star(a) = \delta_{a_\star}(a)$.

The outcome of interest is the average test score $Y$. Since $Y$ is a
subunit variable, we write this estimand as an iterated expectation over units and subunits. The subunit variables are drawn from the
subunit distribution $Q(a,y)$. We have:
\begin{align}
  \label{eq:confounder_estimand}
  \mathbb{E}[Y \s \rmdo(q^a=q^a_\star)] = \mathbb{E}[\mathbb{E}[Y\mid Q] \s \rmdo(q^a=q^a_\star)] =  \EE{\pr}{\EE{Q}{Y} \s \rmdo(q^a = q^{a}_\star)}.
\end{align}
The inner expectation is
\begin{align*}
  \EE{Q}{Y} = \int \int y\, Q(a, y) \, \di a\, \di y = \int y\, Q(y)\, \di y.
\end{align*}
The outer expectation is over the post-intervention distribution of unit-level quantities, including unit variables and subunit distributions. Here,
\begin{align*}
  Q(a, y) &\sim \pr(q(a, y) \s \rmdo(q^a = q^{a}_\star)).
\end{align*}
The double expectation $\EE{\pr}{\EE{Q}{Y} \s \rmdo(q^a = q^{a}_\star)}$ expresses the same quantity as
$\E{Y \s \rmdo(a=a_\star)}$, but expands the expectation to
decompose the unit-level and subunit-level randomness.

\parheadsmall{Step 1: Collapse.} The first step of hierarchical causal ID is to derive a
\textit{collapsed} model from the HCGM in
\Cref{fig:hcm_intro_hpgm}. 
The collapsed model is a flat causal model
that only contains unit-level variables. 
It serves as a mathematical tool to prove identification in the original HCGM.
The collapsed model includes both the original
unit variables and the $Q$ variables, which are now treated as
endogenous causal variables in their own right.  To obtain the
collapsed model, we take $m \rightarrow \infty$, effectively observing
the $Q_i$ variables, and then erase the subunit variables from the
graph.  For the \textsc{unit confounder}, the collapsed graph is in
\Cref{fig:collapse_intro}. It corresponds to the following generative
process,
\begin{equation} \label{eqn:confounder_collapse}
\begin{split}
  U_i &\sim \pr(u)\\
  Q^{a}_i &\sim \pr\big(q^{a} \, \big|\, u_i \big)\\
  Q^{y \mid a}_i &\sim \pr\big(q^{y \mid a}\, \big|\, u_i\big).
\end{split}
\end{equation}
(In other examples, the collapsing step will be more involved.)

What is important about the collapsed model is that its distribution
of unit-level variables is the same as in the hierarchical causal
model, both pre- and post-intervention. (This theory is developed in
general in \Cref{sec:collapse_theory}.)

\parheadsmall{Step 2: Augment.} In the second step we \textit{augment} the
collapsed model, adding new variables that represent quantities which
depend on the subunit distribution $Q(a,y)$. In the estimand of
\Cref{eq:confounder_estimand}, the target outcome
$\EE{Q}{Y}$ is an expectation relative to $Q(y)$, the
marginal distribution over $Y$ within a unit. $Q(y)$ can be written in terms of unit-level $Q$ variables,
\begin{align} \label{eqn:deterministic_qy_mech}
  q(y) = \int q^a(a) q^{y|a}(y \g a) \di a  \triangleq \m(q^a, q^{y|a}).
\end{align}
We augment the collapsed model to include $q(y)$ as an additional variable, denoted $q^y$;
see \Cref{fig:augment_intro}. This new variable is generated according to a
deterministic mechanism $q^y_i = \m(q^a_i, q^{y|a}_i)$, indicated by
double arrows in the graph. In the augmented model, the causal estimand can be written,
\begin{align}
  \label{eq:augmented_estimand}
  \EE{\pr}{\EE{Q^y}{Y} \s \rmdo(q^a=q^a_\star)}.
\end{align}
Here, recall, the outer expectation $\EE{\pr}{\cdot \s \rmdo(q^a=q^a_\star)}$ is over the post-intervention distribution of all unit-level quantities, including unit variables and subunit distributions, observed and hidden.

\parheadsmall{Step 3: Identify.} We reason with the augmented model graph
to identify the causal estimand. The causal estimand concerns the
effect that an intervention on $q^a$ has on $q^y$. We apply do-calculus to
\Cref{fig:augment_intro} to identify the intervention distribution via a backdoor correction,
\begin{equation} \label{eq:collapse_id}
\begin{split}
  \pr(q^y \s \rmdo(q^a = q^a_\star))
  &=
    \int \pr(q^{y|a}) \pr(q^y \g q^a_\star, q^{y|a}) \, \di q^{y|a} \\
  &=
    \int \pr(q^{y|a}) \, \m(q^a_\star, q^{y|a}) \, \di q^{y|a},
\end{split}
\end{equation}
where the second line follows since $q^y$ is a deterministic
function of its parents. The identified causal estimand uses this distribution in the
double expectation of \Cref{eq:augmented_estimand}.

\begin{figure}[t]
\centering
\begin{subfigure}[t]{0.32\textwidth}
        \caption{No confounding} \label{fig:no_confound_sim}
        \includegraphics[width=\textwidth]{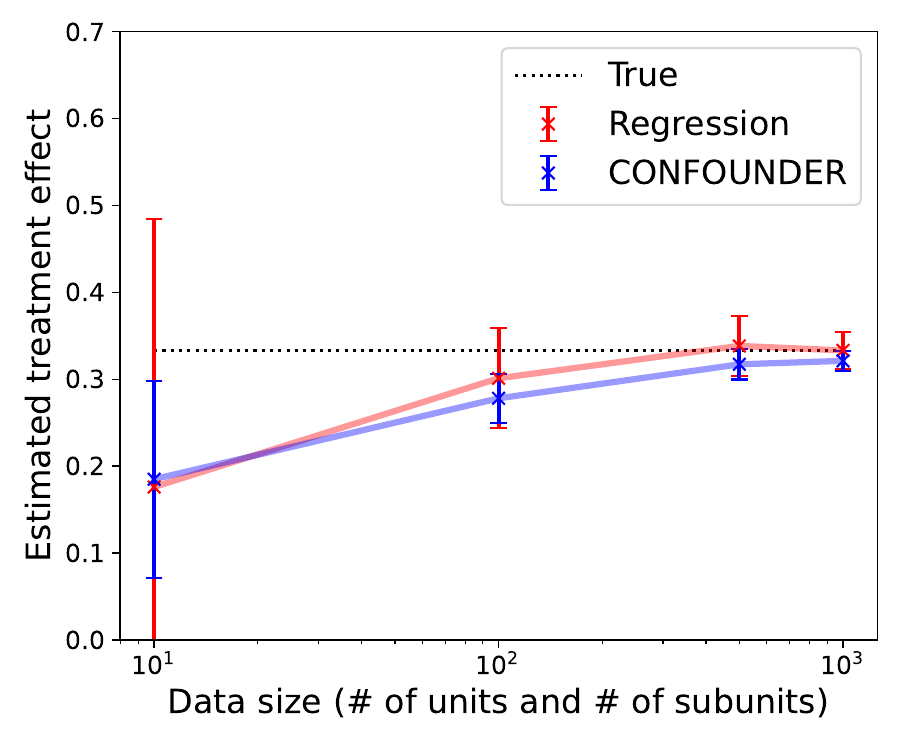}
\end{subfigure}
\begin{subfigure}[t]{0.32\textwidth}
        \caption{Low confounding} \label{fig:low_confound_sim}
        \includegraphics[width=\textwidth]{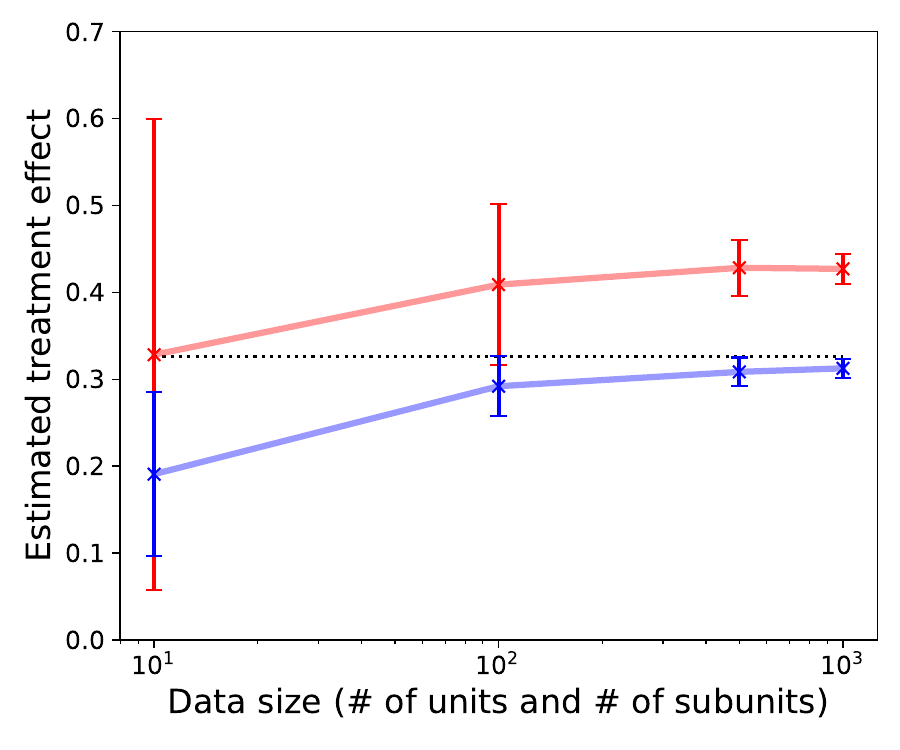}
\end{subfigure}
\begin{subfigure}[t]{0.32\textwidth}
        \caption{High confounding} \label{fig:high_confound_sim}
        \includegraphics[width=\textwidth]{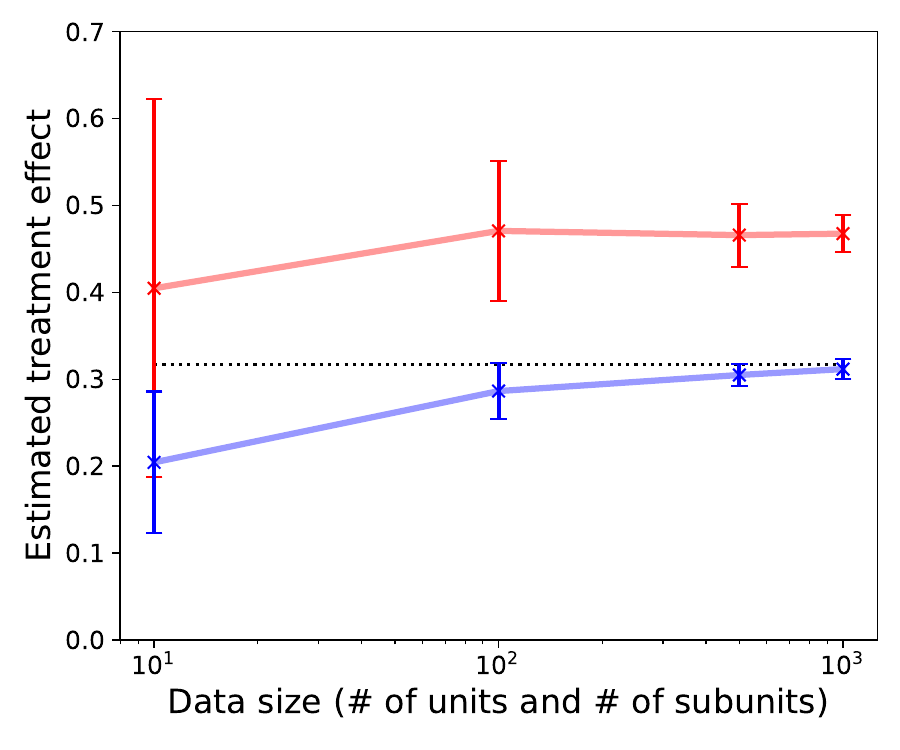}
\end{subfigure}
\caption{\textbf{Estimated effects in a } \textsc{unit confounder} \textbf{simulation.}
  The \textsc{unit confounder} estimate is based on the identification formula for the true HCM.
  The ``Regression'' estimate is based on aggregated data, and so fails to account for confounding. Error
  bars show standard deviation across 20 independent simulations.} \label{fig:unobs_confound_sim}
\end{figure}

\parheadsmall{Step 4: Estimate.} We observe data from $n$ units, and
$m_i$ subunits within each unit, $\{\{a_{ij}, y_{ij}\}_{j=1}^{m_i}\}_{i=1}^n$. We use
this data to approximate the terms of \Cref{eq:collapse_id} and take
the expectation in \Cref{eq:augmented_estimand}.
\begin{enumerate}[leftmargin=*]
\item Estimate the per-unit conditional distribution $\hat{q}^{y|a}_i$
  from $\{a_{ij}, y_{ij}\}_{j=1}^{m_i}$. This step amounts to
  estimating a separate conditional model for each unit, e.g., a set
  of regression models.

\item Calculate the per-unit marginal distribution
  \begin{align*}
    \hat{q}^y_i(y) = \m(q^a_\star, \hat{q}_i^{y|a}) = \int q^a_\star(a) \hat{q}_i^{y|a}(y \g a) \di a.
  \end{align*}
  If $q_\star^a$ is a point mass at a value $a_\star$ then
  $\hat{q}^y_i(y) = \hat{q}^{y|a}_i(y \g a_\star)$. Further calculate the
  expectation with respect to each marginal,
  $\hat{\mu}_i^y \triangleq \EE{\hat{q}_i^y}{Y}$.

\item Estimate the population distribution $\hat{\pr}(q^{y|a})$ with
  the empirical distribution of $\{\hat{q}^{y|a}_i\}_{i=1}^n$.
\end{enumerate}
With these ingredients, the final estimate of
\Cref{eq:augmented_estimand} is simply the average of the per-unit
expectations,
\begin{align}
  \label{eq:confounder_estimator}
  \EE{\pr}{\EE{Q^y}{Y} \s \rmdo(q^a=q^a_\star)} 
  &\approx \frac{1}{n} \sum_{i=1}^{n} \hat{\mu}_i^y.
\end{align}

As a demonstration, we consider binary $A$, $Y$, $U$, and draw
simulated variables from a true \textsc{unit confounder} model (details are in \Cref{sec:unobs_conf_sim}, and code reproducing the full experiment is in the Supplementary Material). Our goal is
to estimate the average treatment effect,
\begin{align}  \label{eq:confounder_effect}
  \EE{\pr}{\EE{Q^y}{Y}; \rmdo(q^a = \delta_{1})} -
  \EE{\pr}{\EE{Q^y}{Y}; \rmdo(q^a = \delta_{0})}.
\end{align}

We observe $\{\{a_{ij}, y_{ij}\}_{j=1}^m\}_{i=1}^n$ for
increasing numbers of units and subunits, and where the number of
units equals the number of subunits in each (i.e., $n=m$).
\Cref{fig:unobs_confound_sim} compares the \textsc{unit confounder} HCM
estimator in \Cref{eq:confounder_estimator} with a regression estimator
that simply models the per-unit average outcome as a linear function
of the per-unit average treatment. 
We consider three different simulations, with different levels of confounding in each. In each case, as the number of units and subunits
increases, the HCM-based estimator converges to the true effect.
The regression model, by contrast, only approaches the true effect if there is no confounding.

In summary, we have derived from the \textsc{unit confounder} graph the methodology of fixed-effect regression models, including nonparametric regression~\citep{Wooldridge2005-bk,Angrist2009-ah,Witty2020-et} (see \Cref{apx:fixed_effect} for further details).
In the following sections, we demonstrate how to derive new causal inference methods for novel hierarchical data settings.

\subsection{The unit confounder and unit interference graph} \label{sec:confound_interfere_id}

We next study the \textsc{unit confounder \& unit interference} graph. The
hierarchical causal model is in \Cref{fig:hcm_interfere_transforms} (repeated from \Cref{fig:hcm_interfere}); the
$Q$ variables are shown explicitly in
\Cref{fig:hcm_interfere_hpgm}. The causal estimand is in
\Cref{eq:confounder_estimand}.

\parheadsmall{Step 1: Collapse.} First we collapse the HCGM, taking
$m \rightarrow \infty$. In this graph, we must consider the edge
between subunit variable $a_{ij}$ and unit variable $z_i$. Earlier, we
required that the mechanism for $z_i$ was invariant to the ordering of
the subunit variables $a_{ij}$. We now make a further requirement: the
mechanism for $z_i$ converges as $m \rightarrow \infty$. Specifically,
$\pr(z \g \{a_{ij}\}_{j=1}^m)$ must converge to a mechanism
$\pr(z \g q_i(a))$, which depends only on the subunit distribution
over $A$.

For example, suppose $z_i$ depends on an empirical average of a
function of $a_{ij}$,
\begin{align*}
  Z_i \sim \textstyle
  \pr\left(z \g \{a_{ij}\}_{j=1}^m\right) = \pr\left(z \mid \frac{1}{m} \sum_{j=1}^{m} h(a_{ij})\right).
\end{align*}
If $h(a_{ij}) = a_{ij}$ then this average is the mean; if
$h(a_{ij}) = \mathbb{I}(a_{ij} > 2))$ then this average is the
fraction of subunits for which $a_{ij} > 2$. As $m \rightarrow \infty$
the average converges,
$\frac{1}{m} \sum_{j=1}^{m} h(a_{ij}) \to \EE{q_i(a)}{h(A)}$. Thus the
mechanism of $z$ converges to one that depends only on the subunit
distribution $q_i(a)$.
(Convergent and divergent mechanisms are discussed in depth in \Cref{sec:exp-family-mechanisms}.)

With this requirement in place, we can write the collapsed model. It is in
\Cref{fig:collapsed_interfere}, and now draws an arrow directly from
$Q_i^a$ to $Z_i$. The generative process is,
\begin{align}
\begin{split}
U_i &\sim \pr(u)\\
Q_i^a &\sim \pr(q^a \mid u_i)\\
Z_i &\sim \pr(z \mid q_i^a)\\
Q_i^{y \mid a} &\sim \pr(q^{y\mid a} \mid z_i, u_i).
\end{split}
\end{align}
Notice the treatment distribution $Q_i^a$ is connected to the
conditional distribution of the outcome $Q^{y|a}_i$ through the
interferer $Z_{i}$.

\parheadsmall{Step 2: Augment.} Next we augment the graph. The causal
estimand involves the within-unit marginal distribution of $Y$, so
again we augment the graph to include $Q_i^y$. Again it is a
deterministic function of $Q_i^{a}$ and $Q_i^{y|a}$, namely $q^y_i = m(q^a_i, q^{y|a}_i)$. The augmented
graph is in \Cref{fig:augment_interfere}.

\parheadsmall{Step 3: Identify.} We now apply do-calculus to the augmented
graph to identify the causal estimand.  We can write the
post-intervention distribution of $Q^y$ as,
\begin{align}
  \label{eq:interfere_intervention}
  \pr(q^y \s \rmdo(q^a =q^a_\star))
  &=
    \int
    \pr(q^y \g q^a_\star, q^{y|a})
    \pr(q^{y|a} \s \rmdo(q^a_\star)) \, \di q^{y|a}.
\end{align}
The first term is a point mass at the marginal,
$q^y(y) = \int q^a_\star(a) \, q^{y|a}(y \g a) \di a$. We identify the
second term with a \textit{front-door adjustment},
\begin{align}
  \label{eq:interfere_front_door}
  \pr(q^{y|a} \s \rmdo(q^a =q^a_\star))
  &=
    \int
    \pr(z \g q^a_\star)
    \underbrace{
    \left(
    \int
    \pr(q^a) \,
    \pr(q^{y|a} | q^a, z)
    \, \di q^a
    \right)}_{\pr(q^{y|a} \s \rmdo(z))}
    \, \di z.
\end{align}
We identify the estimand with the expectation of $Y$ under the
distribution in \Cref{eq:interfere_intervention}.

\parheadsmall{Step 4: Estimate.} We observe $n$ units, each with $m_i$
subunits: $\{z_i, \{a_{ij}, y_{ij}\}_{j=1}^{m_i}\}_{i=1}^n$. We
use the data to form estimates of the elements of
\Cref{eq:interfere_intervention,eq:interfere_front_door}.
\begin{enumerate}[leftmargin=*]
\item Estimate the $i$th term $\hat{q}_i^{a}$ from
  $\{a_{ij}\}_{j=1}^{m_i}$. For example, if $a_{ij}$ is binary then we
  can estimate a Bernoulli parameter.

\item Estimate the $i$th conditional $\hat{q}_i^{y|a}$ from
  $\{y_{ij}, a_{ij}\}_{j=1}^{m_i}$. For example, if $y_{ij}$ is binary
  then this estimate can be a pair of Bernoulli parameters
  $\hat{\pi}_{i,0}$ and $\hat{\pi}_{i,1}$, to parameterize the
  conditional distribution of $Y$ given $A=0$ and $A=1$.

\item Estimate the population conditional $\hat{\pr}(z \g q^a)$ from
  $\{z_i, \hat{q}^a_i\}_{i=1}^n$. For example, if the interference
  variable $z_i$ is binary then this estimate can be a logistic
  regression, conditional on the parameters determining $\hat{q}^a_i$.

\item Estimate the population conditional
  $\hat{\pr}(q^{y|a} \g q^a, z)$ from
  $\{z_i, \hat{q}^a_i, \hat{q}^{y|a}_i\}_{i=1}^n$. If $z_i$ is
  binary then we can use two different regression models, each
  conditional on the parameters determining $\hat{q}^{y|a}_i$.

\item Estimate the population distribution $\hat{\pr}(q^a)$ with the
  empirical distribution of $\{\hat{q}^a_i\}_{i=1}^n$.
\end{enumerate}
We plug these estimates into
\Cref{eq:interfere_intervention,eq:interfere_front_door} to estimate
the intervention distribution.

As a demonstration, we consider binary $A$, $Y$, and $Z$, and continuous $U$, and draw
simulated variables from a true \textsc{unit confounder \& unit interference}
model (details are in \Cref{sec:interference_sim}, and code in the Supplementary Material). We consider three different simulations, each with different
levels of interference, i.e., an increasing effect of $Z$ on $Y$.  Here our
goal is to estimate a difference between two soft interventions,
\begin{align}
  \EE{\pr}
  {\EE{Q^y}{Y}
  \s \rmdo(q^a = \textrm{Bern}(0.75))} -
  \EE{\pr}
  {\EE{Q^y}{Y}
  \s \rmdo(q^a = \textrm{Bern}(0.25))}.
\end{align}
$\textrm{Bern}(\mu)$ is the Bernoulli distribution with mean $\mu$.

\begin{figure}[t]
\centering
\begin{subfigure}[t]{0.32\textwidth}
        \caption{No interference} \label{fig:no_interfere_sim}
        \includegraphics[width=\textwidth]{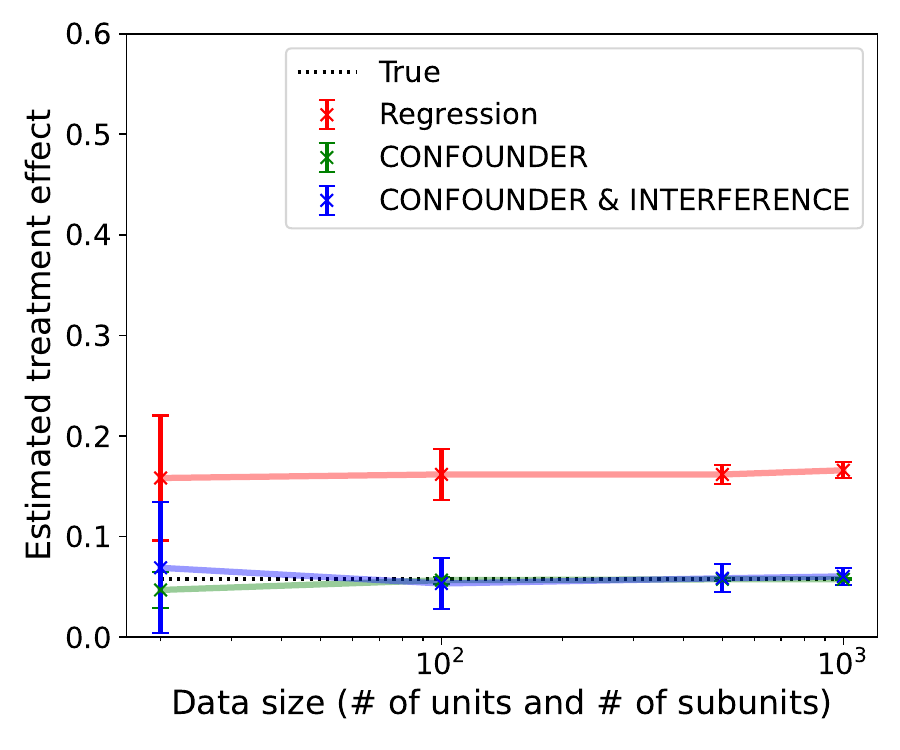}
\end{subfigure}
\begin{subfigure}[t]{0.32\textwidth}
        \caption{Low interference} \label{fig:low_interfere_sim}
        \includegraphics[width=\textwidth]{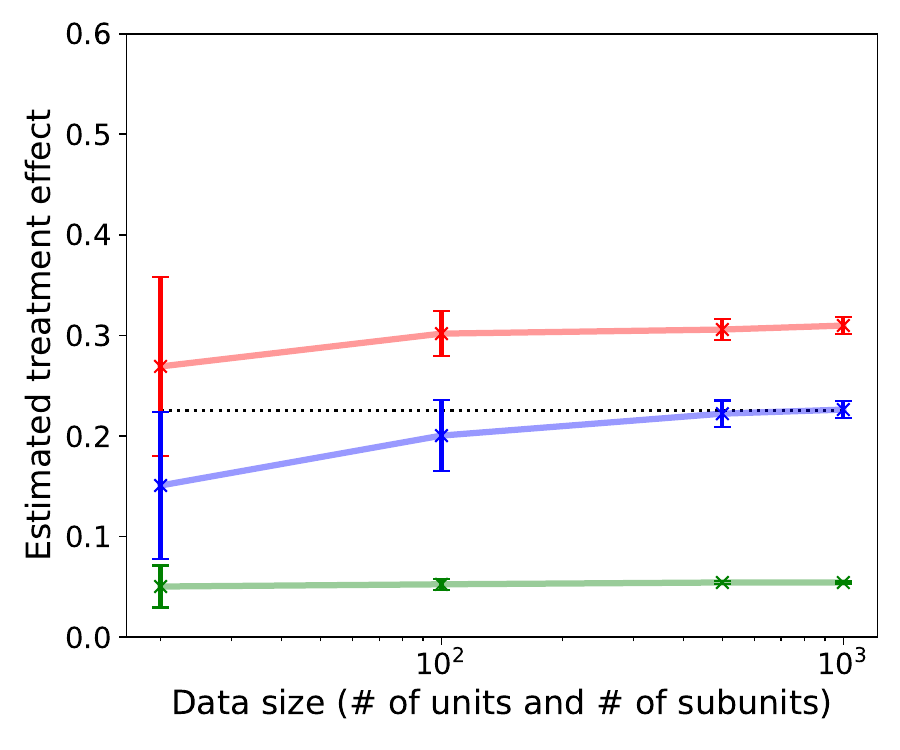}
\end{subfigure}
\begin{subfigure}[t]{0.32\textwidth}
        \caption{High interference} \label{fig:high_interfere_sim}
        \includegraphics[width=\textwidth]{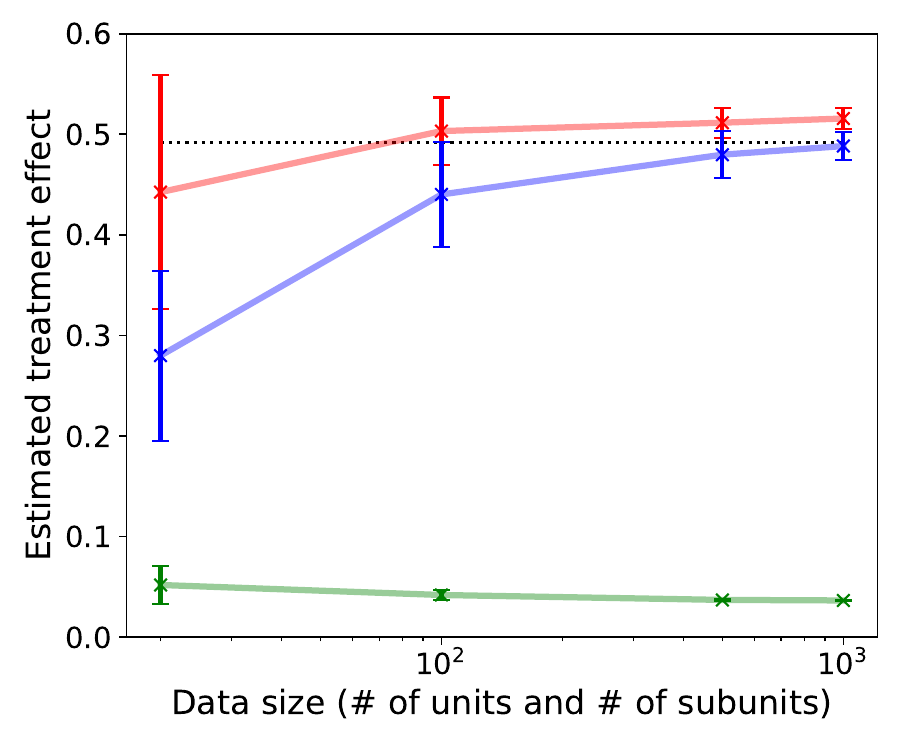}
\end{subfigure}
\caption{\textbf{Estimated effects in a }\textsc{confounder \&
      interference}\textbf{ simulation.} The \textsc{unit confounder \& unit interference} estimate is based on the identification formula for the true HCM. 
      The \textsc{unit confounder} estimate is based on the \textsc{unit confounder} model, which is incorrect in this simulation, as it ignores confounding.
      The ``Regression'' estimate is based on aggregated data, and fails to account for confounding or interference. 
      Error bars show standard deviation across 20 independent simulations.} \label{fig:interfere_sim}
\end{figure}

We observe $\{z_i, \{a_{ij}, y_{ij}\}_{j=1}^m\}_{i=1}^n$ for
increasing numbers of units and subunits, and where the number of
units equals the number of subunits in each. \Cref{fig:interfere_sim}
compares the \textsc{unit confounder \& unit interference} estimator with the
\textsc{unit confounder} estimator from \Cref{sec:confounder_id}, as well as the aggregated linear regression estimator discussed in \Cref{sec:confounder_id}. 
The linear regression estimator does not account for confounding or interference.
The \textsc{unit confounder} estimator does not
account for interference, and we can see its error grow as interference increases. 
Said another way, fixed-effect models fail with interference.
But regardless of the level of interference, the \textsc{unit confounder \& unit interference} estimator
converges to the true effect with increasing data. 
As we increase the number of subunits per unit, the bias at low $N$ decreases (\Cref{fig:interfere_converge_sim_mfrac})~\citep[Appendix B]{Rainforth2018-qp}.

\subsection{The subunit instrument graph}
\label{sec:instrument_id}

Last we study the \textsc{subunit instrument} graph. The HCM is in
\Cref{fig:hcm_instrument_transforms} (repeated from \Cref{fig:hcm_interfere}); the HCGM is in
\Cref{fig:hcm_instrument_hpgm}.  Here the causal estimand involves a
unit-level outcome,
\begin{align}
  \label{eq:inst_estimand}
  \EE{\pr}{Y \s \rmdo(q^a = q^a_\star)}.
\end{align}

\parheadsmall{Step 1: Collapse.} We take $m \rightarrow \infty$, remove the
subunit variables, and correctly connect the $Q$ variables to the
unit-level variables. The collapsed graph is in
\Cref{fig:collapse_instrument}. The generative process is,
\begin{align}
  \label{eqn:instrument_collapse}
  \begin{split}
    U_i &\sim \pr(u)\\
    Q^z_i &\sim \pr(q^z)\\
    Q^{a \mid z}_i &\sim \pr(q^{a \mid z} \mid u_i)\\
    Y_i &\sim \pr(y \mid q^{a \mid z}_i, q^z_i) = \pr\left(y \, \Big|\, q_i(a) = \int q^{a \mid z}_i(a \mid z) q^z_i(z) \di z\right).
  \end{split}
\end{align}
Notice the distribution of $Y$ depends on the marginal distribution of
subunit variable $Q(a)$, which is formed by the two $Q$ variables,
$Q^{z}$ and $Q^{a|z}$. 

\parheadsmall{Step 2: Augment and Marginalize.} Here we augment the
graph to form the distribution of the treatment variable $q^a$, the
variable on which we intervene (\Cref{fig:augment_instrument}). This variable is a deterministic
function of its parents, $q_i^a(a) = m(q_i^z, q_i^{a|z})$ and the
outcome variable $Y$ now depends only on the augmentation variable,
$Y_i \sim \pr(y \g q_i(a))$. Again we require that the mechanism for $y_i$
converges as $m \rightarrow \infty$.

To analyze the \textsc{subunit instrument} graph, we also need a new idea:
\textit{marginalization}.  We marginalize out $Q_i^z$ so that $Q_i^a$ no longer
depends deterministically on its parents; see
\Cref{fig:marginalize_instrument}, and note the double arrows denoting a deterministic mechanism have been replaced by a single arrow from $Q^{a|z}$ to $Q^a$.  In the marginalized graph, $Q_i^a$
depends stochastically on its remaining parent, $Q_i^{z|a}$. This step
ensures positivity. In the marginalized graph, we can have
$\pr(q^a = q^a_\star \g q^{y|a}) > 0$ for all $q^{y|a}$ (we discuss this assumption further in \Cref{sec:completeness}).

\parheadsmall{Step 3: Identify.} With positivity in place, we can use a
backdoor adjustment to identify the intervention distribution,
\begin{align}
  \pr(y \s \rmdo(q^a =q^a_\star))
  =
  \int
  \pr(q^{a|z})
  \, \pr(y \g q^a_\star, q^{a|z})
  \, \di q^{a|z}.
  \label{eq:instrument_backdoor}
\end{align}
We identify the estimand in \Cref{eq:inst_estimand} from the expectation
of $Y$ under the distribution in \Cref{eq:instrument_backdoor}.

\parheadsmall{Step 4: Estimate.}  We observe $n$ units, each with $m_i$
subunits: $\{\{z_{ij}, a_{ij}\}_{j=1}^{m_i}, y_i\}_{i=1}^n$. We
use the data to form estimates of the elements of
\Cref{eq:instrument_backdoor}.
\begin{enumerate}[leftmargin=*]
\item Estimate the $i$th conditional treatment distribution
  $\hat{q}_i^{a|z}$ from $\{z_{ij}, a_{ij}\}_{j=1}^{m_i}$.

\item Estimate the $i$th marginal treatment distribution $\hat{q}_i^a$
  from $\{a_{ij}\}_{j=1}^{m_i}$.

\item Estimate the population outcome distribution
  $\hat{\pr}(y \g q^a, q^{a|z})$ from
  $\{\hat{q}^{a|z}_i, \hat{q}^a_i, y_i\}_{i=1}^n$.

\item Estimate the population distribution of $\hat{\pr}(q^{a|z})$
  with the empirical distribution of
  $\{\hat{q}_i^{a|z}\}_{i=1}^n$.
\end{enumerate}
We plug these estimates into \Cref{eq:instrument_backdoor}, and
calculate the expectation.

\begin{figure}[t]
\centering
\begin{subfigure}[t]{0.32\textwidth}
        \caption{No confounding} \label{fig:iv_low_sim}
        \includegraphics[width=\textwidth]{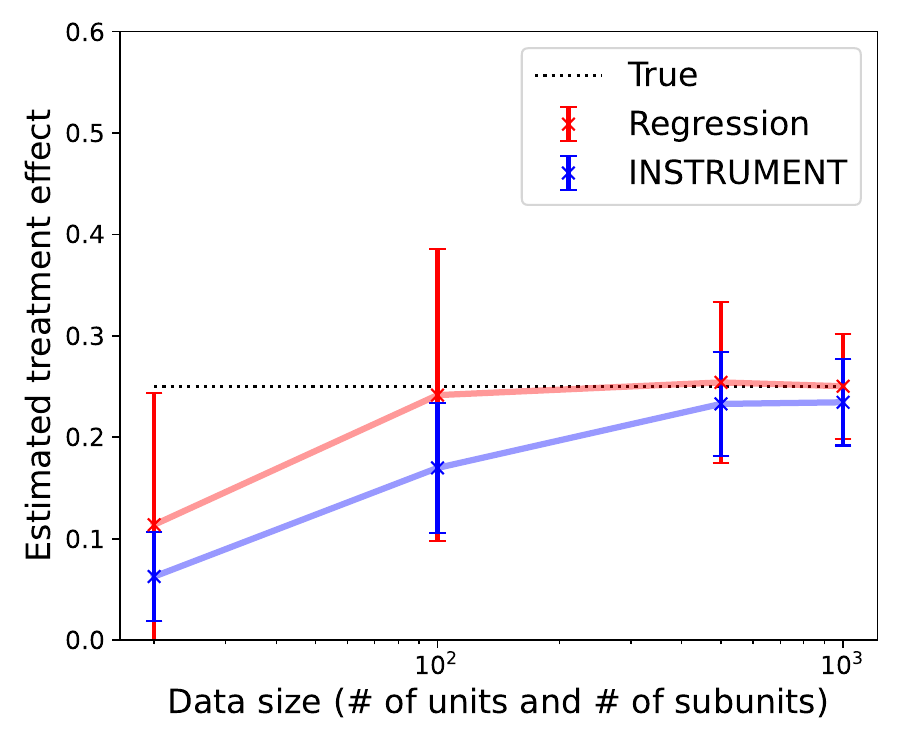}
\end{subfigure}
\begin{subfigure}[t]{0.32\textwidth}
        \caption{Low confounding} \label{fig:iv_medium_sim}
        \includegraphics[width=\textwidth]{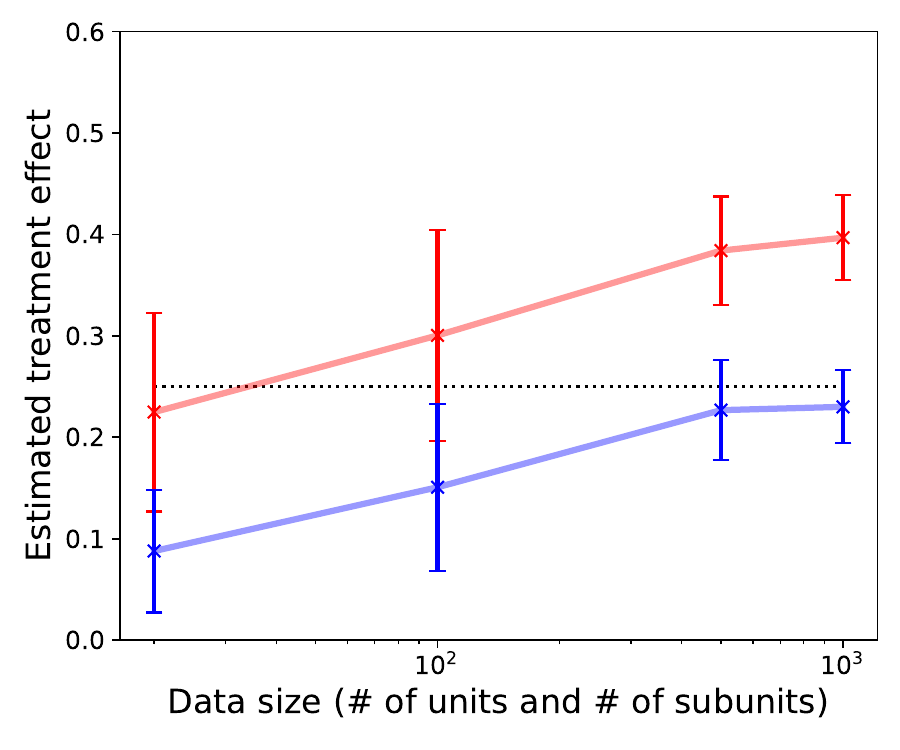}
\end{subfigure}
\begin{subfigure}[t]{0.32\textwidth}
        \caption{High confounding} \label{fig:iv_high_sim}
        \includegraphics[width=\textwidth]{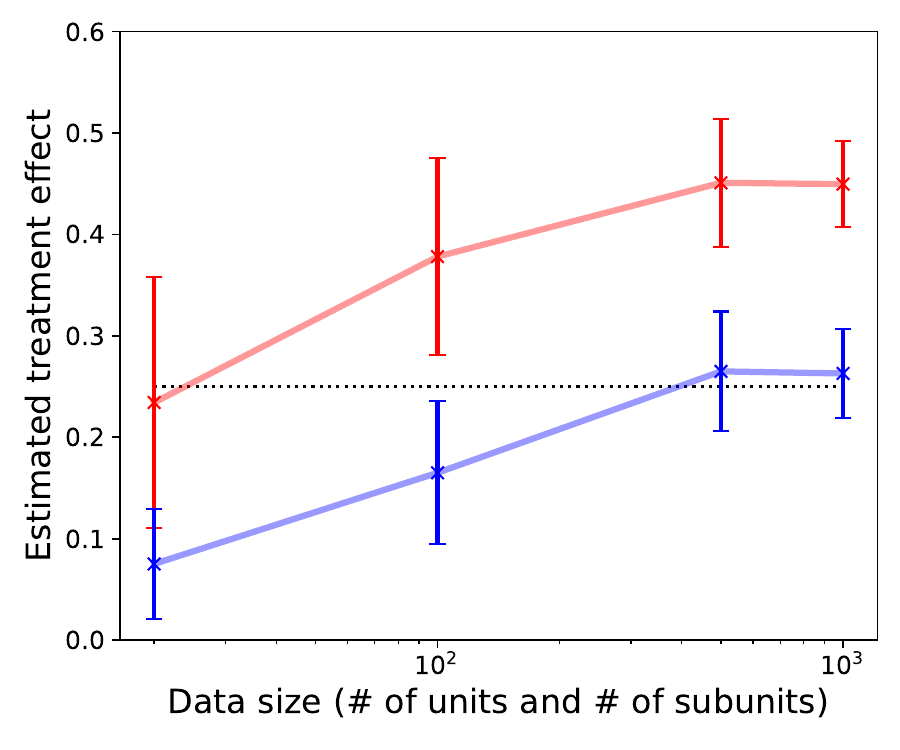}
\end{subfigure}
\caption{\textbf{Estimated effects in an }\textsc{subunit instrument}\textbf{ simulation.}
  The \textsc{subunit instrument} estimate is based on the identification formula for the true HCM. The ``Regression'' estimate ignores the instrument and hence ignores confounding. Error bars show standard deviation across 20 independent simulations. }
\label{fig:instrument_sim}
\end{figure}

As a final demonstration, we consider binary $A$, $Y$, $U$, $Z$, and
draw simulated variables from a true \textsc{subunit instrument} model (details are in \Cref{apx:instrument_sim}, and code in the Supplementary Material). We
consider three different simulations, each with different levels of
confounding.  Our goal is to
estimate a difference between two soft interventions,
\begin{align}
  \EE{\pr}
  {Y
  \s \rmdo(q^a = \textrm{Bern}(0.75))} -
  \EE{\pr}
  {Y
  \s \rmdo(q^a = \textrm{Bern}(0.25))}.
\end{align}

We observe $\{\{z_{ij}, a_{ij}, y_{ij}\}_{j=1}^m\}_{i=1}^n$
for increasing numbers of units and subunits, and where the number of
units equals the number of subunits in each. \Cref{fig:instrument_sim}
compares the \textsc{subunit instrument} estimator with a straightforward regression of
$Y$ on the average of $A$, which does not account for the hidden
confounder.  The error of the regression estimator is larger when
there is more confounding. As the number of units and subunits
increases, the \textsc{subunit instrument} estimator converges to the true effect.
Note that here both estimators use a nonparametric model, a Gaussian process classifier, rather than a well-specified parametric model as in the previous simulations; partially as a result, we see more substantial bias and variance in the estimator than in the previous simulations.
As the number of subunits per unit increases, the bias at low $N$ decreases (\Cref{fig:converge_sim_mfrac})~\citep[Appendix B]{Rainforth2018-qp}.

 \section{Theory} \label{sec:theory}

We demonstrated with the three graphs of \Cref{fig:hcm_motifs} how collapsing, augmenting and marginalizing can be used for hierarchical causal ID. 
In this section we elaborate the key assumptions behind these steps, and explain how they can be used on hierarchical causal models with arbitrary graphs.
Note the notation introduced in this section is also summarized in \Cref{table:notation}.

\subsection{Collapsed models} \label{sec:collapse_theory}

Our central tool for identification in HCMs is the collapsed model.
The collapsed model is a flat CGM that matches a given HCGM, and
in which the $Q$ variables are endogenous variables. Here we derive
the collapsed model for arbitrary HCGMs, and explain the assumptions that justify it.

\begin{figure}[t]
\centering
\begin{tikzpicture}

  \node[obs] (v4) {$X^{4}$};
  \node[obs, above=.8cm of v4] (v1) {$X^{1}$};
  \node[obs, left=2.4cm of v4] (v2) {$X^{2}$};
  \node[obs, left=of v4] (v3) {$X^{3}$};
  \node[obs, right=of v4] (v5) {$X^{5}$};
  \node[obs, below=.8cm of v2] (v6) {$X^{6}$};
  \node[obs, below=.8cm of v4] (v7) {$X^{7}$};

  \edge {v1} {v4} ;
  \edge {v2} {v6} ;
  \edge {v6} {v3} ;
  \edge {v3} {v4} ;
  \edge {v3} {v7} ;
  \edge {v4} {v5} ;
  \edge {v5} {v7};

  \plate[dashed] {in} {(v1)(v2)(v3)(v4)(v5)} {$m$} ;
  \plate {out} {(in)(v6)(v7)} {$n$} ;

\end{tikzpicture}
\caption{\textbf{Direct subunit ancestors and descendants.} In this example, the direct subunit ancestors of $X^{7}$ are $X^{1},X^{3},X^{4}$ and $X^{5}$. Conversely, $X^7$ is the direct unit descendant of $X^{1},X^{3},X^{4}$ and $X^{5}$. The only direct subunit ancestor of $X^6$ is $X^2$, and $X^6$ is the direct unit descendant of $X^2$.} \label{fig:direct_ancestor_ex}
\end{figure}
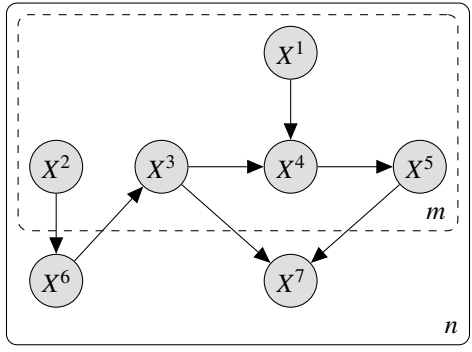

We first define two sets of variables. Let $\da_\mathcal{S}(w)$ denote the indices of the \textit{direct
subunit ancestors} of a unit variable $X^w$, that is, those ancestral subunit
variables which are either parents of $X^w$ or are connected via a
directed path containing only other subunit-level variables; see
\Cref{fig:direct_ancestor_ex}.  Let $\dd_{\mathcal{U}}(v)$ denote the
indices of the \textit{direct unit descendants} of a subunit variable $X^v$, that is, those
unit variables who have $X^v$ as a direct subunit ancestor.

\begin{definition}[Collapsed model] \label{def:um} Consider a
  hierarchical causal graphical model $\Mc^\cgm$, as in
  \Cref{def:hcgm}.  The corresponding collapsed model $\Mc^\col$ is a
  flat causal graphical model. It is
\begin{align}
  \begin{split}
    Q^{v\mid \pa_\mathcal{S}(v)}_i
    &\sim
      \pr\big(q^{v\mid
      \pa_\mathcal{S}(v)} \, \big| \, x^{\pa_\mathcal{U}(v)}_{i}\big)
      \quad \quad \quad \quad \quad
      \text{ for subunit variables } v \in \mathcal{S} \\
    X^{w}_i
    &\sim
      \pr\big(x^{w} \, \big| \, x^{\pa_\mathcal{U}(w)}_{i},
      q_i(x^{\pa_\mathcal{S}(w)})\big)
      \quad \quad \quad
      \text{ for unit variables } w \in \mathcal{U},
\end{split}
\end{align}
for $i \in \{1, \ldots, n\}$. Note the unit variables $X^w$ are drawn
conditional on the subunit marginal of their subunit ancestors
$q_i(x^{\pa_\mathcal{S}(w)})$,
\begin{align}
  \label{eqn:parent_marginal}
  q_i(x^{\pa_\mathcal{S}(w)}) =
  \int \cdots \int
  \prod_{w' \in \da_\mathcal{S}(w)} q_i^{w' \mid
  \pa_\mathcal{S}(w')}(x^{w'} \mid x^{\pa_\mathcal{S}(w')})
  \prod_{w' \in \da_{\mathcal{S}}(w) \setminus \pa_\mathcal{S}(w)}
  \di x^{w'}.
\end{align}
It is through this dependence that $X^w$ connects to
$Q^{v|\pa(v)}$ variables.

In the collapsed model, each endogenous $Q$-variable
$Q_i^{v \mid \pa_\mathcal{S}(v)}$ for $v \in \mathcal{V}$ is observed
if and only if $X^v$ and all its subunit parents are observed in the
original HCGM, i.e. $v \in \Sc_{\obs}$ and
$\pa_\mathcal{S}(v) \subset \mathcal{S}_{\obs}$.  Each endogenous
non-$Q$-variable $X^w_i$ for $w \in \mathcal{U}$ is observed if and
only if it is observed in the original HCGM, i.e.
$w \in \mathcal{U}_{\obs}$.
\end{definition}

\Cref{alg:collapse} is a graphical algorithm for deriving the
collapsed causal graphical model from the hierarchical causal
graphical model. \Cref{fig:hcm_motifs} gives three examples of
collapsed models, which we have discussed. Other examples are in
\Cref{fig:ID_exs} and \Cref{fig:nonID_exs}.

We next show that the effects of interventions in a collapsed model
match the effects of corresponding interventions in the original HCGM.
A hard intervention on a $Q$ variable of a collapsed model, i.e.,
$\rmdo(q^{v\mid \pa_\mathcal{S}(v)} = q_\star^{v\mid
  \pa_\mathcal{S}(v)})$, corresponds to a soft intervention
$X^v \sim q_\star^{v\mid \pa_\mathcal{S}(v)}(x^v \mid
x^{\pa_\mathcal{S}(v)})$ on the subunit variable $X^v$ in the original HCGM.
A hard intervention on a unit variable $X^w$  in the collapsed model
corresponds to the same intervention in the original HCGM.

\begin{algorithm}[t]
\caption{\textbf{Graphical algorithm for collapsing an HCGM.} This algorithm transforms the graph of an HCM into the graph of its collapsed model, following \Cref{def:um}.}\label{alg:collapse}
\begin{algorithmic}
\State \textbf{Input:} a hierarchical causal graphical model $\Mc^\cgm$ (\Cref{def:hcgm})
\State \textbf{Output:} the collapsed flat causal graphical model $\Mc^\col$ (\Cref{def:um})
\For{each subunit variable $v \in \mathcal{S}$}
\State create a unit endogenous variable $Q^{v \mid \pa_\mathcal{S}(v)}$.
\If{$X^v$ and its parent subunit variables are observed ($v \in \Sc_{\obs}, \pa_\mathcal{S}(v) \subset \mathcal{S}_{\obs}$)}
\State mark $Q^{v \mid \pa_\mathcal{S}(v)}$ as observed (filled circle)
\Else
\State mark $Q^{v \mid \pa_\mathcal{S}(v)}$ as hidden (empty circle)
\EndIf
\State disconnect the unit parents $X^{\pa_\mathcal{U}(v)}$ from $X^v$; connect them to $Q^{v \mid \pa_\mathcal{S}(v)}$
\For{each direct unit descendant $w \in \dd_v(\mathcal{S})$}

\State connect $Q^{v \mid \pa_\mathcal{S}(v)}$ to $X^w$ (see \Cref{eqn:parent_marginal})
\EndFor
\State erase the subunit variable $X^v$
\EndFor
\State erase the inner plate
\end{algorithmic}
\end{algorithm}

\parhead{HCGMs converge to collapsed models} We now equate the HCGM to its collapsed model, in the infinite subunit limit. The key assumption is that mechanisms converge. Here, KL denotes the Kullback-Leibler divergence.
\begin{definition}[Mechanism convergence] \label{asm:inf-subunit}
  Consider a unit variable $w \in \mathcal{U}$ in an HCGM, and its
  mechanism
  $\pr(x^w \mid x^{\pa_\mathcal{U}(w)},
  \{x^{\pa_{\mathcal{S}}(w)}\}_{j=1}^m)$.  We say the mechanism
  converges with infinite subunits if there exists a limiting
  conditional distribution
  $\pr(x^{w} \mid x^{\pa_\mathcal{U}(w)}, q(x^{\pa_\mathcal{S}(w)}))$
  such that,
\begin{equation}
  \mathbb{E}_{X_{1:m}^{\pa_\mathcal{S}(w)}\sim q(x^{\pa_\mathcal{S}(w)})}\Big[\mathrm{KL}\Big(\pr\big(x^{w} \mid x^{\pa_\mathcal{U}(w)}, q(x^{\pa_\mathcal{S}(w)})\big)\, \big\| \, \pr\big(x^w \mid x^{\pa_\mathcal{U}(w)}, \{X_{j}^{\pa_\mathcal{S}(w)}\}_{j =1}^m\big)\Big)\Big] \xrightarrow[m\to\infty]{a.s.} 0.
\end{equation}
We define the left hand side as $\mathrm{D}^{w}_m(x^{\pa_\mathcal{U}(w)}, q(x^{\pa_\mathcal{S}(w)}))$.
\end{definition}
\noindent 
Heuristically, if a mechanism depends smoothly on the empirical distribution of subunit variables $\hat{q}_m(x^{\pa_\mathcal{S}(w)}) = \frac{1}{m} \sum_{j=1}^m \delta_{x_j^{\pa_\mathcal{S}(w)}}$ and does not depend on the total number of subunits $m$, we can expect it to converge
(\Cref{sec:exp-family-mechanisms}).
An HCGM converges to its collapsed model so long as its unit variable mechanisms converge.
\begin{theorem}[Collapsing a hierarchical causal model] \label{thm:valid_collapse}
Let $\pr_{\Delta,m}(x^{\mathcal{U}},q(x^{\mathcal{S}}))$ be the joint distribution over $x^{\mathcal{U}}$ and $q(x^{\mathcal{S}})$ given by an HCGM with $m$ subunits, under an intervention $\Delta$ (\Cref{def:intervention}). Let $\pr^{\col}_\Delta(x^{\mathcal{U}}, q(x^{\mathcal{S}}))$ be the distribution given by the corresponding collapsed model, under the corresponding intervention.
Assume each unit variable mechanism converges, such that for all $w \in \mathcal{U}$ we have $\mathbb{E}_{\pr^\col_\Delta}[\mathrm{D}^w_m(X^{\pa_\mathcal{U}(w)}, Q(x^{\pa_\mathcal{S}(w)}))] \to 0$ as $m \to \infty$ a.s..
Then the hierarchical causal graphical model converges to the collapsed model,
\begin{equation}
        \mathrm{KL}(\pr_\Delta^{\col} \, \|\, \pr_{\Delta,m}) \xrightarrow[m\to\infty]{a.s.} 0.
\end{equation}
\end{theorem}
\noindent The proof is in \Cref{appx:collapsing}. This is a key result, as it allows us to analyze identification in hierarchical causal models by analyzing identification in a corresponding flat model. 

\parhead{Do-calculus in the collapsed model} To study
identification, we apply do-calculus to the collapsed model.
Do-calculus rests on the assumption that we know the joint
distribution over observed variables in the model.  This is justified by two key technical assumptions on the HCGM. First,
with infinite data, we can observe the joint distribution over
observable unit variables and the observable subunit distribution.
 \begin{assumption}[Known observable joint] \label{assume:known_joint}
 	The distribution $\pr(x^{\mathcal{U}_\obs},q(x^{\mathcal{S}_\obs}))$ of the HCGM is known.
 \end{assumption}
 \noindent In \Cref{apx:hier_empirical_convergence} we prove that under the conditions of \Cref{thm:valid_collapse}, 
 $\pr(x^{\mathcal{U}_\obs},q(x^{\mathcal{S}_\obs}))$ can be learned
 from data in the limit of infinite units and infinite subunits per unit, sampled from the HCGM.
 In particular, we show that the empirical distribution over units, of the empirical
 distribution over subunits, will converge in Wasserstein distance to
 $\pr(x^{\mathcal{U}_\obs},q(x^{\mathcal{S}_\obs}))$.
 We also characterize the convergence rate in $N$ and $M$.

Knowledge of the joint distribution of subunits, $q(x^{\mathcal{S}_\obs})$, does not always imply full knowledge of the conditional distributions.
For example, in the \textsc{unit confounder} graph, even if we know $q(a, y)$ we do not necessarily know $q^{y| a}(y\mid a')$ for all values of $a'$. The reason is that $q^a(a)$ may put zero probability on some values of $A$, in which case $q^{y| a}(y\mid a')$ will be unobservable for these values.
So, we make the following positivity assumption. Here $\mathcal{X}^v$ is the domain of a variable $X^v$, and $\pr(q(x^{\mathcal{S}}))$ is the distribution over subunit distributions in the HCGM.
\begin{assumption}[Subunit-level positivity] \label{assume:subunit_positive}
        Consider $Q(x^{\mathcal{S}}) \sim \pr(q(x^{\mathcal{S}}))$. With probability one, for all $v \in \mathcal{S}$ and $x^{\pa_\mathcal{S}(v)}\in \mathcal{X}^{\pa_\mathcal{S}(v)}$, we have $Q(x^{\pa_\mathcal{S}(v)}) > 0$.
\end{assumption}
\noindent In short, to use the collapsed model for identification, we
need variability among subunits. (If a subunit variable has the same
value for all subunits then it is effectively a unit variable.)

\Cref{assume:known_joint} and \Cref{assume:subunit_positive} together imply that the distribution over observable variables in the collapsed models is known. Let $q^{\mathcal{Q}} \triangleq \{q^{v | \pa_\Sc(v)}: v \in \Sc\}$ denote the set of $Q$ variables in the collapsed model, and let $q^{\mathcal{Q}_{\obs}}\triangleq \{q^{v | \pa_\Sc(v)} : v \in \mathcal{S}_{\obs}, \pa_\mathcal{S}(v) \subset \mathcal{S}_{\obs}\}$ denote the subset of $Q$ variables that are observed.
\begin{proposition}[Observed collapsed model] \label{prop:observe_collapsed}
        Given the conditions of \Cref{thm:valid_collapse}, \Cref{assume:known_joint} and \Cref{assume:subunit_positive}, the joint distribution over the observed endogenous variables of the collapsed model is known, $\pr^\col(x^{\mathcal{U}_\obs},q^{\mathcal{Q}_{\obs}})$. 
\end{proposition}
\noindent Do-calculus proceeds from the assumption that the joint
distribution over the observed endogenous variables in a flat causal graphical
model is known.  \Cref{prop:observe_collapsed} says that this
distribution is known for the collapsed model.  So,
\Cref{prop:observe_collapsed} implies we can apply do-calculus to
identify the effects of interventions in the collapsed model.  Then,
by \Cref{thm:valid_collapse}, we can equate effects in the collapsed
model to effects in the original HCGM.

Do-calculus also rests on assumptions about unit-level positivity
~\citep{Shpitser2006-jg}.  Most salient is that the intervention we
are studying always has non-zero probability.
\begin{assumption}[Unit-level positivity] \label{assume:unit_positive}
  Let $\tilde{Z}$ be a variable in a collapsed model (it may be a unit
  variable $X^w$ or a $Q$ variable).  Let $Z^{\pa(\tilde{z})}$ denote
  the parents of $\tilde{Z}$ in the collapsed model graph.  For any
  hard intervention $\rmdo(\tilde{z} = \tilde{z}_\star)$, we require
  $\pr^\col(\tilde{z}_\star \mid z^{\pa(\tilde{z})}) > 0$ a.s. for
  $Z^{\pa(\tilde z)} \sim \pr^\col(z^{\pa(\tilde z)})$.
\end{assumption}
\noindent When applying do-calculus, one sometimes finds that identifying one effect in the model depends on first identifying the effects of other interventions. In such cases, positivity most hold for those other interventions as well (\Cref{asm:positivity_2}, \Cref{apx:positivity}).

\subsection{Augmentation and marginalization}
\label{sec:augment_theory}

Augmentation and marginalization are graphical proof techniques that
help establish identification in HCMs.  We use them when do-calculus
on the collapsed model does not immediately yield identification.  We
saw examples of augmented and marginalized models in
\Cref{sec:identification_problem}.  Here we outline the
approach; more details are in \Cref{apx:augment_marginalize}.

When we augment a collapsed model, we add an additional endogenous
variable, which describes some quantity of interest.  For the
augmentation to be valid, we must recover the original model when we
marginalize out the augmentation variable (details on marginalization are in \Cref{appx:marginalization}).
\begin{definition}[Valid augmented model]
  \label{def:valid_augment}
  Consider a collapsed model with distribution
  $\pr^\col(x^{\Uc},q^{\mathcal{Q}})$ over endogenous variables
  $X^{\Uc},Q^{\mathcal{Q}}$.  An augmented model $\Mc^\aug$ includes
  an additional endogenous variable $\tilde{Q}$ generated from parents
  $Q^{\pa(\tilde{q})}$ according to a chosen deterministic mechanism
  $\tilde{q}_i =\tilde{\f}(q^{\pa(\tilde{q})}_i)$.

  This augmentation variable is observed so long as it can always be
  computed from observed variables in the original collapsed model.
  The augmentation is valid so long as if $\tilde{Q}$ is marginalized
  out of the model, we recover the original collapsed model.
\end{definition}

As we have seen, we focus on augmentation variables that describe
marginal or conditional distributions over subunit variables, i.e. new
$Q$ variables. \Cref{alg:augment} provides a graphical algorithm for
augmenting a collapsed model.  \Cref{fig:augment_intro}, \Cref{fig:augment_interfere} and \Cref{fig:augment_instrument} give three
examples of augmented models, which we have discussed. Other examples
are in \Cref{fig:ID_exs} and \Cref{fig:nonID_exs}.

\begin{algorithm}[t]
\caption{\textbf{Graphical algorithm for augmenting a collapsed model.} This algorithm adds an augmentation variable to a collapsed HCGM, following \Cref{def:valid_augment}.} \label{alg:augment}
\begin{algorithmic}
\State \textbf{Input:} a collapsed model $\Mc^\col$ and an augmentation variable $\tilde Q$
\State \textbf{Output:} an augmented model $\Mc^\aug$
\State add the augmentation variable $\tilde Q_i$ to the graph, with a mechanism $\tilde q_i = \tilde{\f}(q^{\pa(\tilde{q})}_i)$
\State if $\tilde{\f}(q^{\pa(\tilde{q})})$ can be computed from
$q(x^{\mathcal{S}_{\obs}})$, mark $\tilde Q_i$ as observed (filled
circle)
\For {each parent of $\tilde Q_i$}
\State connect the parent to $\tilde Q_i$ via a double arrow
\EndFor
\For{each variable $X^w$ for $w \in \mathcal{U}$ in the collapsed model}
\If{$\tilde{\f}(q_i^{\pa(\tilde{q})})$ appears in the mechanism for $X^w_i$}
\State connect $\tilde Q$ to $X^w$
\State in $X^w$'s mechanism, replace $\tilde{\f}(q_i^{\pa(\tilde{q})})$ with $\tilde{q}_i$
\If{one or more parents of $\tilde{Q}$ no longer appear in $X^w$'s mechanism}
\State erase the arrow from the parent(s) to $X^w$
\EndIf
\EndIf
\EndFor
\end{algorithmic}
\end{algorithm}

We finally turn to marginalization. 
We marginalize an augmented model to identify the effects of interventions on its augmentation variable.
Note we may be interested in these effects for their own sake, or as part of our effort to identify the effects of other interventions in the model.
Once we drop a parent of the augmentation variable from the model, it depends stochastically rather than deterministically on its remaining parents.
This allows interventions on the augmentation variable to satisfy positivity (\Cref{assume:unit_positive}).
For example, we used this approach to achieve identification in the \textsc{subunit instrument} graph in \Cref{sec:instrument_id}.

In detail, we can marginalize out any variable with one or zero children
(\Cref{appx:marginalization}).
Graphically, we erase the variable and
connect its parents directly to its child. We modify the mechanism for the child accordingly, absorbing the mechanism for the marginalized variable.
\Cref{alg:marginalize} provides a graphical algorithm for marginalizing an augmented model.
In addition to the \textsc{subunit instrument} model (\Cref{fig:marginalize_instrument}), \Cref{fig:ID_exs} and \Cref{fig:nonID_exs} give further examples of marginalized models.

\begin{algorithm}[t]
	\caption{\textbf{Graphical algorithm for marginalizing an augmented model.} This algorithm marginalizes out parent(s) of an augmentation variable (\Cref{sec:augment_theory}).} \label{alg:marginalize}
\begin{algorithmic}
\State \textbf{Input:} an augmented collapsed model $\Mc^\aug$ with augmentation variable $\tilde Q$
\State \textbf{Input:} a set $Q^\mathcal{C} \subseteq Q^{\pa(\tilde{q})}$ of one or more parents of $\tilde Q$, for whom $\tilde Q$ is their only child
\State \textbf{Output:} a marginalized model $\Mc^\mar$
\For {each variable $Q^c$ in the set of parents $Q^\mathcal{C}$}
\State connect any parents of $Q^c$ to $\tilde{Q}$
\State erase $Q^c$
\EndFor
\State replace all double arrows into $\tilde Q$ with single arrows
\end{algorithmic}
\end{algorithm}

\parhead{Augmented and marginalized models match the original model} \label{sec:augment_marg_matches} The following propositions justify the use of augmented and marginalized models. They show that we can equate effects in these models to effects in the original collapsed model, and hence to effects in the original HCM.
For simplicity, we state these results under the assumption that the collapsed model has been augmented with just one augmentation variable. 

The first result equates causal effects in an augmented model to causal effects in the original collapsed model. It applies to interventions on any variable besides the augmentation variable itself.
Let $\pr(x^{\Uc},q^{\mathcal{Q}})$ denote the distribution of the original collapsed model, and let $\pr^{\mathrm{aug}}(x^{\Uc}, q^{\mathcal{Q}}, \tilde{q})$ denote the distribution of the augmented model.
\begin{proposition}[Augmented model matches original model] \label{prop:augment_valid}
For any intervention $\Delta$ (soft or hard) on one or more endogenous variable in the original collapsed model, we have, for a valid augmented model,
 $\pr_\Delta^{\mathrm{aug}}(x^{\Uc}, q^{\mathcal{Q}}, \tilde{q}) = \pr^\col_\Delta(x^{\Uc}, q^{\mathcal{Q}}, \tilde{f}(q^{\pa(\tilde{q})}))$ a.e..
\end{proposition}
\noindent The result follows immediately from \Cref{def:valid_augment}. It says the post-intervention distribution over the augmentation variable in the augmented model matches the post-intervention distribution over the quantity the augmentation variable describes in the original model.
For instance, in \Cref{fig:augment_intro} the augmentation variable is $q^y$, and we can conclude $\pr^\aug(q^y \s \rmdo(q^a = q^a_\star)) = \pr^\col(\int q^{y \mid a}(y \mid a) q^a(a) \di a \s  \rmdo(q^a = q^a_\star))$. 

In marginalized models, we are interested in interventions on augmentation variables themselves.
We now show that we can equate the effects of interventions on these augmentation variables to the effects of the corresponding intervention in the original HCGM. 
We will focus on  augmentation variables that describe the conditional distribution of one subunit variable given some (or none) of its parents, i.e. augmentation variables $Q^{v|\mathcal{R}}$ where $\mathcal{R} \subset \pa_\Sc(v)$ (see \Cref{apx:aug_var_form} for a full definition).

The following result equates the effects of the intervention $\rmdo(q^{v | \mathcal{R}} = q_\star^{v | \mathcal{R}})$ in a marginalized model to the effects of the intervention $\rmdo(X^{v} \sim q_\star^{v | \mathcal{R}}(x^v|x_{ij}^\mathcal{R}))$ in the original HCGM.
Let $\pr^{\mar}$ denote the marginalized model distribution.
\begin{proposition}[Augmentation interventions match original model] \label{prop:intervene_valid}
	Consider an augmentation variable $Q^{v | \mathcal{R}}$ where $\mathcal{R} \subset \pa_\Sc(v)$ (with mechanism given by \Cref{eqn:augment_mech} in the appendix).
	Recall $Q^{v | \pa_\Sc(v)}$ is the original $Q$ variable describing $X^{v}$.
	Let $Y$ denote one or more outcome variables.
	Assume either (a) $Q^{v | \pa_\Sc(v)}$ does not appear in the marginalized model, or (b) all directed paths from $Q^{v | \pa_\Sc(v)}$ to $Y$ go through $Q^{v | \mathcal{R}}$.
	Then, $p^{\mar}(y \s \rmdo(q^{v | \mathcal{R}} = q_\star^{v|\mathcal{R}})) = \pr^{\col}(y \s \rmdo(q^{v | pa_\Sc(v)} = q_\star^{v|\mathcal{R}}))$ a.e..
\end{proposition}
\noindent A proof is in \Cref{apx:augment_marginalize}. 
As an example, in \Cref{fig:marginalize_instrument}, the augmentation variable is $Q^a$, while the original $Q$ variable describing $X^a$ is $Q^{a \mid z}$.
The only directed path from $Q^{a \mid z}$ to the outcome $Y$ goes through $Q^a$.
So, we can equate the effect $\pr^\mar(y \s \rmdo(q^a = q_\star^a))$ in the marginalized model to $\pr^\col(y \s \rmdo(q^{a |z} = q_\star^a))$ in the collapsed model, which in turn corresponds to $\pr(y \s \rmdo(A \sim q_\star^a(a)))$ in the original HCGM.

\subsection{When does hierarchy enable identification?} \label{sec:hierarchy_enables_id}

We have described a procedure for proving identification in hierarchical causal graphical models. \Cref{fig:ID_exs} and \Cref{fig:nonID_exs} give examples of graphs where it does and does not lead to identification.
In this section, we investigate general features of HCGM graphs that enable identification, and compare them to flat causal models.
Our results reveal when and where hierarchy enables identification.

We build on the bi-directed path criterion for flat causal models~\citep[Chap. 3]{Tian2002-xr,Pearl2009-fh}.
A path is \textit{bi-directed} if it follows the pattern $Z^1 \gets U^1 \to Z^2 \gets U^2 \to \ldots \gets U^k \to Z^{k+1}$ where the $U^v$ variables are hidden and the $Z^v$ are observed.\footnote{The term \textit{bi-directed} comes from the graphical notation in which one draws a dashed, bidirectional arc between every pair of observed variables affected by the same confounder. In this case, a bi-directed path is one in which every edge is bidirectional.}
For example, in \Cref{fig:nonID_ex2u}, there is a bi-directed path $A_i \gets U_i \to Q_i^{y \mid w} \gets U'_i \to Q_i^{w}$ from $A_i$ to $Q_i^{w}$.
Let $X^\obs$ denote the set of observed endogenous variables in the flat model.
\begin{theorem}[Bi-directed path criterion~\citep{Tian2002-xr}, Thm. 3] \label{thm:bidirected}
The effect $\pr(x^{\obs} \s \rmdo(a =a_\star))$ is identified if and only if there is no bi-directed path between $A$ and any of its children.
\end{theorem}

We now develop similar criteria for hierarchical causal models, by applying the bi-directed path criterion to collapsed/augmented/marginalized models.
Call a node $X^{v'}$ a \textit{subunit instrument} of $X^v$ if both $X^v$ and $X^{v'}$ are subunit-level, $X^v$ is the only child of $X^{v'}$, and $X^v$ has no parents.
\begin{theorem}[Sufficient conditions for identification in hierarchical models] \label{thm:sufficient_ID}
        Consider an HCGM with no hidden subunit-level confounders, and assume the treatment variable $A$ is subunit-level.
        We are interested in the effect $\pr(y \s \rmdo(A \sim q^a_\star))$ if $Y$ is unit-level or $\pr(q(y) \s \rmdo(A \sim q^a_\star))$ if $Y$ is subunit-level.
        Delete from the graph any variable that is not either an ancestor of $Y$ or $Y$ itself.
        The effect is identifiable if (1) there is no bi-directed path from $A$ to a direct unit descendant of $A$, or (2) $A$ has a subunit-level instrument.
\end{theorem}
\noindent The proof is in \Cref{proof:sufficient_ID}. 
Part 1 says that in hierarchical causal models, we can ignore unit-level confounding between subunit variables, except insofar as it leads (via a bi-directed path) to confounding with a direct descendant outside the inner plate 
(examples: \Cref{fig:hcm_confounder}, \Cref{fig:hcm_interfere}, \Cref{fig:ID_ex3h}, \Cref{fig:ID_ex2h} and \Cref{fig:ID_ex6h}).
Part 2 says that subunit-level instruments are a license to ignore unit-level confounding entirely 
(examples: \Cref{fig:hcm_instrument}, \Cref{fig:ID_ex5h} and \Cref{fig:ID_ex4h})

\Cref{thm:sufficient_ID} tells us broadly about the advantages of hierarchy. 
When we disaggregate some quantity and make fine-grained measurements instead, we change it from a unit-level variable to a subunit-level variable (e.g. instead of a school's average test score, we have the per-student test score).
This disaggregation allows us to ignore confounding (bi-directed paths) between the treatment variable and some or all of its children.
In each of the examples in \Cref{fig:ID_exs}, the effect is identified in the HCGM, but would not be if all the subunit variables were unit-level.

There are also general situations where disaggregation is not helpful, that is, making subunit-level rather than unit-level measurements does not aid identification. Below, we compare an HCGM directly to a flat causal model where all the subunit variables are unit-level, but the graph and the observed variables are the same. We refer to this as the \textit{erased inner plate} model, with distribution $\pr^\ep(x^\obs)$ over the observed endogenous variables.
\begin{theorem}[No benefits of hierarchy for unit treatments] \label{thm:unit-level-noID}
  Consider an HCGM with no hidden subunit-level confounders, and assume the treatment variable $A$ is unit-level. 
  If the effect $\pr^\ep(x^\obs \s \rmdo(a = a_\star))$ is not identified in the erased inner plate model, then the effect $\pr(x^{\Uc_\obs}, q(x^{\Sc_\obs}) \s \rmdo(a = a_\star))$ is not identified in the HCGM.
\end{theorem}
\noindent The proof is in \Cref{proof:unit-level-noID}.
\Cref{thm:unit-level-noID} tells us, for example, that the effect of $A$ on $W$ and $Y$ in \Cref{fig:nonID_ex2h} is not identified.
A caveat, however, is that the result only deals with effects on all the observed endogenous variables, not a specific outcome variable $Y$~\citep{Tian2002-xr}.

Naively, we might expect that by measuring in finer detail the mechanisms by which a unit-level treatment affects an outcome, we might better be able to infer the treatment's effects. However, \Cref{thm:unit-level-noID} suggests the benefits of hierarchy for causal identification only accrue when we can measure the treatment itself in finer detail.
Intuitively, subunit-level data is useful for causal identification because it provides information about a natural experiment in which subunit treatments are randomized within each unit.
There is no such natural experiment for unit-level treatments, regardless of whether or not other subunit-level variables are observed.

 \section{Application: Eight Schools}
\label{sec:eight_schools}

We now illustrate the use of hierarchical causal models on a
real-world problem. We analyze data from a well-known study describing a
set of randomized experiments conducted at eight secondary schools in
the United States in 1977~\citep{Alderman1979-on}. At each school,
students were randomly assigned to attend special test preparation
programs or to not attend; the scores of each student on the SAT
verbal component were measured at the end of the program. The goal was
to understand the effects of test preparation programs on test scores.

This ``eight schools'' study is used for textbook illustrations of the
principles of hierarchical Bayesian modeling and
inference~\citep{Rubin1981-ob,Gelman2013-cv}. Here we reanalyze the
data in the framework of hierarchical causal models. We first show how
the standard textbook analysis can be derived as estimation under a
hierarchical causal model. We then show how we can account for a
plausible source of interference and refine the inferences of the
standard analysis.

In this data, each student is randomly assigned to the treatment, the
test preparation program. Let $a_{ij} = 1$ if student $j$ in school
$i$ is treated, and $a_{ij} = 0$ otherwise. At the end of the
preparation program, each student takes the SAT verbal component; let
$y_{ij}$ indicate the score of student $j$ in school $i$. Besides the
test preparation program, a student's pre-treatment academic ability
likely contributes to their outcome $y_{ij}$. For this reason,
researchers also recorded each student's scores on several tests taken
before the program began: PSAT verbal, PSAT mathematics and Test of
Standard Written English. Let $x_{ij}$ denote the scores of student
$j$ in school $i$ on these earlier tests.

We are interested in the average treatment effect on test scores if
all students were enrolled in the special preparation program versus
if the special preparation program were discontinued,
\begin{align}
  \textsc{ate} =
  \EE{\pr}{Y \s \rmdo(a = 1)} -
  \EE{\pr}{Y \s \rmdo(a = 0)}.
\end{align}
We will consider several hierarchical causal models. All the details
of this study are in \Cref{apx:schools_details}, and code reproducing the analysis is in the Supplementary Material.

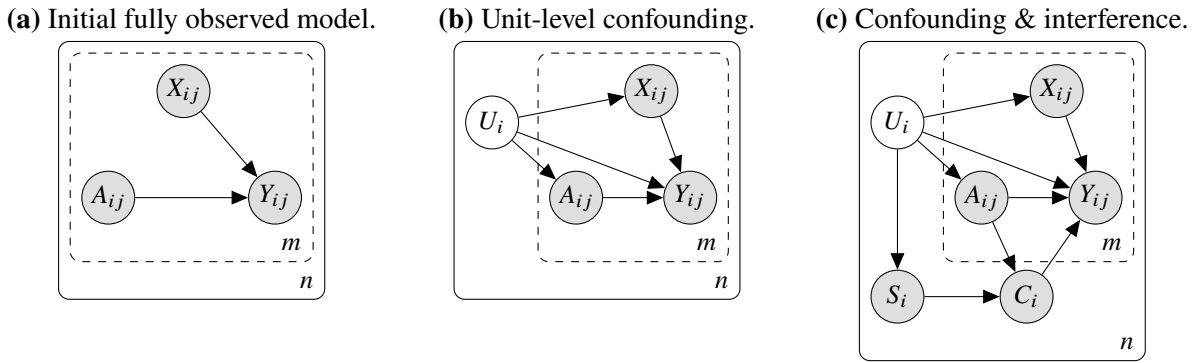
\begin{figure}
\centering
\begin{subfigure}[t]{0.32\textwidth}
\caption{Initial fully observed model.} \label{fig:school_initial_hcm}
\centering
\begin{tikzpicture}

  \node[obs]                  (y) {$Y_{ij}$};
  \node[obs, left=1.5cm of y] (a) {$A_{ij}$};
  \node[obs, above=.7cm of a, xshift=1cm] (x) {$X_{ij}$};

  \edge {a} {y} ;
  \edge {x} {y} ;

  \plate[dashed] {in} {(a)(y)(x)} {$m$} ;
  \plate {out} {(in)} {$n$} ;

\end{tikzpicture}
\end{subfigure}
\begin{subfigure}[t]{0.32\textwidth}
\caption{Unit-level confounding.} \label{fig:school_confound_hcm}
\centering
\begin{tikzpicture}

  \node[obs]                               (y) {$Y_{ij}$};
  \node[obs, left=.8cm of y] (a) {$A_{ij}$};
  \node[obs, above=.7cm of a, xshift=1cm] (x) {$X_{ij}$};
  \node[latent, left=.4cm of a, yshift=1cm] (u) {$U_i$};

  \edge {u} {y} ;
  \edge {u} {x} ;
  \edge {u} {a} ;
  \edge {a} {y} ;
  \edge {x} {y} ;

  \plate[dashed] {in} {(a)(y)(x)} {$m$} ;
  \plate {out} {(in)(u)} {$n$} ;

\end{tikzpicture}
\end{subfigure}
\begin{subfigure}[t]{0.32\textwidth}
\caption{Confounding \& interference.} \label{fig:school_interfere_hcm_0}
\centering
\begin{tikzpicture}

  \node[obs]                                (y) {$Y_{ij}$};
  \node[obs, left=.8cm of y] (a) {$A_{ij}$};
  \node[obs, above=.7cm of a, xshift=1cm] (x) {$X_{ij}$};
  \node[latent, left=.4cm of a, yshift=1cm] (u) {$U_i$};
  \node[obs, below=1.6cm of u] (s) {$S_{i}$};
  \node[obs, right=1cm of s] (c) {$C_{i}$};

  \edge {u} {y} ;
  \edge {u} {x} ;
  \edge {u} {a} ;
  \edge {a} {y} ;
  \edge {x} {y} ;
  \edge {u} {s} ;
  \edge {s} {c} ;
  \edge {a} {c} ;
  \edge {c} {y} ;

  \plate[dashed] {in} {(a)(y)(x)} {$m$} ;
  \plate {out} {(in)(u)(s)(c)} {$n$} ;

\end{tikzpicture}
\end{subfigure}
\caption{\textbf{Models for the eight schools data.} (a) The initial model, in which all variables are observed (\Cref{sec:schools_initial}). (b) An extended model that includes an unobserved unit-level confounder (\Cref{sec:schools_confounding}). (c) An extended model that also includes an observed interferer (\Cref{sec:schools_interfere}).}
\end{figure}

\subsection{A fully observed model} \label{sec:schools_initial}

We first study a fully-observed hierarchical causal model,
\Cref{fig:school_initial_hcm}. Since treatment is randomized, there
are no confounders inside the inner plate, and nor is there an arrow
from $X_{ij}$ to $A_{ij}$. \Cref{fig:school_initial} in the appendix
shows the collapsed, augmented and marginalized models. Applying
do-calculus, we can identify the effect as
\begin{align}
  \label{eq:8-schools_id}
  \textsc{ate} = \mathbb{E}_\pr[\mathbb{E}_{Q^{y|a}}[Y \mid A = 1]] -
  \mathbb{E}_\pr[\mathbb{E}_{Q^{y|a}}[Y \mid A = 0]].
\end{align}

We estimate this effect with hierarchical Bayesian methods. We
parameterize $q^{y|a}_i$ with a linear regression, and parameterize
$\pr(q^{y|a})$ as a normal distribution over its coefficients, with
unknown mean and variance. We perform Bayesian inference on all the
unknown parameters. (There are some subtleties, since the
eight-schools study does not make public its per-student data; see the
details in \Cref{apx:eight_schools_initial}.) The resulting Bayesian
model matches the textbook eight schools model~\citep{Gelman2013-cv}.
We compute the posterior over the ATE using MCMC, specifically the
No-U-turn Hamiltonian Monte Carlo sampler (NUTS) in
NumPyro~\citep{Hoffman2014-ic,Phan2019-om,Bingham2019-aa}.
\Cref{fig:school_ATE_posterior} shows the results (blue distribution).
This analysis suggests the treatment is likely to increase test scores a modest amount: each question on the SAT verbal is worth an average of 7 points, and the posterior mean and standard deviation of the ATE are 4.4 points and 3.4 points respectively.

\subsection{Unit-level confounding}
\label{sec:schools_confounding}

Whether each student attends the test preparation program is
randomized within each school, so there are no subunit-level
confounders between treatment and outcome. But there may still be
unit-level confounders. For example, each school's financial and
administrative resources may affect both student test scores and
student enrollment in the program. Indeed, \citet{Alderman1979-on}
report that ``where student interest far exceeded the program's
capacity [...], a larger number of students went into the control
group than into the treatment group.'' In the data, the fraction of students treated varies between 25\% and 55\% across different schools (\Cref{fig:frac_treated_v_effect}).
It seems plausible that schools
with greater financial resources could have larger and more
effective programs.

\Cref{fig:school_confound_hcm} considers the possibility of unit-level
confounders. Note the graph also allows confounders to impact $X$, the
pre-treatment test scores. 
To identify the causal effect, we first collapse, augment, and marginalize (\Cref{fig:school_confound}). Then, applying do-calculus recovers the same identification
formula for the \textsc{ate} as above in \Cref{eq:8-schools_id}. 
Intuitively, the impact of the confounder is through the conditional distribution of $Y$ given $A$ within each school, so to correct for confounding it suffices to control for this conditional. 
In short, the estimation method of the previous section is robust to unobserved unit-level confounding.

\subsection{Confounding \& interference}
\label{sec:schools_interfere}

There is evidence that increased class size can negatively impact
students' academic performance~\citep[e.g.][]{Angrist1999-hm}. Since
the level of enrollment in each school's tutoring programs presumably
influences their class size, this leads to the possibility of
interference. In the data, \citet{Alderman1979-on} report the class
sizes for each test preparation program at each school, and we can
visually examine its relationship to the per-school estimated
treatment effect; see \Cref{fig:class_v_effect}. It
seems plausible that class size impacts test scores.

To address this formally, we consider the hierarchical causal model in
\Cref{fig:school_interfere_hcm_0}. Here $C_i$ is the class size and
$S_i$ is the total number of students who expressed interest in the
tutoring program. This model allows for the possibility of
interference: enrolling more students in the test preparation program
may drive up class size, which in turn may drive down students' test
scores.

The collapsed, augmented, and marginalized models are shown in
\Cref{fig:school_interfere} in the appendix. Applying do-calculus, the
interventional expectation
$\EE{\pr}{Y \s \rmdo(q^{a} = \delta_{a_\star})}$ can be identified as
\begin{align}
  \label{eqn:schools_interfere_id}
  \int \int
  \pr(s) \,
  \pr(c \mid q^{a} = \delta_{a_\star}, s) \,
  \di s \,
  \int
  \pr({q}^{a}, s') \,
  \EE{\pr}{\EE{Q^{y|a}}{Y \g A = a_\star} \mid {q}^{a}, s', c} \,
  \di {q}^{a} \,
  \di s' \,
  \di c.
\end{align}
We identify the ATE by identifying the expectation for $a_\star = 1$
and $a_\star = 0$.

We develop a hierarchical Bayesian estimation strategy. Again we
parameterize $q^{y|a}_i$ with a linear model. We parameterize $q^a_i$
with a Bernoulli distribution with unknown mean. We parameterize
$\pr(c \g q^a, s)$ with a linear regression that predicts $c$ from $s$
and the mean of $q^a$. We parameterize $\pr(q^{y|a} \g q^a, s, c)$
with a linear regression that predicts the coefficient of the linear
model of $q^{y|a}$ based on $s$, $c$ and the mean of $q^a$. We place
priors on all parameters and perform Bayesian inference, again using
MCMC. We use each sample from the MCMC procedure to form a Monte Carlo
approximation of \Cref{eqn:schools_interfere_id}. All the details of
this procedure are in \Cref{apx:eight_schools_interfere}.

\begin{figure}
\centering
\includegraphics[width=0.4\textwidth]
{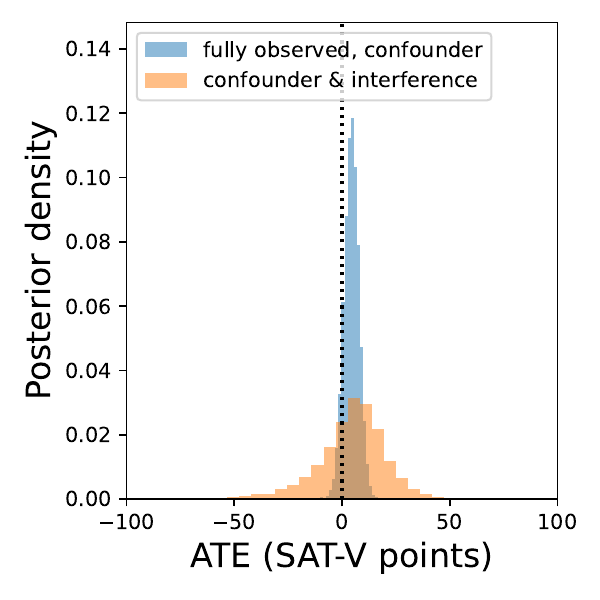}
\caption{\textbf{Posterior over average treatment effect for eight
    schools study.} In blue is the posterior over the ATE for the models discussed in \Cref{sec:schools_initial} and \Cref{sec:schools_confounding}. In orange is the posterior for the model in \Cref{sec:schools_interfere}. We find, when we take into account possible interference, more uncertainty in the effect of the preparation program on test scores.} \label{fig:school_ATE_posterior}
\end{figure}

With this estimation strategy, \Cref{fig:school_ATE_posterior} plots the
posterior ATE (orange). The posterior mean of the ATE ($4.2$ points)
is similar to that for the classical analysis of the
initial fully observed/confounder model ($4.4$ points). But there is substantially
more uncertainty when we account for interference. The 5th percentile
of the ATE posterior is $-26$ points and the 95th percentile is $28$
points, whereas for the classic analysis they are $-1.2$ points and
$9.8$ points respectively. In summary, the standard eight schools
analysis suggests that the test preparation program is likely to be
modestly effective, if rolled out to all students. But when we allow
for the possibility of school-level confounders and interference
through class size, the program's effectiveness is more
uncertain.

 \section{Discussion} \label{sec:discussion}

We proposed hierarchical causal models and studied identification and estimation. HCMs are a general
tool for addressing causal questions using hierarchical data. We
developed proof techniques for identifying causal effects in arbitrary
HCMs, without parametric assumptions on causal mechanisms. We
developed estimation methods based on hierarchical probabilistic models,
and found deep connections to hierarchical Bayesian methods.
These methods enable causal inference in novel hierarchical data settings with novel causal graphs.

A popular rule-of-thumb for handling hierarchical data in causal inference is to include the index of the environment or group as a dummy variable in the regression. 
This approach is well justified when the unit just contributes hidden confounding (the \textsc{unit confounder} graph). But in many real world scenarios, there is more complexity at the unit level, such as sources of interference or outcomes of interest. 
To model what is happening at the unit level, we cannot simply replace all these unit variables with a dummy variable in a regression. 
This is reflected in the distinct identification formula of our three HCM examples.

\subsection{Assumptions and limitations}

Our methods have several important assumptions and limitations.
A defining aspect of the HCM framework is that units are exchangeable and subunits are exchangeable within units.
This assumption is what connects HCMs to hierarchical probabilistic graphical models with inner plates, ensuring that any HCM reduces to such a model in the absence of intervention.
However, exchangeability is not always appropriate, for instance when different units or subunits correspond to nearby points in time or space~\citep{Bertrand2004-sg}.
An important area of investigation is how to relax the exchangeability assumption.
As one example in this direction, \citet{Christiansen2022-ip} present a
nonparametric causal model similar to the \textsc{unit confounder} model that
accounts for spatiotemporal correlation.

A key restriction of our identification results is that they hold only in the limit of infinite subunits, and only for HCMs whose mechanisms converge in this limit.
An advantage of the infinite subunit limit is that it allows us to avoid placing parametric assumptions on the distribution over subunit variables.
In reality, however, there may not be very many subunits within each unit.
To help address such situations, it will be important to explore reasonable assumptions that enable identification from small numbers of subunits.

Our identification theory addresses unobserved unit-level confounders and interferers, but there are many other threats to valid causal inference in HCMs.
Going forward, it will be important to investigate issues such as unobserved subunit-level confounders, selection bias among units or subunits, and more general forms of interference among units and subunits.

\subsection{Outlook}

Broadly speaking, HCMs help formalize the question of when
reductionism---in the sense of analyzing an aggregate phenomena in
terms of its individual component parts---enables understanding of
cause and effect. On one hand, we see that reductionism can be an
enabler for causal inference. By looking at individual subunits,
instead of aggregate unit-level variables, we can effectively hold
unit-level confounders fixed while randomizing subunit treatments. On
the other hand, HCMs also show that reductionism offers no advantage
when we are interested in unit-level treatments. Without randomness at
the subunit level, we are left with the usual requirements for causal
inference.

HCMs provide data analysis methods that leverage technological progress in measurement and intervention methods.
Across many scientific domains, technological advances lead to unit-level data being supplemented or supplanted by subunit-level data.
Consider, for example, a political scientist interested in the impact of news consumption on political behavior.
In the past, they may have had to rely data such as the subscription levels of different newspapers in different cities~\citep{Gentzkow2014-lm}.
Modern media apps, however, enable measurement of the exact news articles read by individuals~\citep{Gonzalez-Bailon2023-ya}.
So, unit-level data about groups of citizens can be replaced by subunit-level data about individual citizens.
These apps further enable interventions on the articles recommended to individuals, a targeted intervention on subunits~\citep{Guess2023-uv}.
HCMs thus offer one potential tool for leveraging this novel technology to better understand the effects of news media.
Moreover, because our HCM identification results are nonparametric, they do not only apply to binary or continuous variables. 
Instead, they can be applied to complex structured data, such as that recorded by apps. For example, one can treat the entire text of news articles as a treatment or an outcome variable~\citep{Feder2022-tk,Egami2022-kk}.

Analogous technological advances are also being made in fields outside social science.
Consider a biologist interested in the effects of gene expression on disease progression~\citep{Tejada-Lapuerta2023-wj}.
In the past, they would have had to rely on bulk gene expression measurements, taking the average expression levels within a tissue. 
With the development of single cell RNA sequencing, expression levels can be measured in individual cells~\citep{Klein2015-kg}.
So, unit-level data about tissues can be replaced by subunit-level data about their constituent cells.
Meanwhile, advances in synthetic biology enable targeted modification of gene expression levels in specific cell types, i.e. conditional soft subunit interventions~\citep{Hrvatin2019-zy}.
HCMs thus offer a possible tool for leveraging single cell data to inform emerging therapeutic strategies.

Theoretically, a central open problem is
finding an identification method for HCMs that is \textit{complete}, in the
sense that if an effect cannot be identified via the method then it is
not identified. The do-calculus is
complete~\citep{Shpitser2006-jg,Huang2006-ba}, and our identification
method rests on application of do-calculus to the collapsed model. But many HCMs have collapsed models that are not fully nonparametric, even though the HCM itself is nonparametric. For example, in the collapsed \textsc{subunit instrument} graph,
the outcome variable $Y_i$ depends on its parents $Q^{a|z}$ and $Q^z$
only through the marginal that they induce. Consequently, there can be
effects that are identified even when do-calculus says they are not.

Another open direction is to advance estimation. We gave some estimation methods, including both hierarchical point estimators (\Cref{sec:identification_problem}) and hierarchical Bayesian models, which account for statistical uncertainty (\Cref{sec:eight_schools}). To improve sample efficiency and robustness, it may be valuable to develop influence function-based corrections, along the lines of doubly robust estimators or targeted learning~\citep{Kennedy2022-ml}.
It is an open question how to extend these and other causal estimation approaches to estimands from hierarchical causal models.

\section*{Acknowledgments}
We thank the anonymous reviewers for suggestions that improved the manuscript, and Jesse Geneson for alerting us to an issue in \Cref{prop:exp-family-converge} in \Cref{sec:exp-converge}.
\bibliography{references}

\pagebreak
\appendix

{
\centering
\Large Supplementary Material
}
\renewcommand{\thefigure}{A\arabic{figure}}
\setcounter{figure}{0}
\renewcommand{\thesection}{\Alph{section}}
\setcounter{section}{0}
\renewcommand{\thetable}{A\arabic{table}}
\setcounter{table}{0}

\begin{table}[h!]
\centering
\begin{tabular}{c c c}
	\textbf{Symbol} & \textbf{Description} & \textbf{Definition}\\
	\hline\\
	$\pr$ & Distribution of an HCM. & \Cref{def:hscm,def:hcgm}\\
	$\f$ & Deterministic mechanism in an HSCM. & \Cref{def:hscm}\\
	$\mathcal{G}$ & Graph of an HCM. & \Cref{def:hscm} \\
	$\mathcal{V} = \{1, \ldots, V\}$ & Indices of the endogenous variables. & \Cref{sec:general-hscm}\\
	$X^v, X^w$ & The $v$th, $w$th endogenous variables. & \Cref{sec:general-hscm} \\
	$\Sc \subseteq \mathcal{V}$ & Indices of subunit endogenous variables. & \Cref{sec:general-hscm} \\
	$\Uc = \mathcal{V} \setminus \Sc$ & Indices of unit endogenous variables. & \Cref{sec:general-hscm} \\
	$\pa(v) \subseteq \mathcal{V}$ & Indices of the parents of $X^v$ in the graph $\mathcal{G}$. & \Cref{sec:general-hscm} \\
	$\pa_\Sc(v), \pa_\Uc(v)$ & Indices of the subunit, unit parents of $X^v$. & \Cref{sec:general-hscm} \\
	$m$ & Number of subunits. & \Cref{def:hscm}\\
	$n$ & Number of units. & \Cref{def:hscm}\\
	$\gamma_i^v$ & Unit noise affecting $X^v$. & \Cref{def:hscm} \\
	$\epsilon_{ij}^v$ & Subunit noise affecting $X^v$. & \Cref{def:hscm} \\
	$\{x_{j}\}_{j=1}^m$ & Shorthand for the multiset $\{x_{1}, \ldots, x_m\}$. & \Cref{sec:hscm-intro} \\
	$\Mc^\scm$ & Hierarchical structural causal model (HSCM). & \Cref{def:hscm} \\
	$\Delta$ & Intervention on an HCM. & \Cref{def:intervention} \\
	$\pr_{\Delta}$ & Post-intervention distribution of an HCM. & \Cref{thm:valid_collapse}.\\
	$\mathcal{I} \subseteq \mathcal{V}$ & Indices of intervened variables. & \Cref{def:intervention} \\
	$x_\star^v$ & Value of $X^v$ in an intervention. & \Cref{def:intervention} \\ 
	$q_\star^v$ & Distribution of $X^v$ in an intervention. & \Cref{def:intervention}.\\
	$\rmdo(x^v = x_\star)$ & A hard intervention that sets $X^v$ to $x_\star$. & \Cref{def:intervention} \\
	$\rmdo(X^v \sim q_\star^v)$ & A soft intervention that draws $X^v$ from $q_\star$. & \Cref{def:intervention} \\
	$Q(x^\mathcal{S})$ & Within-unit distribution over subunit variables. & \Cref{sec:identification_problem} \\
	$Q^{v}, Q^{v|w}$ & $Q$ variables in an HCGM/collapsed model. & \Cref{def:hcgm,def:um}/ \\
	& & \Cref{eqn:augment_mech}\\
	$\Mc^\cgm$ & Hierarchical causal graphical model (HCGM). & \Cref{def:hcgm} \\
	$\Sc_\obs \subseteq \mathcal{S}$ & Indices of observed subunit variables. & \Cref{sec:identification_problem} \\
	$\Uc_\obs \subseteq \mathcal{U}$ & Indices of observed unit variables. & \Cref{sec:identification_problem} \\
	$\da_\Sc(v)$ & Indices of direct subunit ancestors of $X^v$. & \Cref{sec:collapse_theory} \\
	$\dd_\Uc(v)$ & Indices of direct unit descendants of $X^v$. & \Cref{sec:collapse_theory} \\
	$\Mc^\col$ & Collapsed model. & \Cref{def:um} \\
	$\pr^\col$ & Distribution of collapsed model. & \Cref{def:um}, \Cref{thm:valid_collapse} \\
	$\Mc^\aug$ & Augmented model. & \Cref{def:valid_augment}\\
	$\pr^\aug$ & Distribution of augmented model. & \Cref{def:valid_augment}, \Cref{sec:augment_marg_matches}\\
	$\Mc^\mar$ & Marginalized model. & \Cref{sec:augment_theory}\\
	$\pr^\mar$ & Distribution of marginalized model. & \Cref{sec:augment_theory}\\
	$q^\mathcal{Q}$ & Set of $Q$ variables in a collapsed model. & \Cref{sec:collapse_theory}\\
	$q^{\mathcal{Q}_\obs}$ & Set of observed $Q$ variables in a collapsed model. & \Cref{sec:collapse_theory}\\
	$\mathcal{M}^{\ep}$ & Erased inner plate model. & \Cref{sec:hierarchy_enables_id}\\
	$\pr^{\ep}$& Distribution of erased inner plate model.& \Cref{sec:hierarchy_enables_id}
\end{tabular}
\caption{\textbf{Notation.}} \label{table:notation}
\end{table}

\begin{figure}[h]
\centering

\begin{minipage}{0.45\textwidth}

\begin{tikzpicture}

\node[obs]                               (y) {$Y_{i1}$};
  \node[obs, left=1.5cm of y] (a) {$A_{i1}$};
  \node[obs, below=.3cm of a] (a2) {$A_{i2}$};
  \node[obs, below=.3cm of y] (y2) {$Y_{i2}$};
  \node[obs, below=.3cm of a2] (a3) {$A_{i3}$};
  \node[obs, below=.3cm of y2] (y3) {$Y_{i3}$};
  \node[latent, above=.3cm of a, xshift=1cm]  (u) {$U_i$};
  \node[obs, below=.4cm of a3, xshift=1cm] (z) {$Z_i$};

\edge {a,u} {y} ;
  \edge {u} {a} ;
  \edge {a2,u} {y2} ;
  \edge {u} {a2} ;
  \edge {a3,u} {y3} ;
  \edge {u} {a3} ;
  \edge {a,a2,a3} {z} ;
  \edge {z} {y,y2,y3} ;

\plate {out} {(u)(z)(a)(y)} {$n$} ;

\end{tikzpicture}
\caption{A flat causal model corresponding to the \textsc{unit confounder \& unit interference} hierarchical causal  model (\Cref{fig:hcm_interfere}) with the inner plate expanded, for $m = 3$ subunits. } \label{fig:expanded_inner_plate_ex}

\end{minipage} \quad
\begin{minipage}{0.45\textwidth}
\begin{tikzpicture}

  \node[obs]                               (y) {$Y_{i}$};
  \node[obs, left=1.8cm of y] (qaz) {$\scriptstyle Q^{a\mid z}_i$};
  \node[obs, below=.4cm of qaz, xshift=.2cm] (qz) {$\scriptstyle Q^{z}_i$};
  \node[obs, left=.4cm of y] (qa) {$\scriptstyle Q^{a}_i$};
  \node[latent, above=.5cm of qa]  (u) {$U_i$};

  \edge {u} {qaz, y} ;
  \edge {qa} {y} ;
  \path (qaz) edge [very thick, ->]  (qa) ;
  \path (qaz) edge [color=white, thick] (qa) ;
  \path (qz) edge [very thick, ->]  (qa) ;
  \path (qz) edge [color=white, thick] (qa) ;

  \plate {out} {(qz)(qaz)(u)(y)} {$n$} ;

\end{tikzpicture}
\caption{Augmented \textsc{subunit instrument} model.} \label{fig:augment_instrument}
\end{minipage}

\end{figure} 
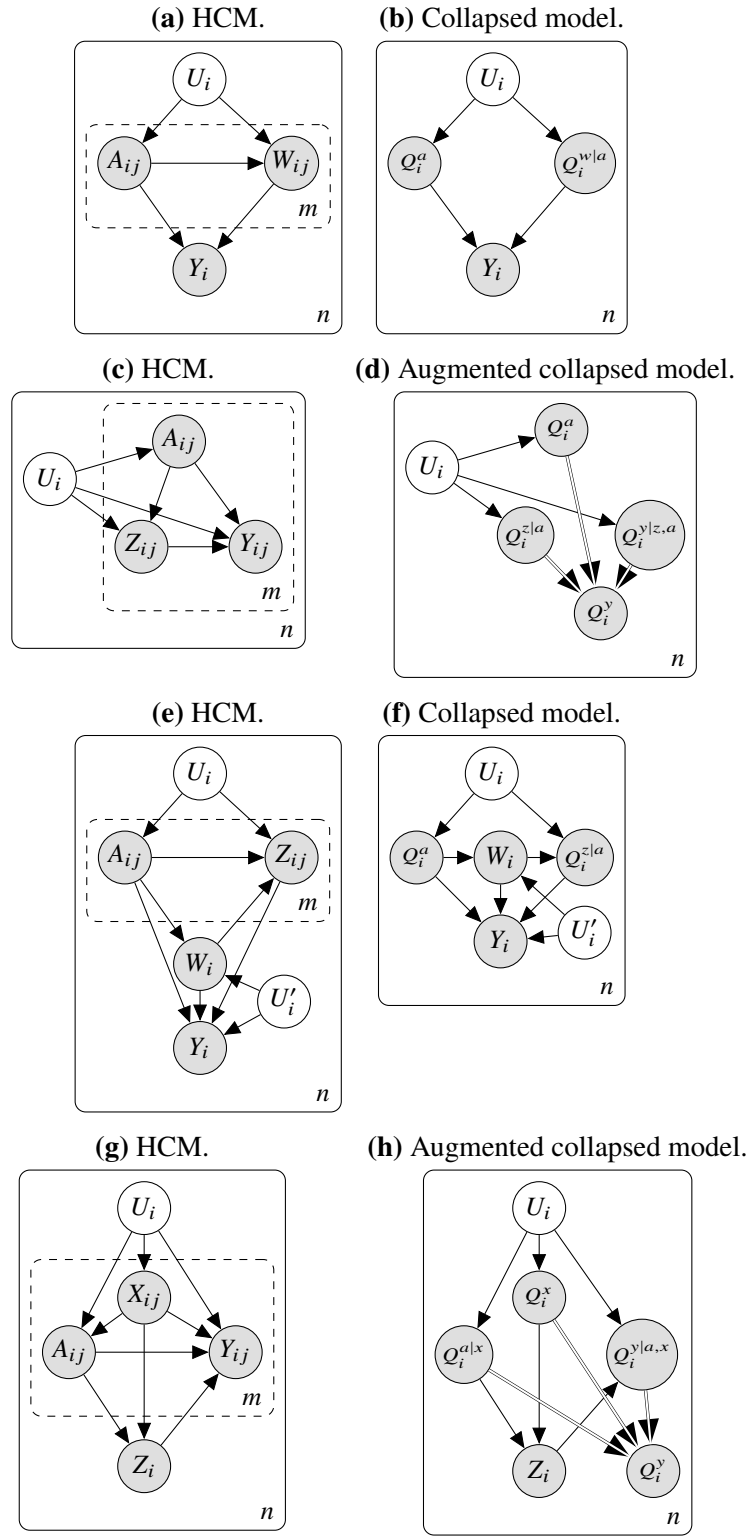
\begin{figure}[p]
\centering
\begin{subfigure}[t]{0.23\textwidth}
\caption{HCM.} \label{fig:ID_ex3h}
\centering
\begin{tikzpicture}

\node[obs]                               (w) {$W_{ij}$};
  \node[obs, left=1.5cm of w] (a) {$A_{ij}$};
  \node[latent, above=.4cm of a, xshift=1cm]  (u) {$U_i$};
  \node[obs, below=0.7cm of a, xshift=1cm] (y) {$Y_i$};

\edge {a,u} {w} ; \edge {u} {a} ;
  \edge {a} {y} ;
  \edge {w} {y} ;

\plate[dashed] {in} {(a)(w)} {$m$} ;
  \plate {out} {(in)(u)(y)} {$n$} ;

\end{tikzpicture}
\end{subfigure}
\begin{subfigure}[t]{0.23\textwidth}
\caption{Collapsed model.} \label{fig:ID_ex3u}
\centering
\begin{tikzpicture}

\node[obs]                               (qwa) {$\scriptstyle Q^{w\mid a}_{i}$};
  \node[obs, left=1.5cm of qwa] (qa) {$\scriptstyle Q^{a}_{i}$};
  \node[latent, above=.4cm of a, xshift=1cm]  (u) {$U_i$};
  \node[obs, below=.7cm of a, xshift=1cm] (y) {$Y_i$};

\edge {u} {qwa,qa} ;
  \edge {qa} {y} ;
  \edge {qwa} {y} ;

\plate {out} {(qa)(qwa)(u)(y)} {$n$} ;

\end{tikzpicture}
\end{subfigure}\\
\begin{subfigure}[t]{0.3\textwidth}
\caption{HCM.}\label{fig:ID_ex2h}
\centering
\begin{tikzpicture}
\node[latent] (u) {$U_i$} ;
\node[obs, right=1cm of u, yshift=0.5cm] (a) {$A_{ij}$};
\node[obs, right=.5cm of u, yshift=-0.9cm] (z) {$Z_{ij}$};
\node[obs, right=.8cm of z] (y) {$Y_{ij}$};

\edge {u} {a,z,y} ;
\edge {a} {z,y} ;
\edge {z} {y} ;
\plate[dashed] {in} {(a)(z)(y)} {$m$};
\plate {out} {(in)(u)} {$n$};

\end{tikzpicture}
\end{subfigure}
\begin{subfigure}[t]{0.31\textwidth}
\caption{Augmented collapsed model.}\label{fig:ID_ex2u}
\centering
\begin{tikzpicture}
\node[latent] (u) {$U_i$} ;
\node[obs, right=1cm of u, yshift=0.5cm] (qa) {$\scriptstyle Q^{a}_{i}$};
\node[obs, right=.5cm of u, yshift=-0.9cm] (qza) {$\scriptstyle Q^{z\mid a}_{i}$};
\node[obs, right=.8cm of qza] (qyaz) {$\scriptstyle Q^{y \mid z, a}_{i}$};
\node[obs, below=.3cm of qza, xshift=1cm] (qy) {$\scriptstyle Q^{y}_i$};

\edge {u} {qa,qza,qyaz} ;
\path (qa) edge [very thick, ->]  (qy) ;
\path (qa) edge [color=white, thick] (qy) ;
\path (qza) edge [very thick, ->]  (qy) ;
\path (qza) edge [color=white, thick] (qy) ;
\path (qyaz) edge [very thick, ->]  (qy) ;
\path (qyaz) edge [color=white, thick] (qy) ;

\plate {out} {(qa)(qza)(qyaz)(u)(qy)} {$n$};

\end{tikzpicture}
\end{subfigure}\\ 
\begin{subfigure}[t]{0.23\textwidth}
\caption{HCM.} \label{fig:ID_ex6h}
\centering
\begin{tikzpicture}

\node[obs]                               (z) {$Z_{ij}$};
  \node[obs, left=1.5cm of z] (a) {$A_{ij}$};
  \node[latent, above=.4cm of a, xshift=1cm]  (u) {$U_i$};
  \node[obs, below=0.7cm of a, xshift=1cm] (w) {$W_i$};
  \node[obs, below=0.4cm of w] (y) {$Y_i$};
  \node[latent, right=.4cm of w, yshift=-0.5cm] (up) {$U'_i$};

\edge {a,u} {z} ; \edge {u} {a} ;
  \edge {a} {w} ;
  \edge {w} {z} ;
  \edge {a} {y} ;
  \edge {w} {y} ;
  \edge {z} {y} ;
  \edge {up} {w,y};

\plate[dashed] {in} {(a)(z)} {$m$} ;
  \plate {out} {(in)(u)(y)} {$n$} ;

\end{tikzpicture}
\end{subfigure}
\begin{subfigure}[t]{0.23\textwidth}
\caption{Collapsed model.} \label{fig:ID_ex6u}
\centering
\begin{tikzpicture}

\node[obs]                               (qza) {$\scriptstyle Q^{z\mid a}_{i}$};
  \node[obs, left=1.5cm of qza] (qa) {$\scriptstyle Q^{a}_{i}$};
  \node[latent, above=.4cm of qa, xshift=1cm]  (u) {$U_i$};
  \node[obs, right=.4cm of qa] (w) {$W_i$};
  \node[obs, below=0.4cm of w] (y) {$Y_i$};
  \node[latent, right=.4cm of w, yshift=-1cm] (up) {$U'_i$};

\edge {u} {qza,qa} ;
  \edge {qa} {w} ;
  \edge {w} {qza} ;
  \edge {qa} {y} ;
  \edge {w} {y} ;
  \edge {qza} {y} ;
  \edge {up} {w,y};

\plate {out} {(qa)(qza)(u)(y)} {$n$} ;

\end{tikzpicture}
\end{subfigure}\\
        \begin{subfigure}[t]{.32\textwidth}
\caption{HCM.} \label{fig:ID_targeted}
\centering
\begin{tikzpicture}

  \node[obs]                               (y) {$Y_{ij}$};
  \node[obs, left=1.5cm of y] (a) {$A_{ij}$};
  \node[obs, above=.01cm of a, xshift=1cm] (x) {$X_{ij}$};
  \node[latent, above=1.2cm of a, xshift=1cm]  (u) {$U_i$};
  \node[obs, below=.8cm of a, xshift=1cm] (z) {$Z_i$};

  \edge {a,u} {y} ;
  \edge {u} {a} ;
  \edge {a} {z} ;
  \edge {z} {y} ;
  \edge {u} {x} ;
  \edge {x} {a} ;
  \edge {x} {y} ;
  \edge {x} {z} ;

  \plate[dashed] {in} {(a)(y)(x)} {$m$} ;
  \plate {out} {(in)(u)(z)(x)} {$n$} ;

\end{tikzpicture}
\end{subfigure}
\begin{subfigure}[t]{0.32\textwidth}
\caption{Augmented collapsed model.}
\centering
\begin{tikzpicture}

  \node[obs]                               (y) {$\scriptstyle Q^{y\mid a, x}_{i}$};
  \node[obs, left=1.5cm of y] (a) {$\scriptstyle Q^{a\mid x}_{i}$};
  \node[obs, above=.01cm of a, xshift=1cm] (x) {$\scriptstyle Q^x_{i}$};
  \node[latent, above=1.2cm of a, xshift=1cm]  (u) {$U_i$};
  \node[obs, below=.8cm of a, xshift=1cm] (z) {$Z_i$};
  \node[obs, right=.8cm of z] (qy) {$\scriptstyle Q^y_{i}$};

  \edge {u} {y,x,a} ;
  \edge {a,x} {z} ;
  \edge {z} {y} ;
\path (x) edge [very thick, ->]  (qy) ;
\path (x) edge [color=white, thick] (qy) ;
\path (a) edge [very thick, ->]  (qy) ;
\path (a) edge [color=white, thick] (qy) ;
\path (y) edge [very thick, ->]  (qy) ;
\path (y) edge [color=white, thick] (qy) ;

  \plate {out} {(in)(u)(z)(x)(qy)(a)} {$n$} ;

\end{tikzpicture}
\end{subfigure}
\caption{\textbf{Examples of hierarchical causal models where the effect of $A$ on $Y$ is identified.} Each row shows a hierarchical causal model (first plot on the left) and graphs derived from it. In each case, the effect of $A$ on $Y$ would not be identifiable if the inner plate were erased. \textit{Figure continues on the next page.}} \label{fig:ID_exs}. 
\end{figure}
\addtocounter{figure}{-1}
\begin{figure}[h]

\begin{subfigure}[t]{0.32\textwidth}
\addtocounter{subfigure}{9}
\caption{HCM.}\label{fig:ID_ex5h}
\centering
\begin{tikzpicture}
\node[obs] (a) {$A_{ij}$};
\node[obs, left=.5cm of a] (z) {$Z_{ij}$};
\node[obs, below=.8cm of a] (x) {$X_{ij}$};
\node[obs, right=.8cm of x] (y) {$Y_i$};
\node[latent, right=.7cm of a, yshift=.4cm] (u) {$U_i$};

\edge {z} {a} ;
\edge {u} {a,x,y} ;
\edge {x} {a,y} ;
\edge {a} {y} ;
\plate[dashed] {in} {(a)(x)(z)} {$m$};
\plate {out} {(in)(u)(y)} {$n$};

\end{tikzpicture}
\end{subfigure}
\begin{subfigure}[t]{0.32\textwidth}
\caption{Augmented collapsed model.}\label{fig:ID_ex5u}
\centering
\begin{tikzpicture}
\node[obs] (qazx) {$\scriptstyle Q^{a\mid z, x}_{i}$};
\node[obs, left=.7cm of a, yshift=-.8cm] (qz) {$\scriptstyle Q^{z}_{i}$};
\node[obs, below=1cm of qazx] (qax) {$\scriptstyle Q^{a \mid x}_{i}$};
\node[obs, right=.5cm of y] (qx) {$\scriptstyle Q^{x}_{i}$};
\node[obs, right=.6cm of qax] (y) {$Y_i$};
\node[latent, right=.7cm of a] (u) {$U_i$};

\edge {u} {y, qazx, qx} ;
\edge{qx} {y};
\edge{qax} {y};

\path (qz) edge [very thick, ->]  (qax) ;
\path (qz) edge [color=white, thick] (qax) ;
\path (qazx) edge [very thick, ->]  (qax) ;
\path (qazx) edge [color=white, thick] (qax) ;

\plate {out} {(qazx)(u)(qx)(y)(qz)} {$n$};

\end{tikzpicture}
\end{subfigure}
\begin{subfigure}[t]{0.32\textwidth}
\caption{Marginalized model.}\label{fig:ID_ex5m}
\centering
\begin{tikzpicture}
\node[obs] (qazx) {$\scriptstyle Q^{a\mid z, x}_{i}$};
\node[obs, below=1cm of qazx] (qax) {$\scriptstyle Q^{a \mid x}_{i}$};
\node[obs, right=.5cm of y] (qx) {$\scriptstyle Q^{x}_{i}$};
\node[obs, right=.6cm of qax] (y) {$Y_i$};
\node[latent, right=.7cm of a] (u) {$U_i$};

\edge {u} {y, qazx, qx} ;
\edge{qx} {y};
\edge{qax} {y};
\edge {qazx} {qax};

\plate {out} {(qazx)(u)(qx)(y)} {$n$};

\end{tikzpicture}
\end{subfigure}\\
\begin{subfigure}[t]{0.32\textwidth}
\caption{HCM.}\label{fig:ID_ex4h}
\centering
\begin{tikzpicture}
\node[obs] (a) {$A_{ij}$};
\node[obs, left=.5cm of a] (z) {$Z_{ij}$};
\node[obs, below=.8cm of a] (y) {$Y_{ij}$};
\node[obs, right=.5cm of a, yshift=-.8cm] (w) {$W_i$};
\node[latent, right=.7cm of a, yshift=.4cm] (u) {$U_i$};

\edge {z} {a} ;
\edge {u} {a,w,y} ;
\edge {a} {w,y} ;
\edge {w} {y} ;
\plate[dashed] {in} {(a)(z)(y)} {$m$};
\plate {out} {(in)(u)(w)} {$n$};

\end{tikzpicture}
\end{subfigure}
\begin{subfigure}[t]{0.32\textwidth}
\caption{Augmented collapsed model.}\label{fig:ID_ex4u}
\centering
\begin{tikzpicture}
\node[obs] (qaz) {$\scriptstyle Q^{a\mid z}_{i}$};
\node[obs, below=.4cm of qaz, xshift=0.8cm] (qa) {$\scriptstyle Q^{a}_{i}$};
\node[obs, left=.6cm of qa] (qz) {$\scriptstyle Q^{z}_{i}$};
\node[obs, right=.5cm of qa] (w) {$W_i$};
\node[latent, right=.7cm of qaz, yshift=.4cm] (u) {$U_i$};
\node[obs, right=.4cm of w] (qya) {$\scriptstyle Q^{y\mid a}_{i}$};
\node[obs, below=.4cm of w] (qy) {$\scriptstyle Q^{y}_i$};

\edge {u} {qaz};
\edge {u} {w} ;
\edge {qa} {w};
\edge {w} {qya};
\edge {u} {qya};
\path (qaz) edge [very thick, ->]  (qa) ;
\path (qaz) edge [color=white, thick] (qa) ;
\path (qz) edge [very thick, ->]  (qa) ;
\path (qz) edge [color=white, thick] (qa) ;
\path (qa) edge [very thick, ->]  (qy) ;
\path (qa) edge [color=white, thick] (qy) ;
\path (qya) edge [very thick, ->]  (qy) ;
\path (qya) edge [color=white, thick] (qy) ;

\plate {out} {(qz)(u)(qya)(qy)} {$n$};

\end{tikzpicture}
\end{subfigure}
\begin{subfigure}[t]{0.32\textwidth}
\caption{Marginalized model.}\label{fig:ID_ex4m}
\centering
\begin{tikzpicture}
\node[obs] (qaz) {$\scriptstyle Q^{a\mid z}_{i}$};
\node[obs, below=.4cm of qaz, xshift=0.8cm] (qa) {$\scriptstyle Q^{a}_{i}$};
\node[obs, right=.5cm of qa] (w) {$W_i$};
\node[latent, right=.7cm of qaz, yshift=.4cm] (u) {$U_i$};
\node[obs, right=.4cm of w] (qya) {$\scriptstyle Q^{y\mid a}_{i}$};
\node[obs, below=.4cm of w] (qy) {$\scriptstyle Q^{y}_i$};

\edge {u} {qaz};
\edge {u} {w} ;
\edge {qa} {w};
\edge {w} {qya};
\edge {qaz} {qa};
\edge {u} {qya};
\path (qa) edge [very thick, ->]  (qy) ;
\path (qa) edge [color=white, thick] (qy) ;
\path (qya) edge [very thick, ->]  (qy) ;
\path (qya) edge [color=white, thick] (qy) ;

\plate {out} {(qz)(u)(qya)(qy)} {$n$};

\end{tikzpicture}
\end{subfigure}\\
\begin{subfigure}[t]{0.32\textwidth}
\caption{HCM.}\label{fig:ID_ex1h}
\centering
\begin{tikzpicture}
\node[obs] (a) {$A_{ij}$};
\node[obs, below=.8cm of a] (w) {$W_{ij}$};
\node[obs, left=.5cm of w] (z) {$Z_{ij}$};
\node[obs, right=.8cm of w] (y) {$Y_i$};
\node[latent, right=.7cm of a, yshift=.4cm] (u) {$U_i$};

\edge {z} {w} ;
\edge {u} {a,w,y} ;
\edge {w} {y} ;
\edge {a} {w} ;
\plate[dashed] {in} {(a)(w)(z)} {$m$};
\plate {out} {(in)(u)(y)} {$n$};

\end{tikzpicture}
\end{subfigure}
\begin{subfigure}[t]{0.32\textwidth}
\caption{Augmented collapsed model.}\label{fig:ID_ex1u}
\centering
\begin{tikzpicture}
\node[obs] (qa) {$\scriptstyle Q^{a}_{i}$};
\node[obs, right=.5cm of qa, yshift=-.3cm] (qwaz) {$\scriptstyle Q^{w \mid a, z}_{i}$};
\node[obs, below=.5cm of qwaz] (qw) {$\scriptstyle Q^{w}_{i}$};
\node[obs, left=.5cm of qw] (qz) {$\scriptstyle Q^{z}_{i}$};
\node[obs, right=.8cm of qw] (y) {$Y_i$};
\node[latent, right=2cm of qa, yshift=.8cm] (u) {$U_i$};

\edge{u} {qa,qwaz,y};
\path (qa) edge [very thick, ->]  (qw) ;
\path (qa) edge [color=white, thick] (qw) ;
\path (qwaz) edge [very thick, ->]  (qw) ;
\path (qwaz) edge [color=white, thick] (qw) ;
\path (qz) edge [very thick, ->]  (qw) ;
\path (qz) edge [color=white, thick] (qw) ;
\edge {qw} {y}
\plate {out} {(u)(y)(qa)} {$n$};

\end{tikzpicture}
\end{subfigure}
\begin{subfigure}[t]{0.32\textwidth}
\caption{Marginalized model.}\label{fig:ID_ex1m}
\centering
\begin{tikzpicture}
\node[obs] (qa) {$\scriptstyle Q^{a}_{i}$};
\node[obs, right=.5cm of qa, yshift=-.3cm] (qwaz) {$\scriptstyle Q^{w\mid a, z}_{i}$};
\node[obs, below=.5cm of qwaz] (qw) {$\scriptstyle Q^{w}_{i}$};
\node[obs, right=.8cm of qw] (y) {$Y_i$};
\node[latent, right=2cm of qa, yshift=.8cm] (u) {$U_i$};

\edge{u} {qa,qwaz,y};
\edge {qa,qwaz} {qw} ;
\edge {qw} {y}
\plate {out} {(u)(y)(qa)} {$n$};

\end{tikzpicture}
\end{subfigure}
\caption{\textit{continued}}
\end{figure} 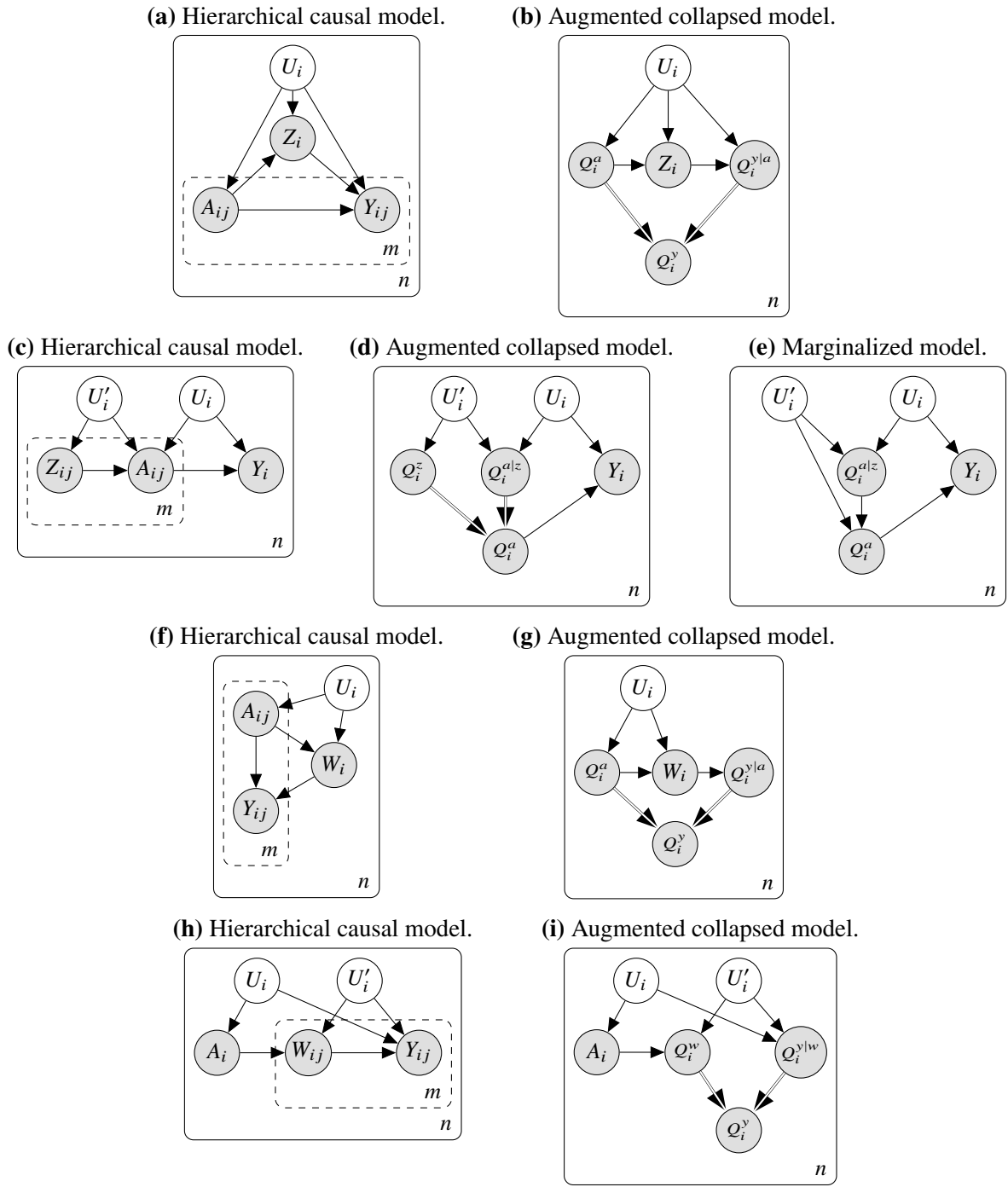
\begin{figure}[t!]
\centering
\begin{subfigure}[t]{0.3\textwidth}
\caption{Hierarchical causal model.} \label{fig:nonID_ex5h}
\centering
\begin{tikzpicture}

\node[obs]                               (y) {$Y_{ij}$};
  \node[obs, left=1.8cm of y] (a) {$A_{ij}$};
  \node[latent, above=1.5cm of a, xshift=1.2cm]  (u) {$U_i$};
  \node[obs, above=.4cm of a, xshift=1.2cm] (z) {$Z_i$};

\edge {a,u} {y} ; \edge {u} {a} ;
  \edge {a} {z} ;
  \edge {z} {y} ;
  \edge {u} {z} ;

\plate[dashed] {ay} {(a)(y)} {$m$} ;
  \plate {ayu} {(ay)(u)(z)} {$n$} ;

\end{tikzpicture}
\end{subfigure}
\begin{subfigure}[t]{0.4\textwidth}
\caption{Augmented collapsed model.} \label{fig:nonID_ex5u}
\centering
\begin{tikzpicture}

\node[obs]                               (qya) {$\scriptstyle Q^{y\mid a}_{i}$};
  \node[obs, left=1.8cm of qya] (qa) {$\scriptstyle Q^a_{i}$};
  \node[latent, above=.8cm of qa, xshift=1.2cm]  (u) {$U_i$};
  \node[obs, left=.6cm of qya] (z) {$Z_i$};
  \node[obs, below=.8cm of z] (qy) {$\scriptstyle Q^y_i$};

\edge {u} {qya,qa} ; \edge {u} {z} ;
  \edge {z} {qya} ;
  \edge {qa} {z} ;

  \path (qya) edge [very thick, ->]  (qy) ;
\path (qya) edge [color=white, thick] (qy) ;
\path (qa) edge [very thick, ->]  (qy) ;
\path (qa) edge [color=white, thick] (qy) ;

\plate {ayu} {(qya)(u)(z)(qa)(qy)} {$n$} ;

\end{tikzpicture}
\end{subfigure}\\
\begin{subfigure}[t]{0.28\textwidth}
\caption{Hierarchical causal model.} \label{fig:nonID_ex4h}
\centering
\begin{tikzpicture}

\node[obs]                               (y) {$Y_{i}$};
  \node[obs, left=1cm of y] (a) {$A_{ij}$};
  \node[latent, above=.4cm of a, xshift=.8cm]  (u) {$U_i$};
  \node[latent, above=.4cm of a, xshift=-.8cm]  (up) {$U'_i$};
  \node[obs, left=0.7cm of a] (z) {$Z_{ij}$};

\edge {a} {y} ; \edge {u} {a,y} ;
  \edge {z} {a} ;
  \edge {up} {a, z};

\plate[dashed] {in} {(z)(a)} {$m$} ;
  \plate {out} {(in)(u)(a)(up)(y)} {$n$} ;

\end{tikzpicture}
\end{subfigure}
\begin{subfigure}[t]{0.38\textwidth}
\caption{Augmented collapsed model.} \label{fig:nonID_ex4u}
\centering
\begin{tikzpicture}

\node[obs]                               (y) {$Y_{i}$};
  \node[obs, left=1cm of y] (qaz) {$\scriptstyle Q^{a\mid z}_{i}$};
  \node[latent, above=.4cm of qaz, xshift=.8cm]  (u) {$U_i$};
  \node[latent, above=.4cm of qaz, xshift=-.8cm]  (up) {$U'_i$};
  \node[obs, left=0.7cm of qaz] (qz) {$\scriptstyle Q^z_{i}$};
  \node[obs, below=0.5cm of qaz] (qa) {$\scriptstyle Q^a_{i}$};

\edge {u} {qaz,y} ;
  \edge {up} {qaz, qz};
  \edge {qa} {y} ;

  \path (qaz) edge [very thick, ->]  (qa) ;
\path (qaz) edge [color=white, thick] (qa) ;
\path (qz) edge [very thick, ->]  (qa) ;
\path (qz) edge [color=white, thick] (qa) ;

\plate {out} {(in)(u)(qa)(up)(y)(qz)} {$n$} ;

\end{tikzpicture}
\end{subfigure}
\begin{subfigure}[t]{0.28\textwidth}
\caption{Marginalized model.} \label{fig:nonID_ex4m}
\centering
\begin{tikzpicture}

\node[obs]                               (y) {$Y_{i}$};
  \node[obs, left=1cm of y] (qaz) {$\scriptstyle Q^{a\mid z}_{i}$};
  \node[latent, above=.4cm of qaz, xshift=.8cm]  (u) {$U_i$};
  \node[latent, above=.4cm of qaz, xshift=-1.2cm]  (up) {$U'_i$};
  \node[obs, below=0.5cm of qaz] (qa) {$\scriptstyle Q^a_{i}$};

\edge {u} {qaz,y} ;
  \edge {up} {qaz, qa};
  \edge {qa} {y} ;
  \edge{qaz} {qa};

\plate {out} {(in)(u)(qa)(up)(y)} {$n$} ;

\end{tikzpicture}
\end{subfigure}\\
\begin{subfigure}[t]{0.3\textwidth}
\caption{Hierarchical causal model.} \label{fig:nonID_ex1h}
\centering
\begin{tikzpicture}
\node[obs] (a) {$A_{ij}$};
\node[obs, below=.8cm of a] (y) {$Y_{ij}$};
\node[obs, right=.5cm of a, yshift=-.8cm] (w) {$W_i$};
\node[latent, right=.7cm of a, yshift=.4cm] (u) {$U_i$};

\edge {u} {a,w} ;
\edge {a} {w,y} ;
\edge {w} {y} ;
\plate[dashed] {in} {(a)(y)} {$m$};
\plate {out} {(in)(u)(w)} {$n$};

\end{tikzpicture}
\end{subfigure}
\begin{subfigure}[t]{0.4\textwidth}
\caption{Augmented collapsed model.} \label{fig:nonID_ex1u}
\centering
\begin{tikzpicture}
\node[obs] (qa) {$\scriptstyle Q^{a}_{i}$};
\node[obs, right=.5cm of qa] (w) {$W_i$};
\node[latent, above=.6cm of w, xshift=-.5cm] (u) {$U_i$};
\node[obs, right=.4cm of w] (qya) {$\scriptstyle Q^{y \mid a}_{i}$};
\node[obs, below=.4cm of w] (qy) {$\scriptstyle Q^{y}_i$};

\edge {u} {w, qa} ;
\edge {qa} {w};
\edge {w} {qya};
\path (qa) edge [very thick, ->]  (qy) ;
\path (qa) edge [color=white, thick] (qy) ;
\path (qya) edge [very thick, ->]  (qy) ;
\path (qya) edge [color=white, thick] (qy) ;

\plate {out} {(qa)(u)(qya)(qy)} {$n$};

\end{tikzpicture}
\end{subfigure}\\ 

\begin{subfigure}[t]{0.35\textwidth}
\caption{Hierarchical causal model.} \label{fig:nonID_ex2h}
\centering
\begin{tikzpicture}

\node[obs]                               (y) {$Y_{ij}$};
  \node[obs, left=1cm of y] (w) {$W_{ij}$};
  \node[latent, above=.4cm of w, xshift=.8cm]  (u) {$U'_i$};
  \node[latent, above=.4cm of w, xshift=-.8cm]  (up) {$U_i$};
  \node[obs, left=0.7cm of w] (a) {$A_i$};

\edge {a} {w} ; \edge {u} {w,y} ;
  \edge {w} {y} ;
  \edge {up} {a, y};

\plate[dashed] {in} {(y)(w)} {$m$} ;
  \plate {out} {(in)(u)(a)(up)} {$n$} ;

\end{tikzpicture}
\end{subfigure}
\begin{subfigure}[t]{0.35\textwidth}
\caption{Augmented collapsed model.} \label{fig:nonID_ex2u}
\centering
\begin{tikzpicture}

\node[obs]                               (qyw) {$\scriptstyle Q^{y\mid w}_{i}$};
  \node[obs, left=1cm of qyw] (qw) {$\scriptstyle Q^w_{i}$};
  \node[obs, below=.5cm of qw, xshift=.8cm] (qy) {$\scriptstyle Q^y_{i}$};
  \node[latent, above=.4cm of qw, xshift=.8cm]  (u) {$U'_i$};
  \node[latent, above=.4cm of qw, xshift=-.8cm]  (up) {$U_i$};
  \node[obs, left=0.7cm of qw] (a) {$A_i$};

\edge {a} {qw} ; \edge {u} {qw,qyw} ;
  \edge {up} {a, qyw};
\path (qw) edge [very thick, ->]  (qy) ;
\path (qw) edge [color=white, thick] (qy) ;
\path (qyw) edge [very thick, ->]  (qy) ;
\path (qyw) edge [color=white, thick] (qy) ;

\plate {out} {(u)(a)(up)(qy)(qyw)(qw)} {$n$} ;

\end{tikzpicture}
\end{subfigure}
\caption{\textbf{Examples of models where the effect of $A$ on $Y$ is \textit{not} identified via our method.} } \label{fig:nonID_exs}
\end{figure}  \section{Marginalizing Hierarchical Causal Models} \label{appx:marginalization}

In this section we describe rules for marginalization in hierarchical causal models.
In particular, we are interested in the question of when we can ignore an endogenous variable, deleting it from the graph, without affecting how the remaining endogenous variables are generated.
These marginalization rules tell us what sorts of variables we can and cannot safely ignore when constructing a hierarchical causal model.
In flat causal models, we can marginalize out any endogenous variable with less than two children~\citep[e.g.][Chap. 9]{Richardson2002-sp,Janzing2022-rn,Peters2017-ww}. For HCMs, we have:

\begin{proposition}[Marginalizing HCMs]
	Consider a variable $Y$ in an HSCM $\mathcal{M}^{\scm}$.
	Assume that (1) $Y$ does not have two or more children ($Y$ is not a confounder), and (2) $Y$ is not a unit level variable with both a subunit parent and a subunit child ($Y$ is not an interferer).
	Then, if $Y$ is marginalized out of $\mathcal{M}^{\scm}$ so that it is no longer an endogenous variable, we obtain a valid HSCM.
\end{proposition}
\begin{proof}
The first assumption prevents us from marginalizing out confounders.
This is necessary since any HCM reduces, as a special case, to a flat causal model with the same graph (namely, the erased inner plate model, \Cref{sec:hierarchy_enables_id}).
Next we analyze variables with one child, on a case-by-case basis.

\parhead{Subunit variable, subunit child} We start by considering the situation where the variable we are interested in marginalizing out, $Y$, is subunit-level, and has a single subunit child $Z$.
We assume without loss of generality that $Y$ has a single subunit parent $X$ and a single unit parent $W$. 
In practice, $Y$ may have more of each kind of parent, but this does not change the analysis, as we can group all the subunit-level parents together and all the unit-level parents together.
We also assume without loss of generality that $X$ and $W$ are parents of $Z$. 
Now, the structural causal model generating $Y$ and $Z$ is,
\begin{equation}
\begin{split}
\gamma_i^{y} \sim \pr(\gamma^{y}) \,\,\,\,\,\,&\,\,\,\,\, \epsilon_{ij}^{y} \sim \pr(\epsilon^{y}) \,\,\,\,\,\,\,\,\,\,\,\, y_{ij} = \f^{y}(w_i, \gamma_i^{y}, x_{ij}, \epsilon_{ij}^{y})\\
\gamma_i^{z} \sim \pr(\gamma^{z}) \,\,\,\,\,\,&\,\,\,\,\, \epsilon_{ij}^{z} \sim \pr(\epsilon^{z}) \,\,\,\,\,\,\,\,\,\,\,\, z_{ij} = \f^{z}(w_i, \gamma_i^{z}, y_{ij}, x_{ij}, \epsilon_{ij}^{z}).
\end{split}
\end{equation}
We can marginalize out $Y$ to obtain,
\begin{equation}
\begin{split}
&\tilde \gamma_i^{z} \sim \tilde \pr(\tilde \gamma^{z}) \,\,\,\,\,\,\,\,\,\,\, \tilde \epsilon_{ij}^{z} \sim \tilde \pr(\tilde \epsilon^{z}) \,\,\,\,\,\,\,\,\,\,\,\, z_{ij} = \tilde \f^{z}(w_i, \tilde \gamma_i^{z}, x_{ij}, \tilde \epsilon_{ij}^{z})\\
\text{ where } & \tilde \gamma_i^{z} = (\gamma_i^{y}, \gamma_i^{z}), \,\, \tilde \epsilon_{ij}^{z} = (\epsilon_{ij}^{y}, \epsilon_{ij}^{z}),\\
\text{ and } &\,\tilde \f^{z}(w_i, \tilde \gamma_i^{z}, x_{ij}, \tilde \epsilon_{ij}^{z}) = \f^{z}(w_i, \gamma_i^{z}, \f^{y}(w_i, \gamma_i^{y}, x_{ij}, \epsilon_{ij}^{y}), x_{ij}, \epsilon_{ij}^{z}).
\end{split}
\end{equation}
The new model for $Z$ meets \Cref{def:hscm}: $\tilde \f^{z}$ depends just on $x_{ij}$ and not on $x_{ij'}$ for $j' \neq j$, the noise $\tilde \gamma^{z}_i$ is i.i.d. across units, and the noise $\tilde \epsilon_{ij}^{z}$ is i.i.d. across subunits and units.

\parhead{Subunit variable, unit child} We next consider the situation where the variable we are interested in marginalizing out, $Y$, is subunit-level, but its child $Z$ is unit-level. We again have a subunit parent $X$ and unit parent $W$ of $Y$ and $Z$. 
When we marginalize out $Y$ we obtain the model,
\begin{equation}
\begin{split}
        \tilde \gamma_i^{z} \sim &\,\, \tilde \pr(\tilde \gamma^{z}) \,\,\,\,\,\,\,\,\,\,\, \tilde \epsilon_{ij}^{z} \sim \tilde \pr(\tilde \epsilon^{z}) \,\,\,\,\,\,\,\,\,\,\,\, z_{i} = \tilde \f^{z}(w_i, \tilde \gamma_i^{z}, \{ (x_{ij}, \tilde \epsilon_{ij}^{z})\}_{j=1}^m)\\
        \text{ where } &\tilde \gamma_i^{z} =  (\gamma_i^{y}, \gamma_i^{z}),\,\, \tilde \epsilon_{ij}^{z} = (\epsilon_{ij}^{y}, \epsilon_{ij}^{z}) \\
        \text{ and } &\, \tilde \f^{z}(w_i, \tilde \gamma_i^{z}, \{ (x_{ij}, \tilde \epsilon_{ij}^{z})\}_{j=1}^m)= \f^{z}\bigg(w_i, \gamma_i^{z}, \Big\{\big(x_{ij}, \f^{y}(x_{ij}, w_i, \gamma_i^{y}, \epsilon_{ij}^{y}), \epsilon_{ij}^{z}\big)\Big\}_{j=1}^m\bigg).
\end{split}
\end{equation}
We can see that $\tilde \f^{z}$ is invariant to permutations of $(x_{i1}, \tilde \epsilon_{i1}^{z}), \ldots, (x_{im}, \tilde \epsilon_{im}^{z})$.
So, the marginalized model meets \Cref{def:hscm}.

\parhead{Unit variable, unit child} We next consider the case where $Y$ is unit-level, and it has a unit child. We again have a subunit parent $X$ and unit parent $W$. 
Marginalizing out $Y$ gives,
\begin{equation}
\begin{split}
        \tilde \gamma_i^{z} \sim &\,\, \tilde \pr(\tilde \gamma^{z}) \,\,\,\,\,\,\,\,\,\,\, \tilde \epsilon_{ij}^{z} \sim \tilde \pr(\tilde \epsilon^{z}) \,\,\,\,\,\,\,\,\,\,\,\, z_{i} = \tilde \f^{z}(w_i, \tilde \gamma_i^{z}, \{ (x_{ij}, \tilde \epsilon_{ij}^{z})\}_{j=1}^m)\\
        \text{ where } &\tilde \gamma_i^{z} =  (\gamma_i^{y}, \gamma_i^{z}),\,\, \tilde \epsilon_{ij}^{z} = (\epsilon_{ij}^{y}, \epsilon_{ij}^{z}), \\
        \text{ and } &\, \tilde \f^{z}(w_i, \tilde \gamma_i^{z}, \{ (x_{ij}, \tilde \epsilon_{ij}^{z})\}_{j=1}^m) = \f^{z}\bigg(\f^{y}\Big(w_i, \gamma_i^{y}, \big\{(x_{ij}, \epsilon_{ij}^{y})\big\}_{j=1}^m\Big), w_i, \gamma_i^{z}, \Big\{\big(x_{ij}, \epsilon_{ij}^{z}\big)\Big\}_{j=1}^m\bigg).
\end{split}
\end{equation}
We see that $\tilde \f^{z}$ is invariant to permutations of $(x_{i1}, \tilde \epsilon_{i1}^{z}), \ldots, (x_{im}, \tilde \epsilon_{im}^{z})$, so \Cref{def:hscm} is met. 

\parhead{Unit variable, subunit child} We now turn to the case where $Y$ is unit-level, and it has a subunit child.
We split this case into two sub-cases. The first is when $Y$ has only unit parents; the second is when $Y$ has at least one subunit parent.

In the first sub-case, when $Y$ only has a unit parent $W$ and no subunit parent, we obtain,
\begin{equation}
\begin{split}
&\tilde \gamma_i^{z} \sim \tilde \pr(\tilde \gamma^{z}) \,\,\,\,\,\,\,\,\,\,\, \tilde \epsilon_{ij}^{z} \sim \tilde \pr(\tilde \epsilon^{z}) \,\,\,\,\,\,\,\,\,\,\,\, z_{ij} = \tilde \f^{z}(w_i, \tilde \gamma_i^{z}, \tilde \epsilon_{ij}^{z})\\
\text{ where } & \tilde \gamma_i^{z} = (\gamma_i^{y}, \epsilon_{i1}^{y}, \ldots, \epsilon_{im}^{y}, \gamma_i^{z}), \,\, \tilde \epsilon_{ij}^{z} = \epsilon_{ij}^{z},\\
\text{ and } & \,\tilde \f^{z}(w_i, \tilde \gamma_i^{z}, \tilde \epsilon_{ij}^{z}) = \f^{z}\bigg(\f^{y}\Big(w_i, \gamma_i^{y}, \big\{\epsilon_{ij}^{y}\big\}_{j=1}^m\Big), w_i, \gamma_i^{z}, \epsilon_{ij}^{z}\bigg).
\end{split}
\end{equation}
This marginalized model meets \Cref{def:hscm}.
Note we have absorbed the subunit-level noise $\epsilon^{y}$ into the unit-level noise $\tilde \gamma^{z}$ on $Z$, rather than the subunit-level noise $\tilde \epsilon^{z}$; this is because $Z_{ij}$ cannot depend on $\tilde \epsilon_{i j'}^{y}$ for $j' \neq j$.

Finally, we arrive at the case where marginalization is impossible: $Y$ is unit-level and has a single subunit child, but it also has a subunit parent $X$.
The original structural causal model for $Y$ and $Z$ is,
\begin{equation}
\begin{split}
        \gamma_i^{y} \sim \pr(\gamma^{y}) \,\,\,\,\,\,&\,\,\,\,\, \epsilon_{ij}^{y} \sim \pr(\epsilon^{y}) \,\,\,\,\,\,\,\,\,\,\,\, y_{i} = \f^{y}(\gamma_i^{y}, \{(x_{ij}, \epsilon_{ij}^{y})\}_{j=1}^m)\\
\gamma_i^{z} \sim \pr(\gamma^{z}) \,\,\,\,\,\,&\,\,\,\,\, \epsilon_{ij}^{z} \sim \pr(\epsilon^{z}) \,\,\,\,\,\,\,\,\,\,\,\, z_{ij} = \f^{z}(y_{i}, \gamma_i^{z}, \epsilon_{ij}^{z}).
\end{split}
\end{equation}
If we attempt to marginalize out $Y$, we obtain,
\begin{equation}
\begin{split}
z_{ij} = \f^{z}\bigg(\f^{y}\Big(\gamma_i^{y}, \big\{\{(x_{ij}, \epsilon_{i1}^{y})\big\}_{j=1}^m\Big), \gamma_i^{z}, \epsilon_{ij}^{z}\bigg).
\end{split}
\end{equation}
This mechanism does not meet \Cref{def:hscm}, because $Z_{ij}$ can depend on $X_{ij'}$ for $j' \neq j$.
Thus, when a variable is an \textit{interferer} -- a unit variable with a subunit child and subunit parent -- it cannot be marginalized out of an HCM.
\end{proof}

Confounding is a central problem in causal inference because it is impossible to ignore: confounders cannot be marginalized out of a causal model.
Variables that have just one child, though, can be ignored in flat causal models.
Intuitively, this implies we do not need to worry about the details of all the intermediate variables between a cause and its effect.

In hierarchical causal models, interference is, like confounding, a central problem: interferers cannot be marginalized out.
So, while we can still ignore the intermediate variables between cause and effect in most cases, we cannot ignore an intermediate variable if it gives rise to interference.
This makes the development of techniques to correct for interference a key issue in the study of hierarchical causal models. \section{Counterfactuals in Hierarchical Causal Models} \label{apx:counterfactuals}

In this section we introduce counterfactuals in hierarchical structural causal models.
We then explain in particular how subunit-level noise contributes to modeling unit variable counterfactuals. This motivates the inclusion of subunit-level noise in the HSCM equations for unit-level variables (\Cref{def:hscm}).

Given a sample of two endogenous variables $A$ and $Y$ from a structural causal model, counterfactuals address the question: what would $Y$ be if $A$ were $a_\star$, all else held equal?
To compute a counterfactual, we (1) fix all the noise variables, (2) set $a$ to $a_\star$, and (3) run the model forward to compute the resulting value of $Y$~\citep[][Chap. 7]{Pearl2009-fh}.

Counterfactuals in HSCMs work the same way. For example, consider an HSCM with a subunit treatment $A$ and unit outcome $Y$. It has equations
\begin{equation} \label{eqn:counterfactual_ex}
\begin{split}
\gamma_i^a \sim \pr(\gamma^y)
  \quad \quad
    \epsilon_{ij}^a \sim \pr(\epsilon^y)
  \quad \quad
    & a_{ij} = \f^a(\gamma_i^a, \epsilon^a_{ij})\\
	\gamma_i^y \sim \pr(\gamma^y)
  \quad \quad
    \epsilon_{ij}^y \sim \pr(\epsilon^y)
  \quad \quad
    & y_{i} = \f^y(\gamma_i^y, \{(a_{ij}, \epsilon^y_{ij})\}_{j=1}^m).
\end{split}
\end{equation}
for $j \in \{1, \ldots, m\}$ and $i \in \{1, \ldots, n\}$.
Say we sample from this model, and the value of $Y$ in unit $i = 1$ is $Y_1 = y_1$. We can ask:
what would $Y_1$ be if $a_{1 1}$ were $a_\star$? This counterfactual is given by,
\begin{equation}
\begin{split}
a_{1 1} &= a_\star\\
 a_{1j} &= \f^a(\gamma_1^a, \epsilon^a_{1j}) \text{ for } j \in \{2,\ldots, m\}\\
y'_1 &= \f^y(\gamma_1^y, \{(a_{1j}, \epsilon^y_{1j})\}_{j=1}^m).
\end{split}
\end{equation}
Here we keep the values of the noise $\gamma_1$, $\epsilon_{12}, \ldots, \epsilon_{1m}$ the same, but plug in the counterfactual value of $a_{1 1}$.

As discussed in \Cref{sec:hcgm_ex_derivation}, the subunit noise in mechanisms for unit variables (e.g. $\epsilon^y$ in~\Cref{eqn:counterfactual_ex}) does not increase the HSCM's expressiveness in describing interventional distributions.
The subunit noise does, however, increase its expressiveness in describing counterfactuals.
That is, 
\begin{equation} \label{eqn:counterfact_subunit_noise}
	\gamma_i^y \sim \pr(\gamma^y)
  \quad \quad
    \epsilon_{ij}^y \sim \pr(\epsilon^y)
  \quad \quad
    y_{i} = \f^y(\gamma_i^y, \{(a_{ij}, \epsilon^y_{ij})\}_{j=1}^m),
\end{equation}
is a more expressive model for describing counterfactuals than,
\begin{equation} \label{eqn:counterfact_no_subunit_noise}
	\gamma_i^y \sim \pr(\gamma^y)
  \quad \quad
  \quad \quad
  \quad \quad
    y_{i} = \f^y(\gamma_i^y, \{a_{ij}\}_{j=1}^m),
\end{equation}
where the subunit noise $\epsilon^y$ has been dropped. To see this, consider the counterfactual scenario where we permute the values of $a_{i1}, \ldots, a_{im}$.
In the model without subunit noise (\Cref{eqn:counterfact_no_subunit_noise}), the counterfactual value of $Y_i$ cannot be different from its actual value, since $\f^y(\gamma_i^y, \{a_{ij}\}_{j=1}^m) = \f^y(\gamma_i^y, \{a_{i\pi(j)}\}_{j=1}^m)$ for any permutation $\pi$.
With subunit noise (\Cref{eqn:counterfact_subunit_noise}), however, we have no such restriction, and the counterfactual value of $Y_i$ may be different.

To illustrate the importance of this expressivity, we consider a simple scenario, where $Y_i$ represents whether school $i$ appears on a ``best schools'' list. Say $Y_i$ depends on whether the fraction of students at the school who pass all their classes is 50\% or more.
However, we only observe $a_{ij}$, which indicates whether student $j$ passes their English class. Then, the subunit noise $\epsilon_{ij}^y$ describes whether student $j$ passes the rest of their classes, and the mechanism generating $Y_i$ can be written,
\begin{equation} \label{eqn:counterfact_concrete}
	y_i = \f^y(\{(a_{ij}, \epsilon^y_{ij})\}_{j=1}^m) = \mathbb{I}\Big(\frac{1}{m}\sum_{j=1}^m a_{ij} \epsilon^y_{ij} \ge 0.5\Big).
\end{equation}
Consider a school $i = 1$ with just two students. One of the students passes all their classes ($a_{1 1} = \epsilon^y_{1 1} = 1$) and the other none of them ($a_{1 2} = \epsilon^y_{1 2} = 0$), so that $y_1 = 1$.
Let us investigate the counterfactual scenario where the English grades of the students are switched, such that the first student fails ($a_{1 1} = 0$) and the second passes ($a_{1 2} = 1$).
From~\Cref{eqn:counterfact_concrete}, we see the counterfactual value of $y_1$ is $y'_1 = \frac{1}{m} (0 \cdot 1 + 1 \cdot 0) = 0$.
That is, in this counterfactual scenario, the school does not appear on the ``best schools'' list, since none of their students passed all their classes.
By contrast, a model that did not include subunit noise (\Cref{eqn:counterfact_no_subunit_noise}) would demand, unreasonably, that $y_1'$ must still be $1$ in this counterfactual scenario. \section{Deriving Hierarchical Causal Graphical Models} \label{apx:hcgm}

In this section, we derive hierarchical causal graphical models (\Cref{def:hcgm}) from hierarchical structural causal models (\Cref{def:hscm}).
The derivation is a generalization of the special case given in \Cref{sec:hcgm_ex_derivation}. 

For each subunit variable $v \in \mathcal{S}$, we have
\begin{equation} \label{eqn:subunit_hcgm_derivation}
\begin{split}
        [g_q^{v | \pa_{\mathcal{S}}(v)}(x^{\pa_{\mathcal{U}}(v)}, \gamma^{v})](X^{v} \in \Xi \mid x^{\pa_\Sc(v)}) & \triangleq \int \mathbb{I}(\f^{v}(x^{\pa_\Uc(v)}, \gamma^{v}, x^{\pa_\Sc(v)}, \epsilon^{v}) \in \Xi) \pr(\epsilon^{v}) \di \epsilon^{v}\\
        \Prob(Q^{v| \pa_\Sc(v)} \in \Pi \mid  x^{\pa_\Uc(v)}) & \triangleq \int \mathbb{I}(g_q^{v | \pa_\Sc(v)}(x^{\pa_{\Uc}(v)}, \gamma^{v}) \in \Pi) \pr(\gamma^{v}) \di\gamma^{v}.
\end{split}
\end{equation}
In the first line, we marginalize out the subunit noise $\epsilon^{v}$ to produce a function $g_q^{v | \pa_{\mathcal{S}}(v)}$ that takes in unit-level variables $X^{\pa_{\mathcal{U}}(v)}$ and $\gamma^{v}$, and returns a conditional distribution over the subunit-level variable $X^{v}$ given the subunit-level variables $X^{\pa_\Sc(v)}$.
In the second line, we marginalize out the unit-level noise $\gamma^{v}$ to produce a conditional distribution over subunit distributions.

For each unit variable $v \in \mathcal{U}$, we have
\begin{equation} \label{eqn:unit_hcgm_derivation}
\begin{split}
        \Prob(x^{v} \in \Xi &\mid x^{\pa_\Uc(v)}, \{x_{j}^{\pa_\Sc(v)}\}_{j=1}^m)\\ &= \int\ldots \int \mathbb{I}\big(\f^{v}(x_i^{\pa_\Uc(v)}, \gamma^{v}, \{(x_{j}^{\pa_\Sc(v)}, \epsilon_{j}^{v})\}_{j=1}^m) \in \Xi\big) \pr(\gamma^{v})\di\gamma^{v} \prod_{j=1}^m \pr(\epsilon^{v}_j) \di\epsilon^{v}_j.
\end{split}
\end{equation}
We can confirm that the stochastic mechanism $\Prob(x^{v} \in \Xi \mid x^{\pa_\Uc(v)}, \{x_{j}^{\pa_\Sc(v)}\}_{j=1}^m)$ is invariant to permutations of the subunits, as for any permutation $\pi$ of $\{1,\ldots,m\}$ we have,
\begin{equation}
=\int\ldots \int \mathbb{I}\big(\f^{v}(x_i^{\pa_\Uc(v)}, \gamma^{v}, \{(x_{\pi(j)}^{\pa_\Sc(v)}, \epsilon_{j}^{v})\}_{j=1}^m) \in \Xi\big) \pr(\gamma^{v})\di\gamma^{v} \prod_{j=1}^m \pr(\epsilon^{v}_j) \di\epsilon^{v}_j,
\end{equation}
since $f^{v}$ is invariant to permutations of $\{(x_{j}^{\pa_\Sc(v)}, \epsilon_{j}^{v})\}_{j=1}^m$ and each $\epsilon^{v}_j$ is drawn i.i.d. for $j \in \{1, \ldots, m\}$.

\section{Details on Simulations} \label{apx:simulations}

\subsection{\textsc{unit confounder}} \label{sec:unobs_conf_sim}

In this section we describe in detail our simulation and estimation procedures for the \textsc{unit confounder} model (\Cref{sec:confounder_id}).
Each endogenous variable ($U$, $A$ and $Y$) is binary, and the data-generating HCGM is,
\begin{equation*} \label{eqn:confound_dgm}
\begin{split}
U_i &\sim \mathrm{Bernoulli}(\omega)\\
\mu^{a}_i \sim \mathrm{Beta}\big(\alpha^{a}(u_i), \beta^a\big) \, \quad \quad \quad \quad \quad \quad \quad \quad  q^{a}_i & = \mathrm{Bernoulli}(\mu_i^{a})
\quad \quad \quad \quad A_{ij} \sim q^{a}_i(a) \\
\mu^{y|a}_i(a) \sim \mathrm{Beta}\big(\alpha^{y|a}(a, u_i), \beta^{y|a}\big) \quad \quad \quad  q^{y|a}_i(\cdot\mid a) & = \mathrm{Bernoulli}(\mu_i^{y|a}(a))
 \quad \quad Y_{ij} \sim q_i^{y|a}(y \mid a_{ij}),
\end{split}
\end{equation*}
for $j \in \{1, \ldots, m\}$ and $i \in \{1, \ldots, n\}$. We set $\alpha^{a}(0) = 0.5, \alpha^{a}(1) = 4$, $\alpha^{y|a}(0, 0) = 0.5, \alpha^{y|a}(1, 0) = 2, \alpha^{y|a}(0, 1) = 1, \alpha^{y|a}(1,1) = 4$, $\beta^a=1$ and $\beta^{y|a}=2$.
We simulate with different values of $\omega$, which governs the amount of confounding.
From the data generating model, we can calculate the true effect as,
\begin{equation*}
        \mathbb{E}_{\pr}[\mathbb{E}_Q[Y] \s \rmdo(q^{a} = \delta_{a_\star})] = (1-\omega) \frac{\alpha^{y | a}(a_\star,0)}{\alpha^{y | a}(a_\star,0) + \beta^{y|a}} + \omega \frac{\alpha^{y | a}(a_\star,1)}{\alpha^{y \mid a}(a_\star,1) + \beta^{y|a}},
\end{equation*}
for $a_\star \in \{0, 1\}$, where we have used the fact that the mean of $\mathrm{Beta}(\alpha, \beta)$ is  $\alpha/(\alpha + \beta)$.

We now turn to estimation. 
\begin{enumerate}[leftmargin=*]
	\item 
For $q^{y | a}_i$ we use the point estimate,
\begin{equation} \label{eqn:QYA_estimate}
        \hat{q}^{y|a}_i(\cdot \mid a) = \mathrm{Bernoulli}\left(\hat{\mu}^{y|a}_i(a) = \frac{\sum_{j=1}^m y_{ij} \delta_{a}(a_{ij}) +1 }{\sum_{j=1}^m \delta_{a}(a_{ij}) + 2}\right),
\end{equation}
for each unit $i \in \{1, \ldots, n\}$ and for $a \in \{0, 1\}$.
Here we have added pseudocounts for regularization, i.e. $\hat{\mu}^{y|a}_i(a)$ corresponds the posterior mean of $\mathbb{E}_{q^{y|a}_i}[Y \mid a]$ under a $\mathrm{Beta}(1, 1)$ prior. 
\item We can compute $\hat{\mu}_i^y(a_\star) = \hat{\mu}^{y|a}_i(a_\star)$ since the intervention distribution $q_\star^a$ is a point mass at $a_\star$.
\end{enumerate}
Applying \Cref{eq:confounder_estimator}, we can estimate the treatment effect as
$\frac{1}{n} \sum_{i=1}^n \hat{\mu}^{y}_i(1) - \frac{1}{n} \sum_{i=1}^n \hat{\mu}^{y}_i(0).$

We simulated data sets of size $n = 1000$ and $m = 1000$ from the data generating model, then constructed our estimate based on subsets of increasing size (observing $10$ subunits and units, $100$ subunits and units, etc.).
Each panel of \Cref{fig:unobs_confound_sim} shows the convergence of our estimator to the true effect with increasing data, across 20 independent simulations.
We use $\omega=0$ for the ``no confounding'' simulations (\Cref{fig:no_confound_sim}), $\omega=0.2$ for ``low confounding'' (\Cref{fig:low_confound_sim}) and $\omega=0.5$ for ``high confounding'' (\Cref{fig:high_confound_sim}).

We compare to a regression estimator that comes from naively applying linear regression to aggregated data. In particular, we estimate $\mathbb{E}_{\pr}[\bar{Y} | \bar{A}=1] - \mathbb{E}_{\pr}[\bar{Y} | \bar{A}=0]$, where $\bar{y} \triangleq \mathbb{E}_{q}[Y] = \int y\, q^y(y) \di y$ is the within-unit average of $y$ and $\bar{a} \triangleq \mathbb{E}_{q}[A] = \int a\, q^a(a) \di a$ is the within-unit average of $a$.
We perform our estimate by running a linear regression predicting $\hat{\bar{y}}_i = \frac{1}{m} \sum_{j=1}^m y_{ij}$ from $\hat{\bar{a}}_i = \frac{1}{m} \sum_{j=1}^m a_{ij}$ for each unit $i$. 

In the presence of confounding, the linear regression estimate does not converge to the true effect (\Cref{fig:low_confound_sim}, \Cref{fig:high_confound_sim}).
Without confounding, the linear regression estimate does converge to the true effect (\Cref{fig:no_confound_sim}).
Note that in general, a linear regression estimate will not always be accurate in the absence of confounding; rather, its success depends on the fact that $A$ and $Y$ are binary in our simulation. 

\begin{figure}[t]
\centering
\begin{subfigure}[t]{0.32\textwidth}
        \caption{Low $M$ ($\rho_m=0.25$).}
        \includegraphics[width=\textwidth]{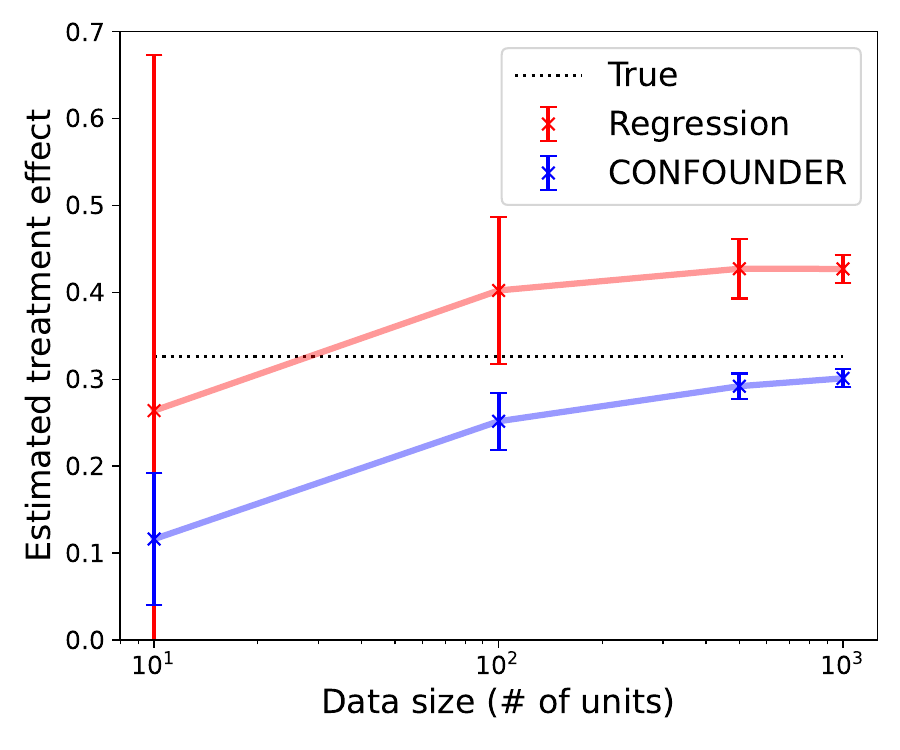}
\end{subfigure}
\begin{subfigure}[t]{0.32\textwidth}
        \caption{Medium $M$ ($\rho_m=1$).}
        \includegraphics[width=\textwidth]{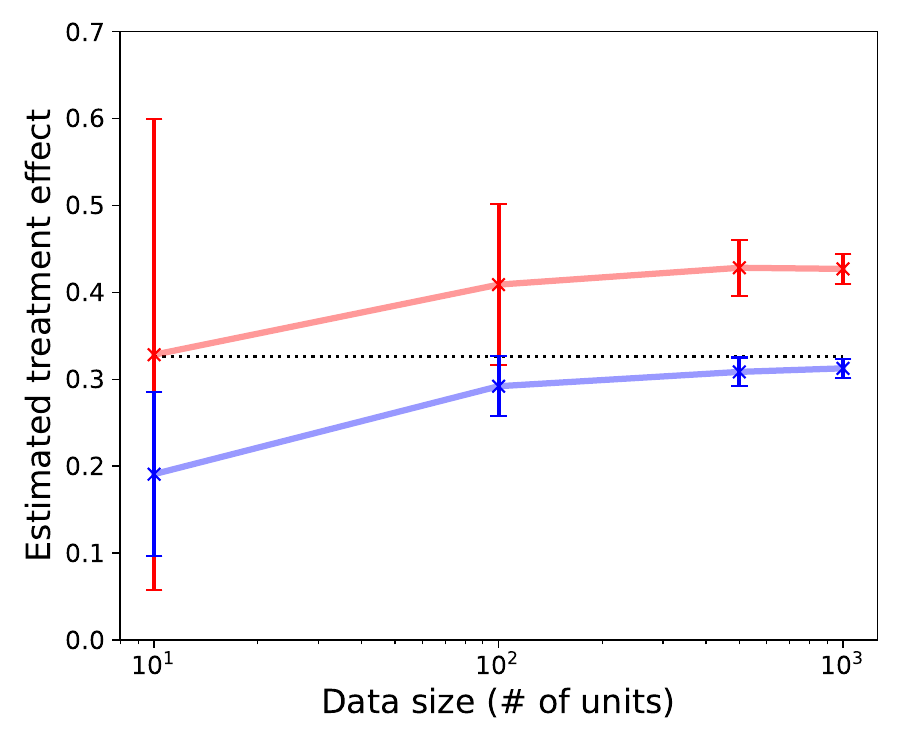}
\end{subfigure}
\begin{subfigure}[t]{0.32\textwidth}
        \caption{High $M$ ($\rho_m=4$).}
        \includegraphics[width=\textwidth]{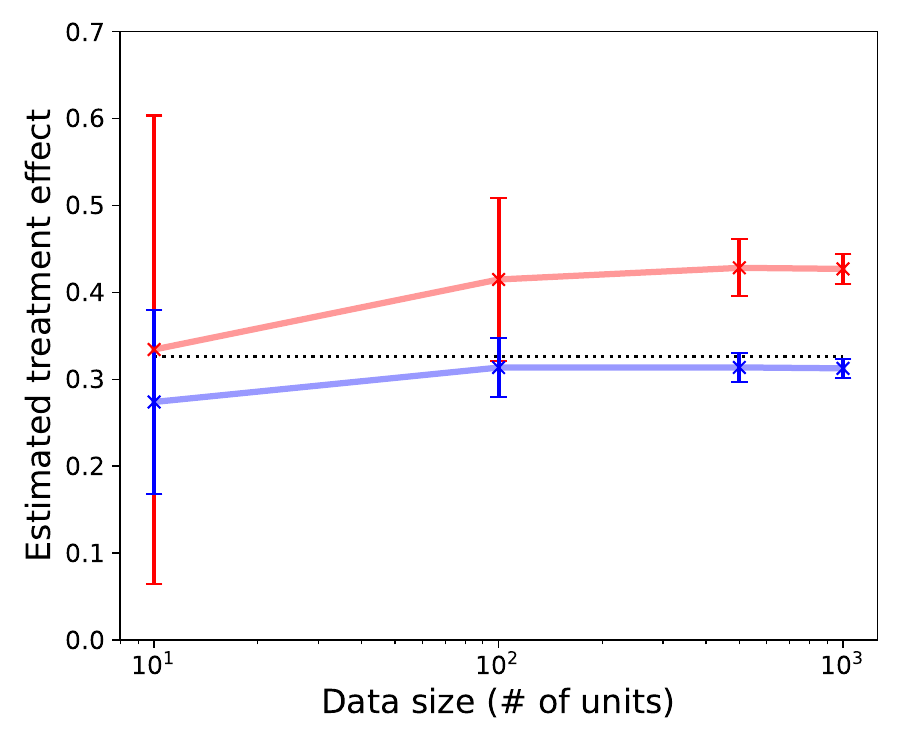}
\end{subfigure}
\caption{\textbf{Estimated effects in a } \textsc{unit confounder} \textbf{simulation for changing number of subunits per unit ($\rho_m = M/N$).}
 Error
  bars show standard deviation across 20 independent simulations.} \label{fig:unobs_confound_sim_mfrac}
\end{figure}

We also evaluated estimation performance as the proportion of subunits to units changes. We set $M = \lceil \rho_m N \rceil$ for different $\rho_m$, and track estimator convergence as $N$ increases (\Cref{fig:unobs_confound_sim_mfrac}). As $\rho_m$ increases, the bias at low $N$ decreases. This is in line with the theory of nested Monte Carlo estimation \citep[Appendix B]{Rainforth2018-qp}.  

\subsection{\textsc{unit confounder \& unit interference}} \label{sec:interference_sim}

We next detail our simulations and estimation procedures for the \textsc{interference} model (\Cref{sec:confound_interfere_id}).
The data generating HCGM for our simulation is,
\begin{equation*}
\begin{split}
        U_i &\sim \mathrm{Normal}(0, 1)\\
        \nu_i^{a}\sim \mathrm{Normal}( 0.5\,  u_i, 1) \,\,\quad\quad\quad q_i^{a} &= \mathrm{Bernoulli}(\sigma(\nu_i^{a}))
        \quad \quad \quad A_{ij} \sim q^{a}_i(a) \\
         Z_i &\sim \mathrm{Bernoulli}\bigg(\sigma\Big(2\, \sigma^{-1}\Big(\frac{1}{m}\sum_{j=1}^m a_{ij}\Big) - 0.8\Big)\bigg) \\
        \nu_i^{y \mid a}(a) \sim \mathrm{Normal}\big(g_\rho(a, z_i, 0.5 u_i),\, 0.1\big)\quad \quad q_i^{y| a}(\cdot \mid a) &= \mathrm{Bernoulli}(\sigma(\nu_i^{y| a}(a))) \quad \quad Y_{ij} \sim q_i^{y|a}(y \mid a_{ij}),
\end{split}
\end{equation*}
for $j \in \{1, \ldots, m\}$ and $i \in \{1, \ldots, n\}$, where $\mathrm{Normal}(\mu,\tau)$ is a normal distribution with mean $\mu$ and variance $\tau^2$, $\sigma(x) = 1/(1 + \exp(-x))$ is the logistic sigmoid function, $\sigma^{-1}(x) = \log(x/(1 - x))$ is its inverse, and $g_\rho(a, z_i, 0.5 u_i) = 0.5 a + \rho (2 z_i - 1) + 0.5 u_i$.
We simulate with different values of $\rho$, which determines the strength of interference.

From the data generating process, we can calculate the true effect of an intervention with $q_\star^a = \mathrm{Bernoulli}(\mu_\star)$ as
\begin{equation*} \label{eq:true_confound_interfere_sim_effect}
\begin{split}
	\mathbb{E}_{\pr}&[\mathbb{E}_{Q^y}[Y] \s \rmdo(q^a = q^a_\star)]
	= \mathbb{E}_{q_\star^a}\bigg[\mathbb{E}_{\pr(z|q^a)}\Big[\mathbb{E}_{\int \pr(q^{y|a}|u, z) \pr(u) \di u}\big[\mathbb{E}_{Q^{y|a}}[Y\mid A]\mid Z\big]\mid q_\star^a\Big]\bigg]\\
	=&\sum_{a=0}^1 (\mu_\star)^a (1 - \mu_\star)^{1-a} \sum_{z=0}^1 \sigma(2 \sigma^{-1}(\mu_\star) - 0.8)^z ( 1 - \sigma(2 \sigma^{-1}(\mu_\star) - 0.8))^{1-z}
	\\ & \quad \int \mathcal{N}(x \mid 0, 1) \sigma(g_\rho(a, z, \sqrt{0.5^2 + 0.1^2}x)) \end{split}
\end{equation*}
Here, $\mathcal{N}(x \mid 0, 1)$ denotes the pdf of a standard normal, and we have used the fact that the distribution $\int \pr(\nu^{y|a}|z, u)\pr(u) \di u$ can be rewritten from a sum of independent normal distributions to depend on a single standard normal $x$.
We compute the Gaussian integral numerically.

We now turn to estimation.
\begin{enumerate}[leftmargin=*]
	\item We estimate $q^{a}_i$ for each unit $i$ as $\hat{q}^{a}_i = \mathrm{Bernoulli}(\hat{\mu}^{a}_i = \frac{1}{m+2} (1 + \sum_{j=1}^m a_{ij}))$, again using pseudocounts for regularization.
	\item We estimate $q^{y| a}_i$ for each unit $i$ with \Cref{eqn:QYA_estimate}.
	\item We estimate $\pr(z \mid q^{a})$ with logistic regression, predicting $z_i$ from $\sigma^{-1}(\hat{\mu}^{a}_i)$. This gives an estimate $\hat{\pr}(z \mid \mu^{a})$.
	\item We estimate $\pr(q^{y|a}\mid q^{a}, z)$ using four separate linear regressions. For each $a \in\{0, 1\}$ and $z \in \{0, 1\}$, we predict $\sigma^{-1}(\hat{\mu}^{y \mid a}_i(a))$ from $\sigma^{-1}(\hat{\mu}^{a}_i)$ for all units $i$ such that $z_i = z$. This gives an estimate $\hat{\pr}(\sigma^{-1}(\mu^{y \mid a}) \mid \mu^{a}, z)$.
	\item We estimate $\pr(q^a)$ with the empirical distribution of $\hat{\mu}^a_i$ for all units $i \in \{1, \ldots, n\}$.
\end{enumerate}
Finally, we combine these estimates following the identification formula,
\begin{equation*}
\begin{split}
\mathbb{E}_\pr[\mathbb{E}_{Q^y}[Y] &\s \rmdo(q^{a} = q_\star^a)]
=\mathbb{E}_{A \sim q_\star^a}\left[\int \pr(z\mid q_\star^a) \int \pr(q^a) \mathbb{E}_\pr[\mathbb{E}_Q[Y \mid A] \mid q^a, z] \di q^a \di z\right] \\ 
\approx \sum_{a = 0}^1 & (\mu_\star)^{a} (1 - \mu_\star)^{1 - a} \sum_{z=0}^1 \hat{\pr}\Big(z | \mu_\star\Big) \frac{1}{n}\sum_{i=1}^n \mathbb{E}_{\hat{\pr}}\Big[\mu^{y | a}(a)\, \Big|\, \hat{\mu}^{a}_i, z\Big].
\end{split}
\end{equation*}
To compute $ \mathbb{E}_{\hat{\pr}}[\mu^{y | a}(a) \mid \hat{\mu}^{a}_i, z]$, we use Monte Carlo integration: we draw 100 samples from $\hat{\pr}(\sigma^{-1}(\mu^{y | a}(a)) \mid \hat{\mu}^{a}_i, z)$, apply $\sigma(\cdot)$ to each sample, then take the average.

As before, we simulated data sets of size $n = 1000$ and $m = 1000$, and constructed our estimate based on subsets of increasing size. Note that here the ground truth has finite but large $m$, while our identification technique makes the approximation $m \to \infty$.
Each panel of \Cref{fig:interfere_sim} shows the convergence of our estimator to the true, analytically computed, $m \to \infty$ effect with increasing data set size, across 20 independent simulations.
We use $\rho=1.5$ for the ``high interference'' simulations (\Cref{fig:high_interfere_sim}), $\rho=0.5$ for the ``low interference'' (\Cref{fig:low_interfere_sim}) and $\rho=0$ for the ``no interference'' (\Cref{fig:no_interfere_sim}).

\begin{figure}[t]
\centering
\begin{subfigure}[t]{0.32\textwidth}
        \caption{Low $M$ ($\rho_m=0.25$).}
        \includegraphics[width=\textwidth]{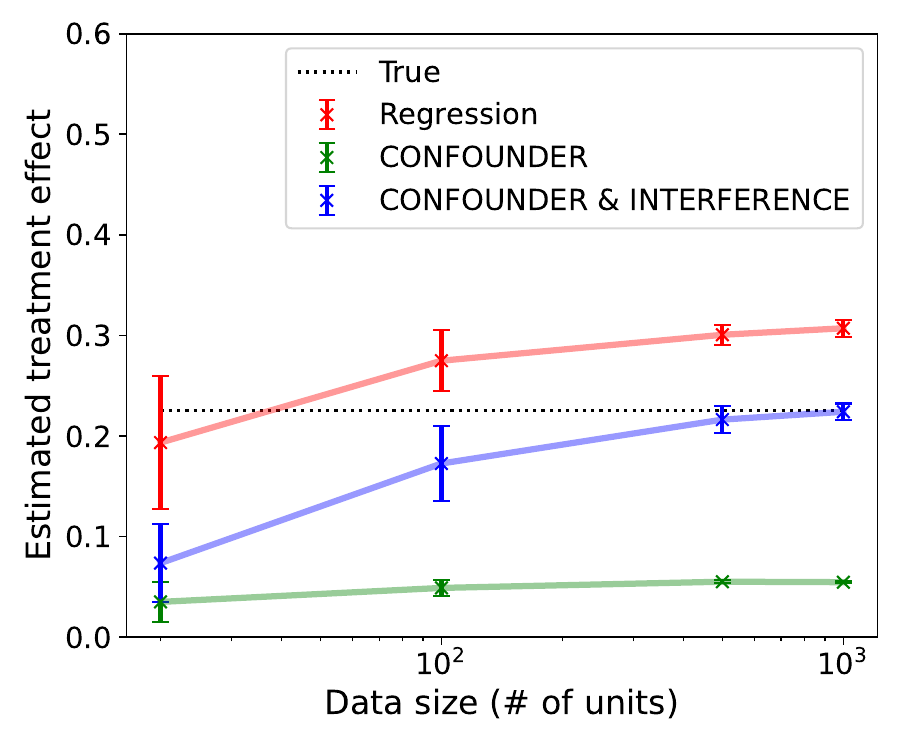}
\end{subfigure}
\begin{subfigure}[t]{0.32\textwidth}
        \caption{Medium $M$ ($\rho_m=1$).}
        \includegraphics[width=\textwidth]{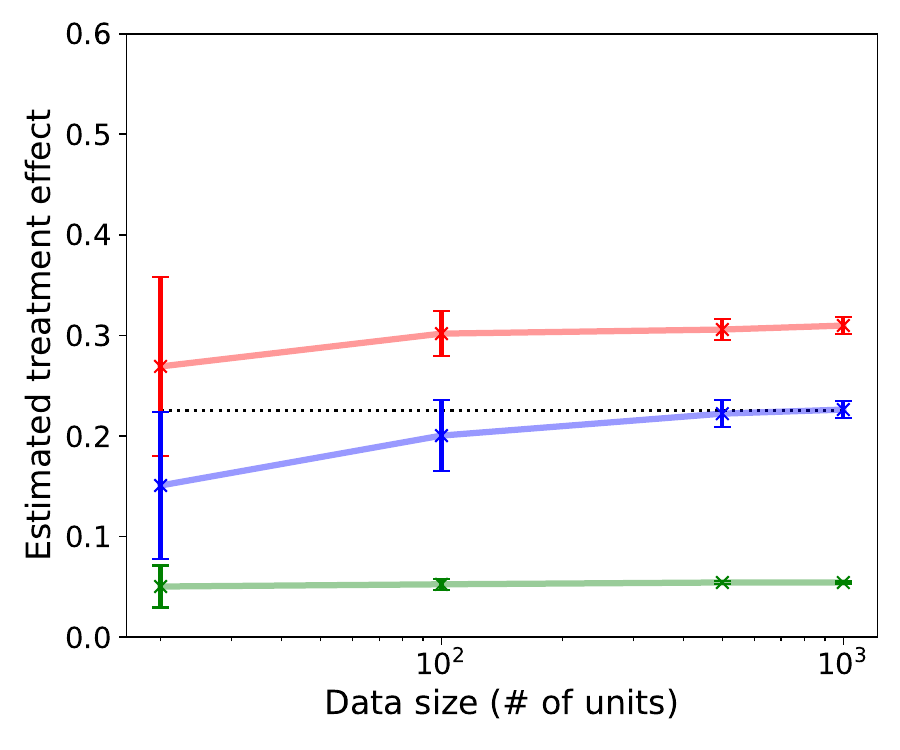}
\end{subfigure}
\begin{subfigure}[t]{0.32\textwidth}
        \caption{High $M$ ($\rho_m=4$).}
        \includegraphics[width=\textwidth]{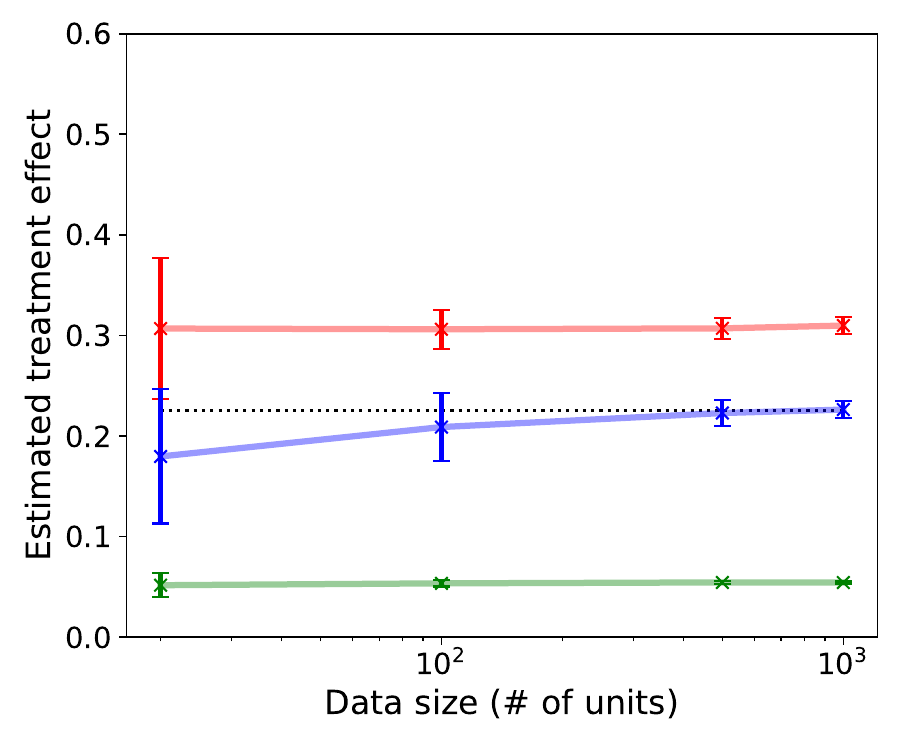}
\end{subfigure}
\caption{\textbf{Estimated effects in a } \textsc{unit confounder \& unit interference} \textbf{simulation for changing number of subunits per unit ($\rho_m = M/N$).}
 Error
  bars show standard deviation across 20 independent simulations.} \label{fig:interfere_converge_sim_mfrac}
\end{figure}

We also evaluated estimation performance as the proportion of subunits to units changes. We set $M = \lceil \rho_m N \rceil$ for different $\rho_m$, and track estimator convergence as $N$ increases (\Cref{fig:interfere_converge_sim_mfrac}). As $\rho_m$ increases, the bias at low $N$ decreases  \citep[Appendix B]{Rainforth2018-qp}.  

\subsection{\textsc{subunit instrument}} \label{apx:instrument_sim}

Finally we detail our simulations and estimation procedures for the \textsc{subunit instrument} model (\Cref{sec:instrument_id}).
In this simulation, the data generating HCGM is,
\begin{equation*}
\begin{split}
        U_i &\sim \mathrm{Bernoulli}(\omega)\\
        \mu_i^{z} \sim \mathrm{Beta}(2, 2)  \, \, \, \, \, \quad \quad q^{z}_i &= \mathrm{Bernoulli}(\mu_i^{z})
        \quad \quad Z_{ij} \sim q^{z}_i(z)\\
        \epsilon_i^{a \mid z}(z) \sim \mathrm{Beta}(2 - 1.8 u_i, 0.2 +  1.8 u_i) \quad \quad q^{a | z}_i(\cdot \mid z) &= \mathrm{Bernoulli}\Big(0.8 z + 0.2 \, \epsilon_i^{a |z }(z)\Big) \quad \quad 
        A_{ij} \sim q^{a |z }_i(a \mid z_{ij}) \\ Y_i &\sim \mathrm{Bernoulli}\Big(0.45 - 0.4\, u_i + 0.5\frac{1}{m}\sum_{j=1}^m a_{ij} \Big),
\end{split}
\end{equation*}
for $j \in \{1, \ldots, m\}$ and $i \in\{1, \ldots,n\}$.
Here, $\omega$ governs the amount of confounding.

We are interested in the effects that soft interventions on $A$ have on the outcome $Y$.
In the infinite $m$ limit,
$\mathbb{E}_\pr[Y \s \rmdo(q^{a} = q_\star)] = 0.45 - 0.4\, \omega + 0.5 \mu_\star$,
where the intervention distribution is $q_\star^a = \mathrm{Bernoulli}(\mu_\star)$.
Note that for the interventions we consider, namely $\mu_\star = 0.25$ and $\mu_\star = 0.75$, the effect is identified, since it satisfies the positivity condition (\Cref{asm:instrument_positive}).
In particular, for any $\mu^{a \mid z}_i$,  with probability 1 there exists a value of $\mu^{z}$ such that $\mu_{\star} = (1 - \mu^z)\mu^{a | z}_i(0) + \mu^{z} \mu^{a|z}_i(1)$. 
This is because in the simulation, $\mu^{a | z}_i(1) \ge 0.8 >  \mu_{\star} > 0.2 \ge \mu^{a \mid z}_i(0)$ with probability one.

We now turn to estimation.
So far, in previous simulations, we have focused on parametric models. Here we explore a nonparametric outcome model (a Gaussian process classifier).
\begin{enumerate}[leftmargin=*]
\item For each unit $i$, we estimate $q_i^{a| z}$ just as we estimated $q_i^{y | a}$ in \Cref{sec:interference_sim}, obtaining $\hat{q}_i^{a |z} = \mathrm{Bernoulli}(\hat{\mu}_i^{a|z})$.
\item For each unit $i$, we estimate $q_i^{a}$ just as we estimated $q_i^{a}$ in \Cref{sec:interference_sim}, obtaining $\hat{q}_i^{a} = \mathrm{Bernoulli}(\hat{\mu}_i^{a})$.
\item We estimate $\pr(y \mid q^{a}, q^{a |z})$ with a Gaussian process classifier, predicting $y_i$ from $\sigma^{-1}(\hat{\mu}^{a}_i)$, $\sigma^{-1}(\hat{\mu}^{a \mid z}_i(0))$ and $\sigma^{-1}(\hat{\mu}^{a |z }_i(1))$. 
This gives an estimate $\mathbb{E}_{\hat{\pr}}[Y \mid \sigma^{-1}(\hat{\mu}^{a}), \sigma^{-1}(\hat{\mu}^{a|z}(0)), \sigma^{-1}(\hat{\mu}^{a|z}(1))]$.
\end{enumerate}
Finally, we combine these estimates following the identification formula,
\begin{equation*}
        \mathbb{E}_\pr[Y \s \rmdo(q^{a} = q_\star)] \approx \frac{1}{n} \sum_{i=1}^n\mathbb{E}_{\hat{\pr}}[Y \mid \sigma^{-1}(\mu_{\star}), \sigma^{-1}(\hat{\mu}_i^{a|z}(0)), \sigma^{-1}(\hat{\mu}_i^{a|z}(1))].
\end{equation*}

We simulate data sets of size $n = 1000$ and $m = 1000$, and constructed our estimate based on subsets of increasing size.
\Cref{fig:instrument_sim} shows the convergence of our estimator to the true $m \to \infty$ effect with increasing data set size, across 20 independent simulations.
We use $\omega=0$ for the ``no confounding'' simulations (\Cref{fig:iv_low_sim}), $\omega=0.2$ for the ``low confounding'' (\Cref{fig:iv_medium_sim}) and $\omega=0.5$ for the ``high confounding'' (\Cref{fig:iv_high_sim}).
As a comparison, we also plot the behavior of an estimator which does not use the backdoor correction, and just predicts $y_i$ from  $\sigma^{-1}(\hat{\mu}^{a}_i)$ using a Gaussian process classifier.
This naive approach is incorrect for this data generating model, as it ignores confounding.

\begin{figure}[t]
\centering
\begin{subfigure}[t]{0.32\textwidth}
        \caption{Low $M$ ($\rho_m=0.25$).}
        \includegraphics[width=\textwidth]{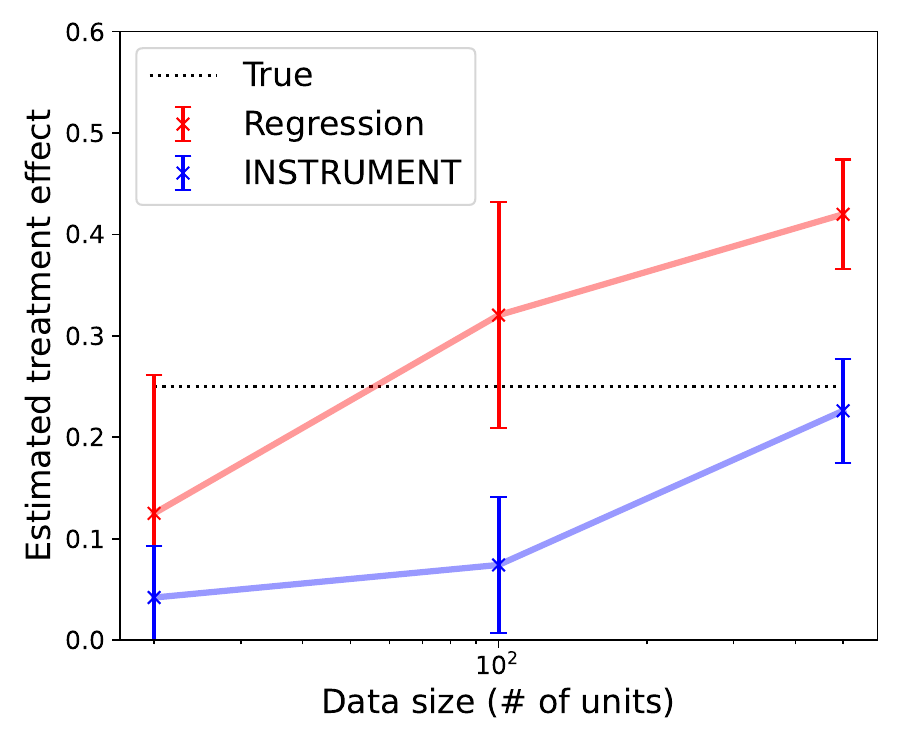}
\end{subfigure}
\begin{subfigure}[t]{0.32\textwidth}
        \caption{Medium $M$ ($\rho_m=1$).}
        \includegraphics[width=\textwidth]{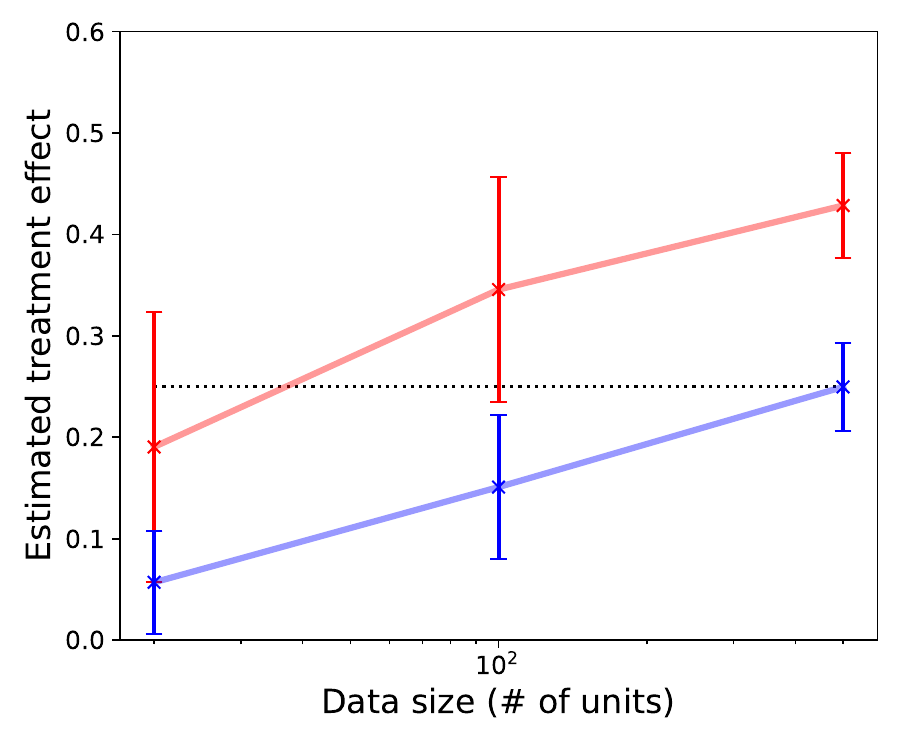}
\end{subfigure}
\begin{subfigure}[t]{0.32\textwidth}
        \caption{High $M$ ($\rho_m=4$).}
        \includegraphics[width=\textwidth]{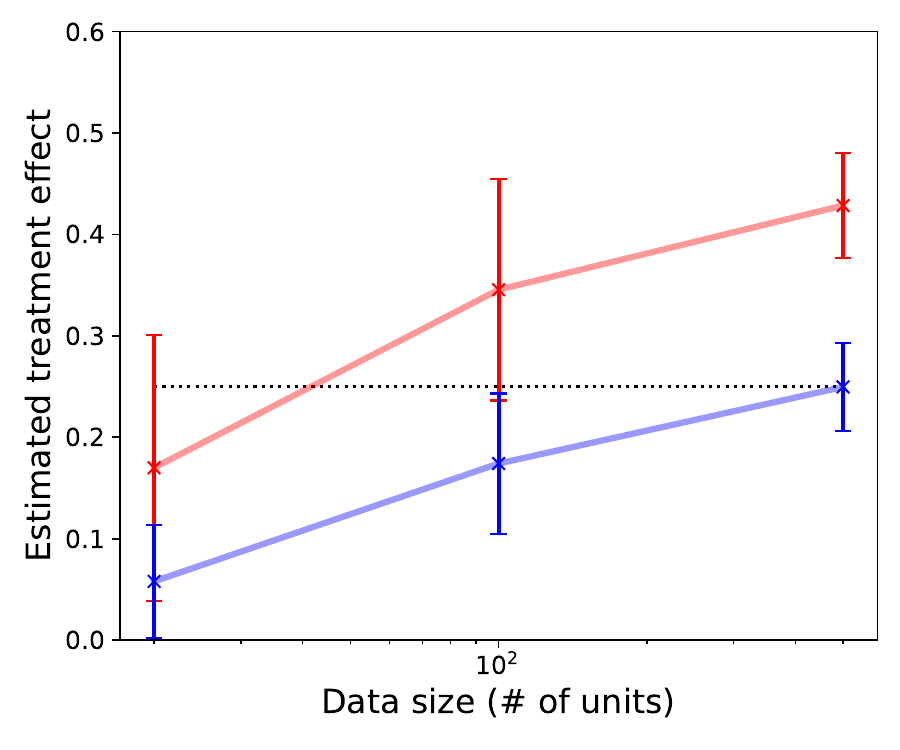}
\end{subfigure}
\caption{\textbf{Estimated effects in a } \textsc{subunit instrument} \textbf{simulation for changing number of subunits per unit ($\rho_m = M/N$).}
 Error
  bars show standard deviation across 20 independent simulations.} \label{fig:converge_sim_mfrac}
\end{figure}

We also evaluated estimation performance as the proportion of subunits to units changes. We set $M = \lceil \rho_m N \rceil$ for different $\rho_m$, and track estimator convergence as $N$ increases (\Cref{fig:converge_sim_mfrac}). As $\rho_m$ increases, the bias at low $N$ decreases  \citep[Appendix B]{Rainforth2018-qp}.  

\section{Convergent and Divergent Mechanisms} \label{sec:exp-family-mechanisms}

In this section we discuss the assumption that causal mechanisms converge (\Cref{asm:inf-subunit}), which is a key assumption on which our identification method rests. 
We describe examples of convergent and divergent mechanisms, and outline their general features.  

\subsection{Convergent mechanisms} \label{sec:exp-converge}
A simple example of a mechanism that converges with infinite subunits is $\pr(y \mid \{a_{j}\}_{j=1}^m) = \mathrm{Normal}(y \mid \frac{1}{m}\sum_{j=1}^m h(a_{j}), \sigma)$.
Note here that the mechanism can be written in terms of the empirical distribution of samples, $\mathrm{Normal}(y \mid \mathbb{E}_{A \sim \hat{q}_{m}(a)}[ h(A)], \sigma)$ with $\hat{q}_{m}(a) = \frac{1}{m} \sum_{j=1}^m \delta_{a_{j}}(a)$.
If $\mathbb{E}_{q(a)}[h(A)]$ is finite, then the mechanism will converge to $\mathrm{Normal}(y \mid \mathbb{E}_{q(a)}[h(A)], \sigma)$.
In particular, plugging in the formula for the KL divergence between two Gaussians, and applying the law of large numbers,
\begin{equation}
	\mathrm{D}^{y}_m(x^{\pa_\mathcal{U}(y)}, q(x^{\pa_\mathcal{S}(y)})) = \frac{1}{2\sigma^2}(\mathbb{E}_{\hat{q}_m(a)}[h(A)] - \mathbb{E}_{q(a)}[h(A)])^2 \to 0.
\end{equation}

This is one example of a broad family of mechanisms based on exponential family distributions that exhibit convergence, namely mechanisms of the form,
\begin{equation} \label{eqn:exp-family-mechanism}
        \pr(x^v \mid x^{\pa_\mathcal{U}(v)}, \{x_{j}^{\pa_\mathcal{S}(v)}\}_{j=1}^m) = r\Big(x^v ; g\Big (\frac{1}{m}\sum_{j=1}^m h(x^{\pa_\mathcal{U}(v)}, x_{j}^{\pa_\mathcal{S}(v)})\Big)\Big),
\end{equation}
where $h$ is a function, $g$ is a continuous function, and
$r(x ; \eta) = \lambda(x) \exp(\eta^\top s(x) - \kappa(\eta))$,
is an exponential family distribution with natural parameter $\eta \in \mathbb{R}^K$.
This class of mechanisms is quite general, as for any finite $m$, any continuous function of $x^{\pa_\mathcal{U}(v)}$ and $\{x_{j}^{\pa_\mathcal{S}(v)}\}_{j=1}^m$ can be written in the form $g\big(\frac{1}{m}\sum_{j=1}^m h(x^{\pa_\mathcal{U}(v)}, x_{j}^{\pa_\mathcal{S}(v)})\big)$ for continuous $g$ and $h$~\citep[Theorem 7]{Zaheer2017-vs}.
Under mild regularity conditions on the exponential family, the mechanism in \Cref{eqn:exp-family-mechanism} will converge.
\begin{proposition} \label{prop:exp-family-converge}
         Assume the exponential family is regular, so the natural parameter space $\Theta \triangleq \{\eta \in \mathbb{R}^K: |\kappa(\eta)| < \infty\}$ is open and $\kappa$ is continuously differentiable on $\Theta$. Assume $g$ is continuous. 
         Let $\eta_m \triangleq g\big (\frac{1}{m}\sum_{j=1}^m h(x^{\pa_\mathcal{U}(v)}, x_j^{\pa_\mathcal{S}(v)})\big) $ denote the finite subunit parameter and let $\eta_0 \triangleq g\big (\mathbb{E}_{q(x^{\pa_\Sc(v)})} \big[h(x^{\pa_\mathcal{U}(v)}, X^{\pa_\mathcal{S}(v)})\big]\big)$ denote the infinite subunit parameter, with $\mathbb{E}_{q(x^{\pa_\mathcal{S}(v)})}\|h(x^{\pa_\mathcal{U}(v)}, X^{\pa_\mathcal{S}(v)})\| < \infty$. 
         Assume both are well-defined, so $\eta_m \in \Theta$ a.s. and $\eta_0 \in \Theta$, and there is a compact set $E \subset \Theta$ for which $\eta_0 \in E$ and $\eta_m \in E$ a.s. for all $m > m_0$. 
         Then, the exponential family mechanism (\Cref{eqn:exp-family-mechanism}) converges with infinite subunits in the sense of \Cref{asm:inf-subunit}.
\end{proposition}
\begin{proof}
The limiting distribution is
$\pr(x^v \mid x^{\pa_\Uc(v)}, q(x^{\pa_\Sc(v)})) \triangleq r(x^v ; \eta_0)$, so consider the divergence
\begin{equation}
	\textsc{kl}\Big(\pr\big(x^{v} \mid x^{\pa_\mathcal{U}(v)}, q(x^{\pa_\mathcal{S}(v)})\big)\, \big\| \, \pr\big(x^v \mid x^{\pa_\mathcal{U}(v)}, \{x_{j}^{\pa_\mathcal{S}(v)}\}_{j =1}^m\big)\Big) = \textsc{kl}(r(x^v ; \eta_0) \| r(x^v ; \eta_m)).
\end{equation}
The KL divergence between two members of an exponential family can be written as a Bregman divergence between their natural parameters \citep[e.g.][]{nielsen2009statistical}, giving
\begin{equation}
        \textsc{kl}(r(x^v ; \eta_0) \| r(x^v ; \eta_m)) = B_\kappa(\eta_m \| \eta_0) = \kappa(\eta_m) - \kappa(\eta_0) - \nabla_\eta \kappa(\eta_0)^\top (\eta_m - \eta_0),
\end{equation}
where $B_\kappa$ is the Bregman divergence. By the strong law of large numbers and the continuous mapping theorem, $B_\kappa(\eta_m \| \eta_0) \to 0$ a.s. as $m \to \infty$. 
Since $\kappa$ is continuously differentiable, $B_\kappa(\eta, \eta')$ is bounded for $\eta,\eta' \in E$, so 
dominated convergence gives $\mathbb{E}[\textsc{kl}(r(x^v ; \eta_0) \| r(x^v ; \eta_m))] \to 0$ a.s. as required by \Cref{asm:inf-subunit}.
\end{proof}

\subsection{Approximating divergent mechanisms} \label{sec:approx_diverge}

Some mechanisms diverge in the infinite subunit limit. 
For example, consider a situation in which $Y_i$ depends on the total rather than an average of its parent subunit variable $A_{ij}$, namely $\pr(y \mid \{a_{ij}\}_{j=1}^m) = \mathrm{Normal}(y \mid \sum_{j=1}^m h(a_{ij}), \sigma)$.
If $h$ is strictly positive then as $m \to \infty$ we have $ \sum_{j=1}^m h(a_{ij}) \to \infty$.
For example, consider a scenario where $Y_i$ is the school budget, $A_{ij}$ the number of classes in which student $j$ enrolls, and the budget depends on the total enrollment.

In such situations, however, it is often reasonable to approximate the original divergent model with an alternative model that does exhibit convergence.
In particular, consider an HCM where the total number of subunits (e.g. the number of students in the school) is represented as a separate unit-level variable $S_i$, and we have the mechanism $\pr(y \mid s_i, \{a_{ij}\}_{j=1}^m) = \mathrm{Normal}(y \mid s_i \frac{1}{m} \sum_{j=1}^m h(a_{ij}), \sigma)$.
This mechanism matches the original mechanism if $s_i = m$, but if we allow $s_i$ to be held fixed as $m$ increases, the new mechanism converges to $\mathrm{Normal}(y \mid s_i \mathbb{E}_{q_i(a)} [h(A)], \sigma)$.
This limit corresponds to an approximation in which we ignore sampling variability among subunits, and replace the empirical mean $\frac{1}{m}\sum_{j=1}^m h(a_{ij})$ with the mean of the underlying distribution, $\mathbb{E}_{q_i(a)} [h(A)]$.
This approximation can be reasonable in settings where the number of subunits is large, according to the law of large numbers.\footnote{Analogous approximations often occur, for example, in statistical physics, when studying how macroscopic quantities (i.e. unit-level variables) depend on microscopic quantities (i.e. subunit-level variables) by considering the limit of infinite particles (i.e. subunits). E.g. rather than use the empirical mean of the squared velocity of the molecules in a gas, one can (in the thermodynamic limit) use the mean under the Maxwell-Boltzmann distribution. }
 \section{Proof of \Cref{thm:valid_collapse} (Collapsing a Hierarchical Causal Model)} \label{appx:collapsing}

In this section we prove \Cref{thm:valid_collapse}, which says that in the limit of infinite subunits, hierarchical causal graphical models converge to collapsed models.

Before beginning the proof, we briefly review two key properties of the KL divergence.
First, the KL divergence can be decomposed into a sum of conditionals: $\textsc{kl}(\pr(x, y) \| \pr'(x, y)) = \mathbb{E}_{\pr}[\log\frac{\pr(X)}{\pr'(X)}] + \mathbb{E}_{\pr}[\log\frac{\pr(Y \mid X)}{\pr'(Y \mid X)}] = \textsc{kl}(\pr(x) \| \pr'(x)) + \textsc{kl}(\pr(y \mid x) \| \pr'(y \mid x))$.
Second, from Jensen's inequality, we have $\textsc{kl}(\pr(y) \| \pr'(y)) = \mathbb{E}_{\pr}[\log \pr(Y)] - \mathbb{E}_{\pr}[\log \int \pr'(Y \mid x) \pr'(x) \di x] \le \mathbb{E}_{\pr}[\log \pr(Y)] - \mathbb{E}_{Y \sim \pr(y)} \mathbb{E}_{X \sim \pr'(x)} [\log \pr'(Y \mid X)] = \mathbb{E}_{\pr'}[\textsc{kl}(\pr(y) \| \pr'(y \mid X))]$.

\begin{proof}
The idea of the proof is to bound the KL divergence between the original HCGM and the collapsed model in terms of the sum of the expected KL divergence between each mechanism.

It will suffice to prove convergence for the observational distribution divergence, i.e.\\ $\mathrm{KL}(\pr^\col(x^\Uc, q(x^\Sc)) \| \pr_m(x^\Uc, q(x^\Sc)))$. The reason is that under the interventions defined by \Cref{def:intervention}, the post-intervention HCGM and collapsed model take the same form as the pre-intervention HCGM and collapsed model, just with a different choice of mechanism for the intervened variables.
For example, the distribution $\pr_{\Delta,m}(x^\Uc, q(x^\Sc))$ under the intervention $\rmdo(X^v \sim q_\star^{v|\pa_\Sc(v)}(x^v \mid X_{ij}^{\pa_\Sc(v)}))$ is identical to the distribution $\tilde{\pr}(x^\Uc, q(x^\Sc))$ under a modified model in which $\pr(q^{v|\pa_\Sc(v)} \mid x^{\pa_\Uc(v)})$ is replaced by $\tilde{\pr}(q^{v|\pa_\Sc(v)} \mid x^{\pa_\Uc(v)}) = \delta_{q_\star^{v|\pa_\Sc(v)}}(q^{v|\pa_\Sc(v)})$.
Likewise for the collapsed model, $\pr_\Delta^\col(x^\Uc, q(x^\Sc))$.
Moreover, in the statement of \Cref{thm:valid_collapse}, the same assumptions on the mechanisms apply for the pre- and post-intervention distributions.
So in short, convergence of the post-intervention distribution will follow as a special case of convergence of the observational distribution.

Also, it is convenient to convert from distributions over subunit joint distributions $q(x^\Sc)$ to distributions over subunit conditionals $q^\mathcal{Q} = \{q^{v|\pa_\Sc(v)}: v \in \Sc\}$, where $q(x^\Sc) = \prod_{v \in \Sc}q^{v|\pa_\Sc(v)}(x^v \mid \pa_\Sc(v))$. 
The mapping from conditional distributions to joint distributions is onto,\footnote{The mapping is not one-to-one because, if a distribution $q(x)$ does not have full support, multiple different values of $q(y\mid x)$ will yield the same joint $q(x, y)$.} which implies,
\begin{equation*}
	\mathrm{KL}(\pr^\col(x^\Uc, q(x^\Sc)) \| \pr_m(x^\Uc, q(x^\Sc))) \le \mathrm{KL}(\pr^\col(x^\Uc, q^\mathcal{Q}) \| \pr_m(x^\Uc, q^\mathcal{Q})).
\end{equation*}
Our task is now to show that the right hand side converges to zero as $m \to \infty$.

We take the indices of the endogenous variables to be causally ordered, so that $\pa(v) \subseteq \{1, \ldots, v-1\}$ for all $v \in \{1, \ldots, V\}$.
Let $v^\Sc_{k}$ denote the index of the $k$th subunit-level variable, i.e. $v^\Sc_{k} \in \Sc$ and $v^\Sc_{k} < v^\Sc_{k+1}$ for all $k \in \{1, \ldots, |\Sc|\}$.
Set $v^\Sc_0 = 0$.
Let $\Sc > v$ denote the set of subunit variables after $v$ in the causal ordering, that is $\{v' \in \mathcal{S}: v' > v\}$, and define $\Uc > v$ analogously, as well as $\Sc \le v$ and $\Uc \le v$.
Let $q^{\Qc > v}$ denote the set of $Q$ variables that describe subunit variables after $v$, that is $q^{\Qc > v} = \{q^{v'|\pa_\Sc(v')} : v' > v\}$.

Our bound will decompose in terms of conditional distributions in the HCGM, over variables after $v^{\Sc}_k$ in the causal ordering given variables up to $v^{\Sc}_k$. In particular, we use,
\begin{equation*}
\begin{split}
	\pr &(x^{\Uc > v^{\Sc}_k}, q^{\Qc > v^{\Sc}_k} \mid x^{\Uc \le v^{\Sc}_k}, \{x_j^{\Sc \le v^{\Sc}_k}\}_{j=1}^m)\\
	& = \int \cdots \int \prod_{k'=k+1}^{|\Sc|}  \left[ \prod_{j=1}^m q^{v_{k'}^\Sc | \pa_\Sc(v_{k'}^\Sc)}(x_j^{v_{k'}^\Sc} | x_j^{\pa_\Sc(v_{k'}^\Sc)}) \right] \pr(q^{v_{k'}^\Sc | \pa_\Sc(v_{k'}^\Sc)} \mid x^{\pa_\Uc(v_{k'}^\Sc)}) \\ 
	& \quad \quad \quad \quad \quad \quad \quad \times \prod_{v'=v^{\Sc}_{{k'}-1}+1}^{v_{k'}^\Sc-1} \pr(x^{v'} \mid x^{\pa_\Uc(v')}, \{x_j^{\pa_\Sc(v')}\}_{j=1}^m) \di x_1^{\Sc > v_k^\Sc}\ldots \di x_m^{\Sc > v_k^\Sc}.
\end{split}
\end{equation*}
Note the full distribution of the HCGM over all $Q$ variables and unit $X^v$ variables, $\pr_m(x^\Uc, q^\mathcal{Q})$, corresponds to the case where $k = 0$.
We compare to the collapsed model, which has the conditional,
\begin{equation*}
\begin{split}
\pr^\col(x^{\mathcal{U} > v^{\Sc}_k}, q^{\Qc > v^{\Sc}_k} \mid x^{\mathcal{U} \le v^{\Sc}_k}, & q^{\Qc \le v^{\Sc}_k}) = \prod_{k'=k+1}^{|\Sc|}  \pr(q^{v_{k'}^\Sc | \pa_\Sc(v_{k'}^\Sc)} \mid x^{\pa_\Uc(v_{k'}^\Sc)}) \prod_{v'=v^{\Sc}_{{k'}-1}+1}^{v_{k'}^\Sc-1} \pr(x^{v'} \mid x^{\pa_\Uc(v')}, q(x^{\pa_\Sc(v')})).
\end{split}
\end{equation*} 
The full distribution of the collapsed model over all $Q$ variables and unit $X^v$ variables, $\pr^\col(q^\mathcal{Q}, x^\Uc)$, corresponds to the case where $k = 0$.

We now bound the KL by decomposing it into a sum over conditionals, and applying Jensen's inequality.
\begin{equation*}
\begin{split}
\mathrm{KL}&(\pr^\col(x^\Uc, q^\mathcal{Q}) \| \pr_m(x^\Uc, q^\mathcal{Q}))\\
=& \mathbb{E}_{\pr^\col(x^{\Uc \le v_1^\Sc}, q^{v_1^\Sc})} [ \mathrm{KL}(\pr^\col(x^{\Uc > v^{\Sc}_1}, q^{\Qc > v^{\Sc}_1} \mid X^{\Uc \le v^{\Sc}_1}, Q^{v^{\Sc}_1}) \| \mathbb{E}_{Q(x^{v_1^\Sc})} [ \pr(x^{\Uc > v^{\Sc}_1}, q^{\Qc > v^{\Sc}_1} \mid X^{\Uc \le v^{\Sc}_1}, \{X_j^{v^{\Sc}_1}\}_{j=1}^m) ] ) ]\\
&  + \mathrm{KL}( \pr(q^{v_{1}^\Sc} \mid x^{\pa_\Uc(v_{1}^\Sc)}) \| \pr(q^{v_{1}^\Sc} \mid x^{\pa_\Uc(v_{1}^\Sc)}) )
 + \sum_{v'=1}^{v_1^\Sc - 1} \mathrm{KL}( \pr(x^{v'} \mid x^{\pa_\Uc(v')}) \| \pr(x^{v'} \mid x^{\pa_\Uc(v')}) ) \\
\le & \mathbb{E}_{\pr^\col(x^{\Uc \le v_1^\Sc}, q^{v_1^\Sc})} [ \mathbb{E}_{Q(x^{v_1^\Sc})} [ \mathrm{KL}(\pr^\col(x^{\Uc > v^{\Sc}_1}, q^{\Qc > v^{\Sc}_1} \mid X^{\Uc \le v^{\Sc}_1}, Q^{v^{\Sc}_1}) \| \pr(x^{\Uc > v^{\Sc}_1}, q^{\Qc > v^{\Sc}_1} \mid X^{\Uc \le v^{\Sc}_1}, \{X_j^{v^{\Sc}_1}\}_{j=1}^m) )]]\\
=& \mathbb{E}_{\pr^\col(x^{\Uc \le v_2^\Sc}, q^{\Qc \le v_2^\Sc})} [ \mathbb{E}_{Q(x^{v_1^\Sc})} [ \mathrm{KL}(\pr^\col(x^{\Uc > v^{\Sc}_2}, q^{\Qc > v^{\Sc}_2} \mid X^{\Uc \le v^{\Sc}_2}, Q^{\Qc \le v^{\Sc}_2}) \| \\ &\quad \quad \quad \quad \quad \quad \quad \quad \quad \quad \quad  \mathbb{E}_{Q^{v_2^\Sc|\pa_\Sc(v_2^\Sc)}(x^{v^\Sc_2}|X^{\pa_\Sc(v_2^\Sc)})} [ \pr(x^{\Uc > v^{\Sc}_2}, q^{\Qc > v^{\Sc}_2} \mid X^{\Uc \le v^{\Sc}_2}, \{X_j^{\Sc \le v^{\Sc}_2}\}_{j=1}^m) ] )]]\\
&+ \mathbb{E}_{\pr^\col(x^{\Uc \le v_2^\Sc})}[\mathrm{KL}(\pr^\col(q^{v_2^\Sc|\pa_\Sc(v_2^\Sc)} \mid X^{\pa_\Uc(v_2^\Sc)}) \| \pr(q^{v_2^\Sc|\pa_\Sc(v_2^\Sc)} \mid X^{\pa_\Uc(v_2^\Sc)}))]\\
&+ \sum_{v'=v_1^\Sc+1}^{v_2^\Sc-1} \mathbb{E}_{\pr^\col(x^{\Uc \le v'}, q^{\Qc < v'})} [ \mathbb{E}_{Q(x^{\Sc < v'})} [ \mathrm{KL}( \pr^\col(x^{v'} \mid X^{\pa_\Uc(v')}, Q(x^{\pa_\Sc(v')})) \| \pr(x^{v'} \mid X^{\pa_\Uc(v')}, \{X^{\pa_\Sc(v')}\}_{j=1}^m))].
\end{split}
\end{equation*}
We can recognize the terms in the final sum as $\mathbb{E}_{\pr^\col}[\mathrm{D}^{v'}_m(X^{\pa_\mathcal{U}(v')}, Q(x^{\pa_\mathcal{S}(v')}))]$.
Since there are a finite number of unit variables $|\mathcal{U}|$, we can choose $m$ large enough such that $\mathbb{E}_{\pr^\col}[\mathrm{D}^{v'}_m(X^{\pa_\mathcal{U}(v')}, Q(x^{\pa_\mathcal{S}(v')}))] < \epsilon$ for all $v' \in \Uc$.
So, continuing the bound, we obtain,
\begin{equation*}
\begin{split}
\le & \mathbb{E}_{\pr^\col(x^{\Uc \le v_2^\Sc}, q^{\Qc \le v_2^\Sc})} [ \mathbb{E}_{Q(x^{\Sc \le v_2^\Sc})} [ \mathrm{KL}(\pr^\col(x^{\Uc > v^{\Sc}_2}, q^{\Qc > v^{\Sc}_2} \mid X^{\Uc \le v^{\Sc}_2}, Q^{\Qc \le v^{\Sc}_2}) \| \\ &\quad \quad \quad \quad \quad \quad \quad \quad \quad \quad \quad \quad\quad\quad\quad  \pr(x^{\Uc > v^{\Sc}_2}, q^{\Qc > v^{\Sc}_2} \mid X^{\Uc \le v^{\Sc}_2}, \{X_j^{\Sc \le v^{\Sc}_2}\}_{j=1}^m) )]]\\
&+ \epsilon(v_2^\Sc - v_1^\Sc - 1)\\
\cdots &  \\
\le & \mathbb{E}_{\pr^\col(x^{\Uc \le v_k^\Sc}, q^{\Qc \le v_k^\Sc})} [ \mathbb{E}_{Q(x^{\Sc \le v_k^\Sc})} [ \mathrm{KL}(\pr^\col(x^{\Uc > v^{\Sc}_k}, q^{\Qc > v^{\Sc}_k} \mid X^{\Uc \le v^{\Sc}_k}, Q^{\Qc \le v^{\Sc}_k}) \| \\ &\quad \quad \quad \quad \quad \quad \quad \quad \quad \quad \quad  \quad\quad\quad\quad\pr(x^{\Uc > v^{\Sc}_k}, q^{\Qc > v^{\Sc}_k} \mid X^{\Uc \le v^{\Sc}_k}, \{X_j^{\Sc \le v^{\Sc}_k}\}_{j=1}^m) )]]\\
&+ \sum_{k'=1}^{k-1} \epsilon(v_{k'+1}^\Sc - v_{k'}^\Sc - 1)\\
\cdots &\\
& \le |\Uc| \epsilon.
\end{split}
\end{equation*}
Thus, the KL divergence between the collapsed model and the HCGM converges to zero as $m \to \infty$ a.s..
\end{proof} \section{Convergence of Hierarchical Empirical Distributions} \label{apx:hier_empirical_convergence}

In \Cref{sec:identification_problem}, we introduced the identification assumption that $\pr(x^{\Uc_\obs}, q(x^{\Sc_\obs}))$ is known (\Cref{assume:known_joint}).
In this section we justify this assumption, by showing that with sufficient data we can infer $\pr(x^{\Uc_\obs}, q(x^{\Sc_\obs}))$.

In \Cref{thm:valid_collapse} we showed that HCGMs converge to collapsed models, in the limit of infinite subunits. So in this limit, data from the HCGM can be modeled as (\Cref{def:um}),
\begin{equation} \label{eqn:hcm_observed_generation}
\begin{split}
	X^{\Uc_\obs}_i, Q_i &\sim \pr(x^{\Uc_\obs}, q(x^{\Sc_\obs}))\\
	X^{\Sc_\obs}_{ij} &\sim Q_i(x^{\Sc_\obs}).
\end{split}
\end{equation}
We assume now that we have a dataset $\{x^{\Uc_\obs}_i, \{x^{\Sc_\obs}_{ij}\}_{j=1}^M\}_{i=1}^N $, with $N$ units and $M$ subunits per unit, drawn from \Cref{eqn:hcm_observed_generation}. 
Here, we make a distinction here between the true underlying number of subunits in the model, $m$, and the number of subunits we actually observe, $M$.
In other words, we assume that while in reality there are an effectively infinite number of subunits -- which justifies the use of the collapsed model to describe the data (\Cref{thm:valid_collapse}) -- in practice we have a finite dataset.
We similarly draw a distinction between the true and observed number of units, $n$ versus $N$.
We will show that as we gather more data, i.e. as $N,M\to\infty$, we can learn $\pr(x^{\Uc_\obs}, q(x^{\Sc_\obs}))$.

In flat causal models, the data $\{x_{i}^\obs\}_{i=1}^N$ is drawn as $X_i^\obs \sim \pr(X^{\obs})$, and identification is studied under the assumption that $\pr(x^{\obs})$ is known.
This assumption is motivated by the fact that as we gather more data, i.e. as $N \to \infty$, we can infer $\pr(x^{\obs})$ arbitrarily well.
More precisely, $p_N \to \pr$ a.s., where $p_N = \frac{1}{N} \delta_{x_i^\obs}$ is the empirical distribution of the data \citep[e.g.][Theorem 11.4.1]{Dudley2002-pd}.
Here, $p_N \to \pr$ denotes weak convergence, though note that convergence holds in many other senses as well (for instance, the classic Glivenko-Cantelli theorem, the ``fundamental theorem of statistics'', describes uniform convergence of the empirical c.d.f.).

In hierarchical causal models, the situation is somewhat different: we do not actually observe an empirical distribution of data from $\pr(x^{\Uc_\obs}, q(x^{\Sc_\obs}))$.
That is, $p_N = \frac{1}{N} \delta_{x^{\Uc_\obs}_i, q_i}$ is unobserved, since the $q_i$ are unobserved.
Instead, we have access to the empirical distribution of empirical distributions $p_{N,M} = \frac{1}{N} \sum_{i=1}^N \delta_{x^{\Uc_\obs}_i,q_{M,i}}$ where $q_{M,i} = \frac{1}{M}\sum_{j=1}^M \delta_{x^{\Sc_\obs}_{ij}}$.
We refer to $p_{N,M}$ as a \textit{hierarchical} empirical distribution.
In this section, we will show that $p_{N,M} \to \pr$ a.s., just like the non-hierarchical empirical distribution. So, despite the noise contributed by variation among subunits, we can still learn about the true distribution without making any parametric assumptions. 

Both $\pr(x^{\Uc_\obs}, q(x^{\Sc_\obs}))$ and $p_{N,M}$ are distributions over $\mathcal{X}^{\Uc_\obs} \times \mathcal{P}(\mathcal{X}^{\Sc_\obs})$, where $\mathcal{P}(\mathcal{X}^{\Sc_\obs})$ is the set of distributions on $\mathcal{X}^{\Sc_\obs}$.
To establish weak convergence of the hierarchical empirical distribution, we will need a metric over this space of distributions.
As in previous studies of the asymptotic behavior of hierarchical probabilistic models, we rely on the Wasserstein distance~\citep{Nguyen2013-tv}.
In particular, we focus on the Wasserstein 1-distance, defined for two measures $\mu,\nu$ on a separable and complete metric space $(\mathcal{X},d)$ as,
\begin{equation*}
	\Was_{1}(\mu,\nu) = \inf_{\gamma \in \Gamma(\mu,\nu)} \mathbb{E}_{X, Y \sim \gamma}[d(X,Y)],
\end{equation*}
where $\Gamma(\mu,\nu)$ denotes the set of all couplings of $\mu,\nu$, i.e. the set of all joint distributions with marginals $\mu,\nu$.

We will use the Wasserstein distance not only to construct a metric on probability distributions but also a metric on probability distributions over probability distributions.
We can do so using \textit{Vershik's tower}~\citep[Chap. 1.1]{Bogachev2012-dz}. Let $\mathcal{P}^1$ be the set of all Borel probability measures on $\mathcal{X}$ with finite first moment, i.e. $\mathbb{E}_{X \sim \mu}[d(X, x_0)] < \infty$ for all $\mu \in \mathcal{P}^1$ and an arbitrary $x_0 \in \mathcal{X}$. 
\begin{proposition}[Vershik's tower]
	Let $(\mathcal{X}, d)$ be complete, separable metric space. Then, $(\mathcal{P}^1(\mathcal{X}), \Was_{1})$ is a complete and separable metric space, as is $(\mathcal{P}^1(\mathcal{P}^1(\mathcal{X})), \Was_{\Was_{1}})$,
	$(\mathcal{P}^1(\mathcal{P}^1(\mathcal{P}^1(\mathcal{X}))), \Was_{\Was_{\Was_{1}}})$, etc.  Moreover, if $(\mathcal{X}, d)$ is compact, so are all the other spaces.
\end{proposition}
\noindent Vershik's tower is a useful tool for analyzing hierarchical probabilistic models, as it allows us to construct metrics on distributions with any level of hierarchy.

We now show that the hierarchical empirical distribution converges with increasing data.
\begin{proposition}[Hierarchical empirical distributions converge]
	Let $(\mathcal{X}^{\Sc_\obs}, d^{\Sc_\obs})$ and $(\mathcal{X}^{\Uc_\obs}, d^{\Uc_\obs})$ be compact, separable metric spaces.
	Assume $\pr(x^{\Uc_\obs}, q(x^{\Sc_\obs})) \in \mathcal{P}^1(\mathcal{X}^{\Uc_\obs} \times \mathcal{P}^1(\mathcal{X}^{\Sc_\obs}))$, where $\mathcal{X}^{\Uc_\obs} \times \mathcal{P}^1(\mathcal{X}^{\Sc_\obs})$ has metric $d^{\Uc_\obs} + \Was_{1,d^{\Sc_\obs}}$.
	In the limit as $M \to \infty $ and then $N \to \infty$,  $p_{N,M}$ converges weakly to $\pr(x^{\Uc_\obs}, q(x^{\Sc_\obs}))$ a.s..
\end{proposition}
\begin{proof}
	Since $(\mathcal{X}^{\Sc_\obs}, d_{\Sc_\obs})$ is compact and separable, so is $(\mathcal{P}^1(\mathcal{X}^{\Sc_\obs}), \Was_{1, d_{\Sc_\obs}})$, and since $(\mathcal{X}^{\Uc_\obs}, d_{\Uc_\obs})$ is also compact and separable, so is the product space $(\mathcal{X}^{\Uc_\obs} \times \mathcal{P}^1(\mathcal{X}^{\Sc_\obs}), d^{\Uc_\obs} + \Was_{1,d^{\Sc_\obs}})$.
	This implies that the Wasserstein distance over this last space is well-defined.
	
	Using the triangle inequality, and the explicit form of the Wasserstein distance for empirical distributions, we can bound the distance between the hierarchical empirical distribution and the true distribution as,
	\begin{align}
		\Was_1(p_{N,M}, \pr) \le & \Was_1(p_{N,M}, p_N) + \Was_1(p_N, \pr) 
		\le \frac{1}{N} \sum_{i=1}^N d(x^{\Uc_\obs}_i, x^{\Uc_\obs}_i) + \Was_1(q_{M,i}, q_i) + \Was_1(p_N,\pr)\\ = & \frac{1}{N} \sum_{i=1}^N \Was_1(q_{M,i}, q_i) + \Was_1(p_N,\pr). \label{eqn:wass_decomp}
	\end{align}
	Each term in the final expression compares a (non-hierarchical) empirical distribution to the distribution it is sampled from.
	For any Borel measure $\mu$ over a compact and separable metric space, the empirical distribution $\mu_N$ satisfies $\Was_1(\mu_N,\mu) \to 0$ a.s. \citep[e.g.][]{Weed2019-mk}.
	Therefore each term of \Cref{eqn:wass_decomp} converges and we have $\Was_1(p_{N,M}, \pr) \to 0$ a.s. when $M \to \infty$ and then $N \to\infty$.
	
	Since the Wasserstein 1-distance metrizes weak convergence \citep[e.g.][Theorem 6.9]{Villani2008-gw}, the conclusion follows.
\end{proof}
\noindent This result tells us that with sufficient data, it is possible to learn $\pr(x^{\Uc_\obs}, q(x^{\Sc_\obs}))$, without making any parametric assumptions about this distribution.

\paragraph{Convergence rate}
We can also obtain a rate of convergence in $N$ and $M$. \citet{Weed2019-mk} show that the Wasserstein distance between a distribution $\mu$ and its empirical distribution of samples $\mu_N$ satisfies $\EE{}{W_1(\mu, \mu_N)} = O(n^{-1/d_\mu})$ where $d_\mu$ is a constant describing the effective dimension of the distribution. (If $\mu$ is a distribution on $\mathbb{R}^d$ then $d_\mu \le d$.) 
From \Cref{eqn:wass_decomp} we can obtain the convergence rate,
\begin{equation}
	\EE{}{\Was_1(p_{N,M}, \pr)} = \frac{1}{N} \sum_{i=1}^N \EE{}{\Was_1(q_{M,i}, q_i)} + \EE{}{\Was_1(p_N,\pr)} = O(M^{-1/d_q} + N^{-1/d_\pr})
\end{equation}
where $d_q$ is the maximum effective dimension of $q$ (under $\pr$), and $d_\pr$ is the effective dimension of $\pr$. See \citet{Weed2019-mk} for details on the effective dimension.

Another way of quantifying convergence is to look at the expected value of a test function, e.g. how quickly $\frac{1}{N} \sum_{i=1}^N g(x_i)$ converges to $\EE{\pr}{g(X)}$. For hierarchical distributions, this can be generalized to examine the mismatch between,
\begin{align}
A \triangleq \EE{\pr}{g(X^{\Uc_\obs}, \EE{Q}{h(X^{\Sc_\obs}, X^{\Uc_\obs})})}\\
A_{N,M} \triangleq \frac{1}{N}\sum_{i=1}^N g(x_i^{\Uc_\obs}, \frac{1}{M}\sum_{j=1}^M h(x_i^{\Uc_\obs}, x^{\Sc_\obs}_{ij})) .
\end{align}
for any test functions $g$ and $h$.
Under regularity conditions--when $g$ and $h$ are Lipschitz and output finite-dimension vectors, and $A_{N, M}$ has finite variance--\citet[Theorem 1]{Rainforth2018-qp} implies,
\begin{equation}
	\|A - A_{N,M}\|^2 = O(1/N + 1/M).
\end{equation} 
\section{Positivity for Do-Calculus} \label{apx:positivity}

Do-calculus rests on positivity assumptions, which ensure the post-intervention distribution can be computed from the pre-intervention distribution. 
It is common, for the sake of simplicity, to assume that the joint distribution over all the endogenous variables is strictly positive. 
However, this assumption is stronger than necessary, and indeed can block identification in some HCMs, such as in the \textsc{subunit instrument} model (\Cref{sec:completeness}).
We therefore employ the weaker positivity assumptions developed in \citet{Shpitser2006-jg}.

The first of these assumptions is that the intervention has positive probability (\Cref{assume:unit_positive}).
This assumption ensures that the intervention we are considering is well-defined.
The second positivity assumption stems from the fact that do-calculus often provides identification formulae with terms of the form, 
\begin{equation} \label{eqn:positive_term}
	\int \pr(y \mid x) \tilde{\pr}(x) \di x,
\end{equation}
where $\tilde{\pr}(x)$ denotes a non-observational distribution over $x$ (here, $x$ and $y$ may each represent one or more endogenous variables).
For example, if we are performing a hard intervention on $x$ we could have $\tilde{\pr}(x) = \delta_{x_\star}(x)$ (or see e.g. \Cref{eq:interfere_front_door} for another example).
To compute \Cref{eqn:positive_term}, we need to be able to estimate $\pr(y \mid x)$ for all values of $x$ on which $\tilde{\pr}(x)$ has support, using observational data.
\begin{assumption}[Unit-level positivity, part two~\citep{Shpitser2006-jg}] \label{asm:positivity_2}
	For each term of the form \Cref{eqn:positive_term} that appears in the identification formula provided by do-calculus, we require that $\pr(x)$ is positive wherever $\tilde{\pr}(x)$ is positive, i.e. $\tilde{\pr}(x) \ll \pr(x)$, where $\ll$ denotes absolute continuity.
\end{assumption}

There is one additional subtlety: while do-calculus is typically studied under the simplifying assumption that the variables are discrete, in collapsed models the $Q$ variables are necessarily non-discrete, even when all the endogenous variables of the HCGM are discrete.
The above positivity assumption only applies to the discrete variable case, since in the continuous case, when $\tilde{\pr}(x)$ involves a delta function, we do not have $\tilde{\pr}(x) \ll \pr(x)$ in general.
However, \Cref{asm:positivity_2} can be relaxed with some technical regularity assumptions.
Here we give a relaxation that is general enough to apply to the distribution-valued endogenous variables that appear in collapsed models.
Intuitively, this assumption extends \Cref{asm:positivity_2} to consider positivity in a neighborhood of the intervention.
Let $\mathcal{X}$ and $\mathcal{Y}$ denote the domains of $x$ and $y$, and let $\mathcal{P}(\mathcal{Y})$ denote the set of distributions over $\mathcal{Y}$.
\begin{assumption}[Positivity for general variables] \label{asm:positivity_general}
        Consider each term in the identification formula of the form of \Cref{eqn:positive_term}.
        Assume there exists a known sequence of distributions $\nu_1, \nu_2, \cdots$ converging weakly to $\tilde{\pr}(x)$ such that $\nu_k(x) \ll \pr(x)$ for all $k$.
        Assume for all measurable $\mathcal{A} \subseteq \mathcal{Y}$, $\Pr(y \in \mathcal{A} \mid x)$ is a continuous function of $x$.
\end{assumption}
\begin{proposition} \label{prop:positivity_general}
        Under \Cref{asm:positivity_general}, and given that do-calculus identifies $\tilde{\pr}(x)$, \Cref{eqn:positive_term} is identified.
\end{proposition}
\begin{proof}
From the observational distribution $\pr(x, y)$, we can compute the expected value $\mathbb{E}_{X \sim \nu_k}[\Pr(y \in \mathcal{A} \mid X)]$ using the importance sampling formula,
$\mathbb{E}_{X \sim \nu_k}[\Pr(y \in \mathcal{A} \mid X)] = \mathbb{E}_{X \sim \pr(x)}\Big[\Pr(y \in \mathcal{A} \mid X)\frac{\nu_k(X)}{\pr(X)}\Big]$,
since $\nu_k(x) \ll \pr(x)$.
Since $\Pr(y \in \mathcal{A} \mid x)$ is continuous and bounded, and $\nu_k$ converges weakly to $\tilde{\pr}(x)$, we have
$\mathbb{E}_{X \sim \nu_k}[\Pr(y \in \mathcal{A} \mid X)] \to \mathbb{E}_{X \sim \tilde{\pr}(x)}[\Pr(y \in \mathcal{A} \mid X)]$.
Since each term $\mathbb{E}_{X \sim \nu_k}[\Pr(y \in \mathcal{A} \mid X)]$ is identified, and since the sequence converges, we can identify $\int \Pr(y \in \mathcal{A} \mid x) \tilde{\pr}(x) \di x$ as its limit. 
Since this holds for any $\mathcal{A}$, we can identify the entire distribution $\int \pr(y \mid x) \tilde{\pr}(x) \di x$
\end{proof}

 \section{Instrumental Variable Assumptions} \label{sec:completeness}

 Here we discuss further the assumptions needed for identification in the \textsc{subunit instrument} graph (\Cref{sec:instrument_id}), and explain how they relate to the identification assumptions used in conventional, flat instrumental variable models.
 
 The key assumptions for applying do-calculus and achieving identification in the \textsc{subunit instrument} graph are the positivity assumptions, \Cref{assume:subunit_positive} and \Cref{assume:unit_positive}.
 First, there must always be within-unit variation in the instrument, i.e. we must have $Q^z(z) > 0$ a.s. for $Q^z \sim \pr(q^z)$ (\Cref{assume:subunit_positive}). 
 Second, there must always be a non-zero probability of $q^a = q^a_\star$ given $q^{a \mid z}$, that is, $\pr(q^a = q^a_\star \mid q^{a \mid z}) > 0$ a.s. for $Q^{a \mid z} \sim \pr(q^{a \mid z})$ (\Cref{assume:unit_positive}).
Said another way:
\begin{assumption}[Unit-level positivity for the \textsc{subunit instrument} graph]\label{asm:instrument_positive} 
For $Q^{a \mid z} \sim \pr(q^{a \mid z})$, there must exist a.s. a solution $q^z$ to the integral equation
\begin{equation} \label{eqn:iv_integral_eqn}
        \int q^{a \mid z} (a \mid z) q^{z} (z) \di z = q_\star^a(a),
 \end{equation}
 such that $\pr(q^z) > 0$. 
 \end{assumption}
\noindent In brief, for any value of $q^{a \mid z}$, there must exist some value of $q^z$ that produces the marginal $q^a = q^a_\star$.

The positivity requirements in the hierarchical \textsc{subunit instrument} model are related to the relevance and completeness assumptions that appear in flat instrumental variable models~\citep{Newey2003-dl}.
Intuitively, in both the hierarchical and flat settings we need the instrument to (1) vary and (2) actually affect the treatment, i.e. it cannot be a completely unrelated quantity.
To see this in the hierarchical case, note that if $q^{a \mid z}(a \mid z)$ were constant with respect to $z$, then $\int q^{a \mid z} (a \mid z) q^{z} (z) \di z$ would be constant with respect to $q^{z}$.
This will in general violate \Cref{asm:instrument_positive}.\footnote{One implication is that, to achieve identification, we must have $\pr(q^{a|z}=\tilde{q}(a)) = 0$ for all distributions $\tilde{q}(a)$ that are constant with respect to $z$. This conflicts with a common positivity assumption made in do-calculus, that the observational distribution is positive everywhere. Hence, we employ the weaker do-calculus positivity assumptions proposed by \citet{Shpitser2006-jg}, and described in \Cref{apx:positivity}.}

Despite these similarities, the positivity assumptions in the hierarchical instrumental variable model are distinct from the assumptions made in flat instrumental variable models.
To see this, we compare to the completeness assumption that is widely used in flat nonparametric instrumental variable models~\citep{Newey2003-dl}.
In a flat instrumental variable model with $Q^z$ as the instrument and $Q^a$ as the treatment, the completeness assumption can be stated as: There must exist a unique solution $\mathrm{g}(\cdot)$ to the equation,
\begin{equation} \label{eqn:flat-iv}
        \int \pr(q^{a}\mid q^{z}) \mathrm{g}(q^{a}) \di q^{a} = \mathbb{E}_\pr[Y \mid q^{z}].
\end{equation}
While both \Cref{eqn:iv_integral_eqn} and \Cref{eqn:flat-iv} are Fredholm integral equations of the first kind, where the kernel describes a conditional distribution over $A$ given $Z$, they are otherwise quite distinct~\citep{Carrasco2007-cv}.
For example, \Cref{eqn:iv_integral_eqn} involves a within-unit conditional distribution of $A$ given $Z$, whereas \Cref{eqn:flat-iv} involves a between-unit conditional distribution.
Moreover, the solution to \Cref{eqn:iv_integral_eqn} just needs to exist, whereas in \Cref{eqn:flat-iv} the solution must be unique.

An important advantage of the hierarchical instrument variable model is that identification does not require any assumptions on the causal mechanism generating the outcome.
In particular, the mechanism generating $Y$ from $Q^{a}$ does not need to have additive noise or to be monotonic with respect to the noise, as is required in the flat instrumental variable setting~\citep{Imbens2002-tk,Newey2003-dl,Saengkyongam2022-se}.
Moreover, in the hierarchical model we can identify the entire post-intervention distribution $\pr(y \s \rmdo(q^{a} = q_*^a))$, whereas in the flat model we can only identify the mean $\mathbb{E}_\pr[Y \s \rmdo(q^{a} = q_*^a)]$.
This is especially relevant for problems with structured outcome variables, for example if $Y$ is a text, graph, or molecule. In these cases, additive noise may be ill-defined, limiting the application of flat instrumental variable methods.

\section{Details on Augmentation} \label{apx:augment_marginalize}

In this section we provide further details on our augmentation approach, and prove \Cref{prop:intervene_valid}.

\subsection{General form of augmentation variables} \label{apx:aug_var_form} We employ augmentation variables of a particular form, namely those generated as,
\begin{equation}\label{eqn:augment_mech}
        q_i^{\mathcal{L} \mid \mathcal{R}}(x^{\mathcal{L}} \s \rmdo(x^{\mathcal{R}})) = \int \cdots \int \prod_{v \in \mathcal{L} \cup \da_\mathcal{S}(\mathcal{L}) \setminus \mathcal{R}} q_i^{v \mid \pa_\mathcal{S}(v)}(x^v \mid x^{\pa_\mathcal{S}(v)})\prod_{w \in \da_\mathcal{S}(\mathcal{L}) \setminus \mathcal{R}} \di x^{w},
\end{equation}
where $\mathcal{L} \subseteq \mathcal{S}$ is a set of subunit-level variables and $\mathcal{R} \subseteq \da_\mathcal{S}(\mathcal{L}) = \big(\bigcup_{v \in \mathcal{L}} \da_\mathcal{S}(v)\big)\setminus \mathcal{L}$ is a set of subunit direct ancestors of $\mathcal{L}$.
\Cref{eqn:augment_mech} describes the within-unit distribution over $X^{\mathcal{L}}$ after an intervention on $X^{\mathcal{R}}$, holding fixed the unit variables. In other words, it is the interventional effect derived from the subunit variable graph, with the unit variables and outer plate ignored.
For example, in \Cref{fig:ID_ex5u}, when we erase the unit variables and outer plate we obtain a graph $Z \rightarrow A \leftarrow X$, so the augmentation variable $q_i^{a \mid x}$ is given by $q_i^{a \mid x}(a \s \rmdo(x)) = \int q_i^{a \mid z, x}(a \mid z, x) q_i^z(z) \di z$.

\subsection{Proof of \Cref{prop:intervene_valid}}
Fundamentally, the purpose of \Cref{prop:intervene_valid} is to allow us to take advantage of constraints in the mechanisms of collapsed models, in order to establish identification.
Even when we place no parametric restrictions on mechanisms in a hierarchical causal model, they appear in the collapsed model.
For example, in the \textsc{subunit instrument} HCGM, we did not constrain the mechanism generating $Y$, but we found that in the collapsed model $Y$ can only depend on its parents $Q^{a|z}$ and $Q^{z}$ through their marginal $\int Q^{a|z}(a \mid z) Q^z(z) \di z$ (\Cref{eqn:instrument_collapse}, \Cref{sec:instrument_id}).
Do-calculus operates under the assumption that there are no parametric constraints on the mechanisms in a causal graphical model.
So, when we apply do-calculus directly to the collapsed model, we cannot take advantage of the model's constraints to prove identification.
Augmentation, together with \Cref{prop:intervene_valid}, allows us to use these constraints effectively.

\begin{proof}
	Note the augmentation variable $Q^{v|\mathcal{R}}$ follows \Cref{eqn:augment_mech} with $\mathcal{L} = \{v\}$. In this proof, we use $\mathcal{L}$ in place of $v$ to make clear we are discussing the intervened variable rather than a generic variable.
	
	We will show that $\pr^\col(y \s \rmdo(\qLp = \qsLR)) = \pr^\aug(y \s \rmdo(\qLR = \qsLR))$. This implies the result $\pr^\col(y \s \rmdo(\qLp = \qsLR)) = \pr^\mar(y \s \rmdo(\qLR = \qsLR))$, since marginalizing out a variable from the model cannot change the effect.
	Note that, in the augmented model, all directed paths from $\QLp$ to $Y$ must go through $\QLR$. 
	
	We use $\an(Z)$ to denote the ancestors of a variable $Z$, with $\an^\col(Z)$ and $\an^\aug(Z)$ denoting the ancestors in the collapsed and augmented models respectively.
	We use $\ch(Z)$ to denote children, with $\ch^\col(Z)$ and $\ch^\aug(Z)$ defined analogously.
	We use $\overline{\an}(Z)$ to denote the ancestors of $Z$ inclusive of $Z$, that is $\overline{\an}(Z) = \{Z\} \cup \an(Z)$, and likewise for $\overline{\ch}(Z)$.
	Finally, we use $Z^v$ to denote a generic endogenous variable in the collapsed or augmented model; it can be either a unit variable $X^v$ or a $Q$ variable. 
	
We can write the effect in the collapsed model as,
\begin{equation*}
\begin{split}
	\pr^\col (y \s \rmdo( &\qLp = \qsLR))\\ = \int \cdots \int & \prod_{v \in \overline{\an}^\col(Y) \setminus \overline{\ch}^\col(\QLp)} \pr^\col(z^v \mid z^{\pa^\col(v)})\\ & \times \prod_{v \in \ch^\col(\QLp)} \pr^\col(z^v \mid \qLp = \qsLR, z^{\pa^\col(v)}\setminus \qLp) \prod_{v \in \an^\col(Y) \setminus \QLp} \di z^v.
\end{split}
\end{equation*}

We will analyze the terms in the first and second product separately, equating them to terms in the augmented model.
Consider first the terms in the first product, which describe the mechanisms generating $Z^v$ for $v \in \overline{\an}^\col(Y) \setminus \overline{\ch}^\col(\QLp)$. 
We will argue that these mechanisms are the same in the augmented model, i.e. $\pr^\col(z^v \mid z^{\pa^\col(v)}) = \pr^\aug(z^v \mid z^{\pa^\aug(v)})$. 
Since the augmentation is valid (\Cref{def:valid_augment}), marginalizing out $\QLR$ from the augmented model must recover the collapsed model. So, the only way for a mechanism to differ in the augmented model is if $Z^v$ is a child of $\QLR$ in the augmented model.
However, in that case, marginalizing out $\QLR$ would make $Z^v$ a child of $\QLp$ in the collapsed model. This violates the condition that $v \in \overline{\an}^\col(Y) \setminus \overline{\ch}^\col(\QLp)$.

Next, consider the terms in the second product, which describe the mechanisms generating $Z^v$ for $v \in \ch^\col(\QLp)$. 
Since all paths from $\QLp$ to $Y$ go through $\QLR$ in the augmented model, to satisfy validity (\Cref{def:valid_augment}) each of these $Z^v$ must be a child of $\QLR$ in the augmented model, and not a child of $\QLp$.
Moreover, validity further implies that, 
\begin{equation*}
\begin{split}
	\pr^\col & (z^v \mid \qLp = \qsLR, z^{\pa^\col(v)}\setminus \qLp)\\ & = \pr^\aug(z^v \mid \qLR=\fLR(\qLp = \qsLR, z^{\pa^\aug(\QLR)} \setminus \qLp), z^{\pa^\aug(v)}\setminus \qLR),
\end{split}
\end{equation*}
where $\fLR$ is the mechanism generating the augmentation variable (\Cref{eqn:augment_mech}).
Examining \Cref{eqn:augment_mech}, we can see that if $\qLp = \qsLR$, then it must be the case $\qLR=\qsLR$, regardless of the value of the other parents of $\QLR$ in the augmented model.
Hence,
\begin{equation}
	\pr^\col(z^v \mid \qLp = \qsLR, z^{\pa^\col(v)}\setminus \qLp) = \pr^\aug(z^v \mid \qLR =\qsLR, z^{\pa^\aug(v)}\setminus \qLR).
\end{equation}

Now, again using the assumption that all paths from $\QLp$ to $Y$ go through $\QLR$, we can rewrite the effect in the collapsed model as,
\begin{equation*}
\begin{split}
	\pr^\col (y \s \rmdo( \qLp & = \qsLR))\\ = \int \cdots \int & \prod_{v \in \overline{\an}^\col(Y) \setminus \overline{\ch}^\col(\QLp)} \pr^\aug(z^v \mid z^{\pa^\aug(v)})\\ & \times \prod_{v \in \ch^\col(\QLp)} \pr^\aug(z^v \mid \qLR =\qsLR, z^{\pa^\aug(v)}\setminus \qLR) \prod_{v \in \an^\col(Y) \setminus \QLp} \di z^v\\
	=\int \cdots \int & \prod_{v \in \overline{\an}^\aug(Y) \setminus \overline{\ch}^\aug(\QLR)} \pr^\aug(z^v \mid z^{\pa^\aug(v)})\\ & \times \prod_{v \in \ch^\aug(\QLR)} \pr^\aug(z^v \mid \qLR =\qsLR, z^{\pa^\aug(v)}\setminus \qLR) \prod_{v \in \an^\aug(Y) \setminus \QLR} \di z^v\\
	=\pr^\aug(y \s \rmdo(&\qLR = \qsLR)).
\end{split}
\end{equation*}
\end{proof}
  
	 \section{Proofs of General Identification Conditions}

\subsection{Proof of \Cref{thm:sufficient_ID}} \label{proof:sufficient_ID}

\begin{proof}
\textbf{Condition 1} In the collapsed model, the only children of $Q^{a|\pa_\Sc(a)}$ are its direct unit descendants (\Cref{def:um}). By assumption, there is no bi-directed path to these variables. 
If $Y$ is subunit-level, we can augment with $Q^y$ to identify the effect of interest, but there will again be no bi-directed path from $Q^{a|\pa_\Sc(a)}$ to $Q^y$. 
So from \Cref{thm:bidirected}, the effect of $\rmdo(q^{a|\pa_\Sc(a)}=q_\star^a)$ on $Y$ is identified.

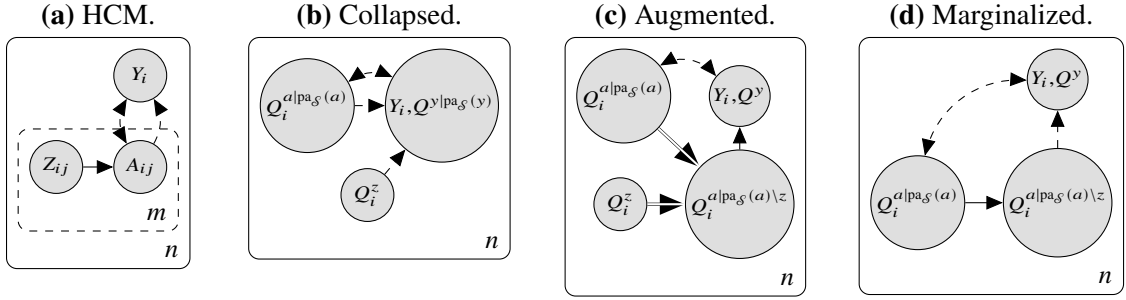
\begin{figure}[t]

\begin{subfigure}[t]{0.2\textwidth}
\caption{HCM.} \label{fig:gen_iv_hcm}
\centering
\begin{tikzpicture}

  \node[obs]                               (y) {$\scriptstyle Y_{i}$};
  \node[obs, below=0.5cm of y] (a) {$\scriptstyle A_{ij}$};
  \node[obs, left=0.4cm of a] (z) {$\scriptstyle Z_{ij}$};

  \edge {z} {a} ;
  
  \draw[<->, dashed, bend left] (a) to (y);
  \draw[->, dashed, bend right] (a) to (y);

  \plate[dashed] {in} {(a)(z)} {$m$} ;
  \plate {out} {(in)(y)} {$n$} ;

\end{tikzpicture}
\end{subfigure}
\begin{subfigure}[t]{0.24\textwidth}
\caption{Collapsed.} \label{fig:gen_iv_collapse}
\centering
\begin{tikzpicture}

  \node[obs]                               (y) {$\scriptstyle Y_{i},{\scriptstyle Q^{y|\pa_\Sc(y)}}$};
  \node[obs, left=0.4cm of y] (qaz) {$\scriptstyle Q^{a|\pa_\Sc(a)}_i$};
  \node[obs, below=.2cm of qaz, xshift=0.8cm] (qz) {$\scriptstyle Q^{z}_i$};

\draw[<->, dashed, bend left] (qaz) to (y);
\draw[->, dashed] (qaz) to (y);
\draw[->, dashed] (qz) to (y);

  \plate {out} {(qaz)(qz)(y)} {$n$} ;
\end{tikzpicture}
\end{subfigure}
\begin{subfigure}[t]{0.24\textwidth}
\caption{Augmented.} \label{fig:gen_iv_augment}	
\centering
\begin{tikzpicture}

  \node[obs]                               (y) {$\scriptstyle Y_{i},{\scriptstyle Q^{y}}$};
  \node[obs, left=.5cm of y] (qaz) {$\scriptstyle Q^{a|\pa_\Sc(a)}_i$};
  \node[obs, below=.3cm of y] (qa) {$\scriptstyle Q^{a|\pa_\Sc(a)\setminus z}_i$};
  \node[obs, left=.5cm of qa] (qz) {$\scriptstyle Q^{z}_i$};

  \draw[->, dashed] (qa) to (y);
  \draw[<->, dashed, bend left=40] (qaz) to (y);
  \path (qaz) edge [very thick, ->]  (qa) ;
  \path (qaz) edge [color=white, thick] (qa) ;
  \path (qz) edge [very thick, ->]  (qa) ;
  \path (qz) edge [color=white, thick] (qa) ;

  \plate {out} {(qz)(qaz)(y)(qa)} {$n$} ; 

\end{tikzpicture}
\end{subfigure}
\begin{subfigure}[t]{0.24\textwidth}
\caption{Marginalized.} \label{fig:gen_iv_marg}	
\centering
\begin{tikzpicture}

  \node[obs]                               (y) {$\scriptstyle Y_{i},{\scriptstyle Q^{y}}$};
  \node[obs, below=.5cm of y] (qa) {$\scriptstyle Q^{a|\pa_\Sc(a)\setminus z}_i$};
  \node[obs, left=.5cm of qa] (qaz) {$\scriptstyle Q^{a|\pa_\Sc(a)}_i$};

  \draw[->, dashed] (qa) to (y);
  \draw[<->, dashed, bend left=40] (qaz) to (y);
  \edge{qaz}{qa} ;

  \plate {out} {(qa)(qaz)(y)} {$n$} ; 

\end{tikzpicture}
\end{subfigure}
\caption{\textbf{Steps of the proof of \Cref{thm:sufficient_ID}, condition 2.} (a) The original HCM. Here $Y_i$ is shown as unit-level, but it also may be subunit-level. We use a double-sided dashed arrow to denote the possibility of subunit or unit-level parents of $A$, $Y$, or both (including possibly hidden unit confounders). We use a one-sided dashed arrow to denote the possibility of directed paths from $A$ to $Y$, which may be direct or run through subunit or unit-level variables. (b) Collapsed model. If $Y$ is unit-level in the HCGM, rightmost node is also $Y$, but if it is subunit-level in the original HCGM, the rightmost node is $Q^{y|\pa_\Sc(y)}$. (c) Augmented model. (d) Marginalized model.}
\end{figure} 
\textbf{Condition 2} The general setup is shown in \Cref{fig:gen_iv_hcm}, with $Z$ denoting the subunit instrument. In the collapsed model we have the variables $Q^{a | \pa_\Sc(a)}$ and $Q^z$ (\Cref{fig:gen_iv_hcm}).
Now, the children of $Q^{a | \pa_\Sc(a)}$ in the collapsed model are the direct unit descendants of $A$ in the original HCGM.
Since $A$ is the only child of $Z$ in the HCGM, the direct unit descendants of $A$ can only depend on $Q^{a | \pa_\Sc(a)}$ through terms of the form $\int Q^{a | \pa_\Sc(a)}(a \mid z, x^{\pa_\Sc(a)\setminus z}) q^z(z) \di z$ in the collapsed model (to see this, consider collapsing the original HCGM but with $Z$ marginalized out).
If $Y$ is subunit level, we can augment the collapsed model with $Q^y$. Again, since $A$ is the only child of $Z$, $Q^y$ must depend on $Q^{a | \pa_\Sc(a)}$ only through terms of the form $\int Q^{a | \pa_\Sc(a)}(a \mid z, x^{\pa_\Sc(a)\setminus z}) q^z(z) \di z$.

We can therefore augment the collapsed model with $Q^{a | \pa_\Sc(a)\setminus z}$ (\Cref{eqn:augment_mech}) such that $Q^{a | \pa_\Sc(a)\setminus z}$ is the only child of $Q^{a \mid \pa_\Sc(a)}$ (\Cref{fig:gen_iv_augment}).
Moreover, the only child of $Q^z$ will also be the augmentation variable $Q^{a | \pa_\Sc(a)\setminus z}$.
Hence, we marginalize out $Q^z$ such that the positivity assumption $\pr^\mar(q^{a | \pa_\Sc(a)\setminus z} = q_\star^a \mid q^{a|\pa_\Sc(a)})$ can be met (\Cref{assume:unit_positive}; \Cref{fig:gen_iv_marg}).
From \Cref{prop:intervene_valid} the effect of interest is equivalent to $\pr^\mar(y \s \rmdo(q^{a | \pa_\Sc(a)\setminus z} = q_\star^a))$ if $Y$ is unit-level, and to $\pr^\mar(q^y \s \rmdo(q^{a | \pa_\Sc(a)\setminus z} = q_\star^a))$ if $Y$ is subunit-level.

Since there is no bi-directed path in the original HCGM between $A$ and $Z$, there is no bi-directed path in the collapsed model between $Q^{a \mid \pa_\Sc(a)}$ and $Q^z$, and thus no bi-directed path in the marginalized model between $Q^{a \mid \pa_\Sc(a)}$ and its only child $Q^{a | \pa_\Sc(a)\setminus z}$.
So from \Cref{thm:bidirected}, the effect on $Y$ is identified.
\end{proof}

\subsection{Proof of \Cref{thm:unit-level-noID}} \label{proof:unit-level-noID}
\begin{proof}
If the effect is not identified in the erased inner plate model, there must be a bi-directed path between $A$ and at least one of its children (\Cref{thm:bidirected}).
Since the graph of the erased plate model is the same as that of the HCGM, there must also be a bi-directed path between $A$ and at least one of its children in the HCGM.

Now, consider a modified HCGM in which all the outgoing arrows from subunit-level variables are erased. 
This is a special case of the original HCGM, so if the effect of interest is not identified in this modified model, it cannot be identified in the original model.
There remains a bi-directed path between $A$ and at least one of its children, since by assumption there are no subunit-level confounders, and hence all bi-directed paths must go through unit-level confounders.

We now collapse the HCGM. From \Cref{def:um}, there must remain a bi-directed path between $A$ and a child in the collapsed model. Moreover, since there were no outgoing arrows from subunit variables, there are no constraints on the mechanisms in the collapsed model (i.e. it is fully nonparametric).
Hence, from \Cref{thm:bidirected} the effect is not identified in the collapsed model.
Since the collapsed model is equivalent to the original HCGM (\Cref{thm:valid_collapse}) the result follows.
\end{proof}
 
\section{Details on Eight Schools} \label{apx:schools_details}

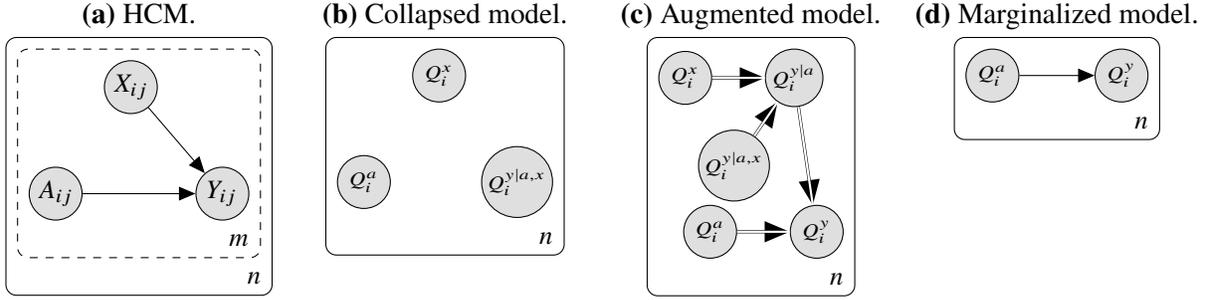
\begin{figure}[t]
\centering
\begin{subfigure}[t]{0.24\textwidth}
\caption{HCM.} \label{fig:school_initial_hcm_2}
\centering
\begin{tikzpicture}

  \node[obs]                               (y) {$Y_{ij}$};
  \node[obs, left=1.5cm of y] (a) {$A_{ij}$};
  \node[obs, above=.7cm of a, xshift=1cm] (x) {$X_{ij}$};

  \edge {a} {y} ;
  \edge {x} {y} ;

  \plate[dashed] {in} {(a)(y)(x)} {$m$} ;
  \plate {out} {(in)} {$n$} ;

\end{tikzpicture}
\end{subfigure}
\begin{subfigure}[t]{0.24\textwidth}
\caption{Collapsed model.} \label{fig:school_initial_collapse}
\centering
\begin{tikzpicture}

  \node[obs]                               (qyax) {$\scriptstyle Q^{y|a, x}_{i}$};
  \node[obs, left=1.2cm of qyax] (qa) {$\scriptstyle Q^{a}_{i}$};
  \node[obs, above=.7cm of qa, xshift=1cm] (qx) {$\scriptstyle Q^{x}_{i}$};

  \plate {out} {(qyax)(qa)(qx)} {$n$} ;

\end{tikzpicture}
\end{subfigure}
\begin{subfigure}[t]{0.24\textwidth}
\caption{Augmented model.} \label{fig:school_initial_augment}
\centering
\begin{tikzpicture}

  \node[obs] (qyax) {$\scriptstyle Q^{y| a, x}_{i}$};

  \node[obs, above=.3cm of qyax, xshift=.8cm] (qya) {$\scriptstyle Q^{y|a}_i$};
  \node[obs, below=1.3cm of qya, xshift=.3cm] (qy) {$\scriptstyle Q^{y}_i$};
  \node[obs, left=.7cm of qy] (qa) {$\scriptstyle Q^{a}_{i}$};
  \node[obs, left=.7cm of qya] (qx) {$\scriptstyle Q^{x}_{i}$};

  \path (qa) edge [very thick, ->]  (qy) ;
  \path (qa) edge [color=white, thick] (qy) ;
  \path (qya) edge [very thick, ->]  (qy) ;
  \path (qya) edge [color=white, thick] (qy) ;
  \path (qx) edge [very thick, ->]  (qya) ;
  \path (qx) edge [color=white, thick] (qya) ;
  \path (qyax) edge [very thick, ->]  (qya) ;
  \path (qyax) edge [color=white, thick] (qya) ;

  \plate {out} {(qyax)(qa)(qx)(qya)(qy)} {$n$} ;

\end{tikzpicture}
\end{subfigure}
\begin{subfigure}[t]{0.24\textwidth}
\caption{Marginalized model.} \label{fig:school_initial_marginalize}
\centering
\begin{tikzpicture}

  \node[obs] (qa) {$\scriptstyle Q^{a}_{i}$};
  \node[obs, right= of qa] (qy) {$\scriptstyle Q^{y}_{i}$};

  \edge {qa} {qy}

  \plate {out} {(qa)(qy)} {$n$} ;

\end{tikzpicture}
\end{subfigure}
\caption{\textbf{Initial model for eight schools data.}} \label{fig:school_initial}
\end{figure}

\begin{figure}[t]
\centering
\begin{subfigure}[t]{0.25\textwidth}
\caption{HCM.} \label{fig:school_confound_hcm_2}
\centering
\begin{tikzpicture}

  \node[obs]                               (y) {$Y_{ij}$};
  \node[obs, left=.8cm of y] (a) {$A_{ij}$};
  \node[obs, above=.7cm of a, xshift=1cm] (x) {$X_{ij}$};
  \node[latent, left=.4cm of a, yshift=1cm] (u) {$U_i$};

  \edge {u} {y} ;
  \edge {u} {x} ;
  \edge {u} {a} ;
  \edge {a} {y} ;
  \edge {x} {y} ;

  \plate[dashed] {in} {(a)(y)(x)} {$m$} ;
  \plate {out} {(in)(u)} {$n$} ;

\end{tikzpicture}
\end{subfigure}
\begin{subfigure}[t]{0.25\textwidth}
\caption{Collapsed model.} \label{fig:school_confound_collapse}
\centering
\begin{tikzpicture}

  \node[obs]                               (qyax) {$\scriptstyle Q^{y|a,x}_{i}$};
  \node[obs, left=.8cm of y] (qa) {$\scriptstyle Q^{a}_{i}$};
  \node[obs, above=.7cm of a, xshift=1cm] (qx) {$\scriptstyle Q^{x}_{i}$};
  \node[latent, left=.4cm of a, yshift=1cm] (u) {$U_i$};

  \edge {u} {qyax} ;
  \edge {u} {qx} ;
  \edge {u} {qa} ;

  \plate {out} {(qa)(qyax)(qx)(u)} {$n$} ;

\end{tikzpicture}
\end{subfigure}
\begin{subfigure}[t]{0.23\textwidth}
\caption{Augmented model.} \label{fig:school_confound_augment}
\centering
\begin{tikzpicture}

  \node[obs]                               (qyax) {$\scriptstyle Q^{y|a,x}_{i}$};
  \node[obs, left=.8cm of y] (qa) {$\scriptstyle Q^{a}_{i}$};
  \node[obs, above=.7cm of a, xshift=1cm] (qx) {$\scriptstyle Q^{x}_{i}$};
  \node[latent, above=.7cm of a, xshift=-.2cm] (u) {$U_i$};
  \node[obs, right=.5cm of qx] (qya) {$\scriptstyle Q^{y|a}_{i}$};
  \node[obs, below=.1cm of qyax, xshift=.8cm] (qy) {$\scriptstyle Q^{y}_{i}$};

  \edge {u} {qyax} ;
  \edge {u} {qx} ;
  \edge {u} {qa} ;
  \path (qx) edge [very thick, ->]  (qya) ;
  \path (qx) edge [color=white, thick] (qya) ;
  \path (qyax) edge [very thick, ->]  (qya) ;
  \path (qyax) edge [color=white, thick] (qya) ;
  \path (qa) edge [very thick, ->]  (qy) ;
  \path (qa) edge [color=white, thick] (qy) ;
  \path (qya) edge [very thick, ->]  (qy) ;
  \path (qya) edge [color=white, thick] (qy) ;

  \plate {out} {(qa)(qyax)(qx)(u)(qya)(qy)} {$n$} ;

\end{tikzpicture}
\end{subfigure}
\begin{subfigure}[t]{0.24\textwidth}
\caption{Marginalized model.} \label{fig:school_confound_marginalize}
\centering
\begin{tikzpicture}

  \node[obs]                               (qya) {$\scriptstyle Q^{y|a}_{i}$};
  \node[obs, left=1cm of qyax] (qa) {$\scriptstyle Q^{a}_{i}$};
  \node[obs, below=.7cm of qa, xshift=1cm] (qy) {$\scriptstyle Q^{y}_{i}$};
  \node[latent, above=1.7cm of qy] (u) {$U_{i}$};

  \edge {u} {qa, qya} ;
  \path (qa) edge [very thick, ->]  (qy) ;
  \path (qa) edge [color=white, thick] (qy) ;
  \path (qya) edge [very thick, ->]  (qy) ;
  \path (qya) edge [color=white, thick] (qy) ;

  \plate {out} {(qyax)(qa)(qy)(u)} {$n$} ;

\end{tikzpicture}
\end{subfigure}
\caption{\textbf{Model for eight schools data with unobserved confounding.}} \label{fig:school_confound}
\end{figure}
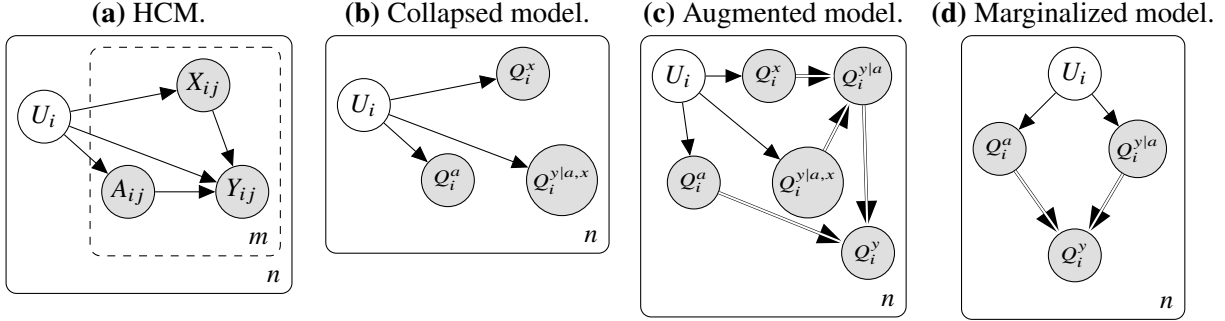

\begin{figure}[t]
\centering
\begin{subfigure}[t]{0.24\textwidth}
\caption{HCM.} \label{fig:school_interfere_hcm}
\centering
\begin{tikzpicture}

  \node[obs]                               (y) {$Y_{ij}$};
  \node[obs, left=.8cm of y] (a) {$A_{ij}$};
  \node[obs, above=.7cm of a, xshift=1cm] (x) {$X_{ij}$};
  \node[latent, left=.3cm of a, yshift=1cm] (u) {$U_i$};
  \node[obs, below=1.6cm of u] (s) {$S_{i}$};
  \node[obs, right=1cm of s] (c) {$C_{i}$};

  \edge {u} {y} ;
  \edge {u} {x} ;
  \edge {u} {a} ;
  \edge {a} {y} ;
  \edge {x} {y} ;
  \edge {u} {s} ;
  \edge {s} {c} ;
  \edge {a} {c} ;
  \edge {c} {y} ;

  \plate[dashed] {in} {(a)(y)(x)} {$m$} ;
  \plate {out} {(in)(u)(s)(c)} {$n$} ;

\end{tikzpicture}
\end{subfigure}
\begin{subfigure}[t]{0.24\textwidth}
\caption{Collapsed model.} \label{fig:school_interfere_collapse_2}
\centering
\begin{tikzpicture}

  \node[obs]                               (qyax) {$\scriptstyle Q^{y|a,x}_{i}$};
  \node[obs, left=.8cm of y] (qa) {$\scriptstyle Q^{a}_{i}$};
  \node[obs, above=.7cm of a, xshift=1cm] (qx) {$\scriptstyle Q^{x}_{i}$};
  \node[latent, left=.3cm of a, yshift=1cm] (u) {$U_i$};
  \node[obs, below=1.6cm of u] (s) {$S_{i}$};
  \node[obs, right=1cm of s] (c) {$C_{i}$};

  \edge {u} {qyax} ;
  \edge {u} {qx} ;
  \edge {u} {qa} ;
  \edge {u} {s} ;
  \edge {s} {c} ;
  \edge {qa} {c} ;
  \edge {c} {qyax} ;

  \plate {out} {(qa)(qyax)(qx)(u)(s)(c)} {$n$} ;

\end{tikzpicture}
\end{subfigure}
\begin{subfigure}[t]{0.24\textwidth}
\caption{Augmented model.} \label{fig:school_interfere_augment}
\centering
\begin{tikzpicture}

  \node[obs]                               (qyax) {$\scriptstyle Q^{y|a,x}_{i}$};
  \node[obs, left=.2cm of qyax] (qa) {$\scriptstyle Q^{a}_{i}$};
  \node[obs, above=.7cm of a, xshift=1cm] (qx) {$\scriptstyle Q^{x}_{i}$};
  \node[latent, above=.7cm of a, xshift=-.3cm] (u) {$U_i$};
  \node[obs, right=.5cm of qx] (qya) {$\scriptstyle Q^{y|a}_{i}$};
  \node[obs, below=.5cm of qyax, xshift=1cm] (qy) {$\scriptstyle Q^{y}_{i}$};
  \node[obs, below=2cm of u] (s) {$S_{i}$};
  \node[obs, right=.6cm of s] (c) {$C_{i}$};

  \edge {u} {qyax} ;
  \edge {u} {qx} ;
  \edge {u} {qa} ;
  \path (qx) edge [very thick, ->]  (qya) ;
  \path (qx) edge [color=white, thick] (qya) ;
  \path (qyax) edge [very thick, ->]  (qya) ;
  \path (qyax) edge [color=white, thick] (qya) ;
  \path (qa) edge [very thick, ->]  (qy) ;
  \path (qa) edge [color=white, thick] (qy) ;
  \path (qya) edge [very thick, ->]  (qy) ;
  \path (qya) edge [color=white, thick] (qy) ;
  \edge {u} {s} ;
  \edge {s} {c} ;
  \edge {qa} {c} ;
  \edge {c} {qyax} ;

  \plate {out} {(qa)(qyax)(qx)(u)(qya)(qy)} {$n$} ;

\end{tikzpicture}
\end{subfigure}
\begin{subfigure}[t]{0.24\textwidth}
\caption{Marginalized model.} \label{fig:school_interfere_marginalize}
\centering
\begin{tikzpicture}

  \node[obs]                               (qya) {$\scriptstyle Q^{y|a}_{i}$};
  \node[obs, left=1.4cm of qyax] (qa) {$\scriptstyle Q^{a}_{i}$};
  \node[obs, below=.1cm of qa, xshift=1.4cm] (qy) {$\scriptstyle Q^{y}_{i}$};
  \node[latent, above=2.2cm of qy, xshift=-.5cm] (u) {$U_{i}$};
  \node[obs, left=.3cm of qyax] (c) {$C_i$};
  \node[obs, above=.3cm of c, xshift=-.2cm] (s) {$S_i$};

  \edge {u} {qa, qya} ;
  \path (qa) edge [very thick, ->]  (qy) ;
  \path (qa) edge [color=white, thick] (qy) ;
  \path (qya) edge [very thick, ->]  (qy) ;
  \path (qya) edge [color=white, thick] (qy) ;
  \edge {u} {s} ;
  \edge {s} {c} ;
  \edge {qa} {c} ;
  \edge {c} {qya} ;

  \plate {out} {(qyax)(qa)(qy)(u)} {$n$} ;

\end{tikzpicture}
\end{subfigure}
\caption{\textbf{Model for eight schools data with unobserved confounding and interference.}} \label{fig:school_interfere}
\end{figure}
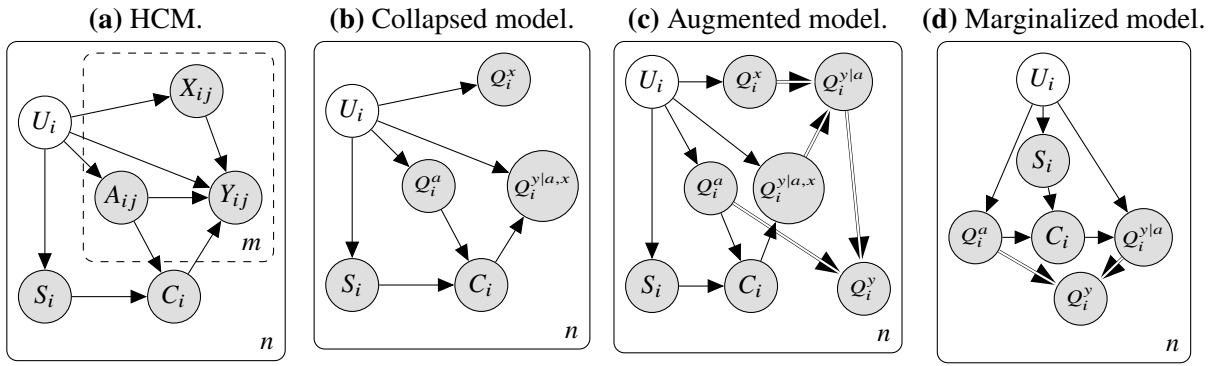

\begin{figure}[t]
\centering
\begin{subfigure}[t]{0.24\textwidth}
        \caption{} \label{fig:class_v_effect}
        \includegraphics[width=\textwidth]{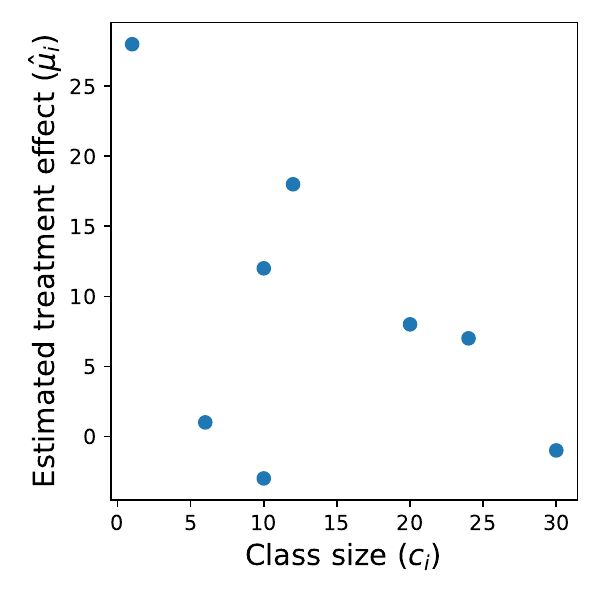}
\end{subfigure}
\begin{subfigure}[t]{0.24\textwidth}
        \caption{}
        \includegraphics[width=\textwidth]{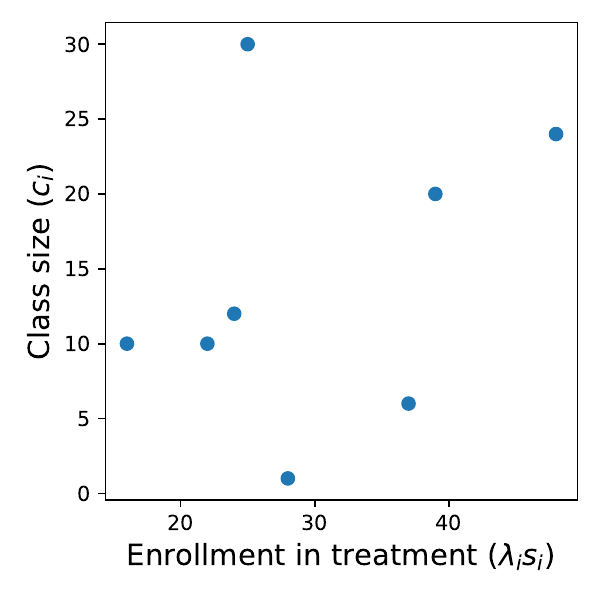}
\end{subfigure}
\begin{subfigure}[t]{0.24\textwidth}
        \caption{}
        \includegraphics[width=\textwidth]{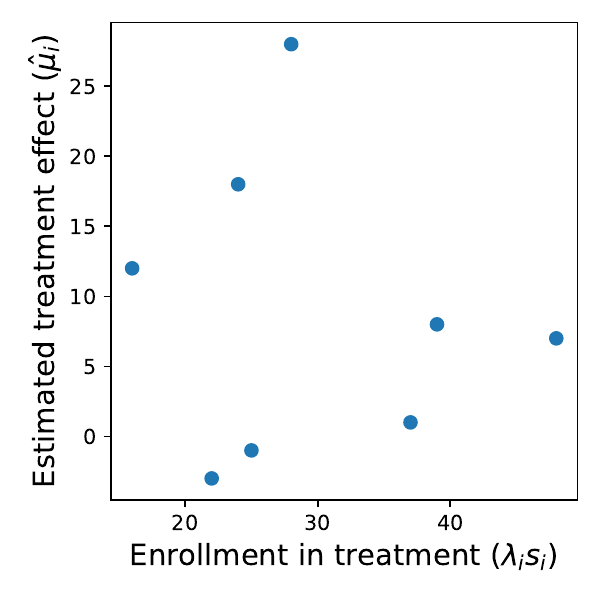}
\end{subfigure}
\begin{subfigure}[t]{0.24\textwidth}
\caption{} \label{fig:frac_treated_v_effect}
        \includegraphics[width=\textwidth]{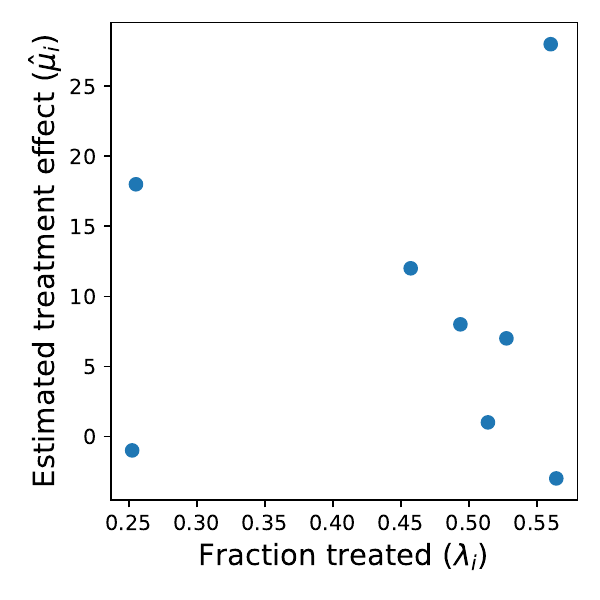}
\end{subfigure}
\caption{\textbf{Scatter plots of eight schools data.} Each point represents a school.} \label{fig:eight_school_scatter}
\end{figure}

\subsection{Fully observed and confounder models} \label{apx:eight_schools_initial}

In this section we provide further details on the estimation methods used for the initial analysis of the eight schools data, based on the fully observed (\Cref{sec:schools_initial}) and hidden unit confounder (\Cref{sec:schools_confounding}) models.

The eight schools study does not make public its data at the student level.
However, for each school $i$, the authors ran a linear regression predicting $Y_{ij}$ from $A_{ij}$ and $X_{ij}$, and reported the estimated coefficient on the treatment, $\hat{\mu}_i$.
We can understand this per-school treatment effect $\hat{\mu}_i$ as a parametric estimate of $\mu_i = \mathbb{E}_{q^{y|a}_i}[Y \mid A = 1] - \mathbb{E}_{q^{y|a}_i}[Y \mid A = 0]$.
Since there are only a finite number of students $m_i$ per school, the estimate $\hat{\mu}_i$ comes with some uncertainty; \citep{Alderman1979-on} report the standard error, $\sigma_i$.
We can model $\hat{\mu}_i$ as a sample from $\mathrm{Normal}(\mu_i, \sigma_i)$.
To obtain the average treatment effect, we also need to estimate $\pr(q^{y|a})$, or, at minimum, its marginal $\pr(\mu)$.
A simple parametric approach is to assume that $\pr(\mu)$ takes the form of a normal distribution, with unknown mean $\nu$ and standard deviation $\tau$. The mean of this distribution is then the treatment effect we are interested in,
$\nu = \mathbb{E}_\pr[\mu] =  \mathbb{E}_\pr[\mathbb{E}_{Q^{y|a}}[Y \mid A = 1] - \mathbb{E}_{Q^{y|a}}[Y \mid A = 0]] = \textsc{ate}.$
Placing a diffuse prior on $\nu$ and $\tau$, we obtain a hierarchical Bayesian model,
\begin{equation*}
\begin{split}
\nu &\sim \mathrm{Normal}(0, 5) \quad \quad \quad \tau \sim \mathrm{HalfCauchy}(5)\\
\mu_i &\sim \mathrm{Normal}(\nu, \tau) \quad \quad \quad 
\hat{\mu_i} \sim \mathrm{Normal}(\mu_i, \sigma_i),
\end{split}
\end{equation*}
where HalfCauchy is the half Cauchy distribution with support on only positive values.
We compute the posterior over $\nu$, the estimate of the ATE, using MCMC (as described in \Cref{sec:schools_initial}).
We have thus recovered, from a hierarchical causal model, the classic eight schools hierarchical Bayesian analysis.

\subsection{Confounding \& interference} \label{apx:eight_schools_interfere}

In this section we provide further details on the model used for the analysis of the eight schools data that accounts for confounding and interference (\Cref{sec:schools_interfere}).

In constructing the model, we treat the number of subunits as a separate unit variable, following the strategy described in \Cref{sec:approx_diverge}. 
In particular, the number of students interested in the tutoring program, $S_i$, corresponds to the sum of the number of students in the treatment and control groups for school $i$, that is $m_i$. (Note also that here we can have different numbers of subunits per unit.)

In our model, we assume that class size only impacts the average test scores of the treated students, such that $\mathbb{E}_\pr[\mathbb{E}_Q[Y \mid A=0] \s \rmdo(q^{a} = \delta_1)] = \mathbb{E}_\pr[\mathbb{E}_Q[Y \mid A=0] \s \rmdo(q^{a} = \delta_0)]$.
Then, using \Cref{eqn:schools_interfere_id},
\begin{equation} \label{eqn:eight_schools_interfere_simplified_ate}
\begin{split}
	\textsc{ate} &= \mathbb{E}_\pr[\mathbb{E}_{Q^{y|a}}[Y \mid A = 1] - \mathbb{E}_{Q^{y|a}}[Y \mid A = 0] \s \rmdo(q^{a} = \delta_1)]
	 = \mathbb{E}_\pr[\mu \s \rmdo(q^{a} = \delta_1)]\\
	& = \int \int \pr(c \mid q^{a} = \delta_1, s) \pr(s) \di s \int \mathbb{E}_\pr\big[\mu \mid {q}^{a}, \tilde{s}, c\big] \pr({q}^{a}, \tilde{s}) \di {q}^{a}\, \di\tilde{s}\, \di c.
\end{split}
\end{equation}
So, this assumption allows us to make use of the summary results reported by \citet{Alderman1979-on}: we do not need information about the actual value of the test scores of the treated and untreated, only about the difference between treated and untreated.

We treat $q^a_i$ as observed, and equal to the empirical mean of $\{a_{ij}\}_{j=1}^m$, i.e. $q^a_i = \mathrm{Bernoulli}(\lambda_i = \frac{1}{m_i} a_{ij})$.
This modeling choice reflects the fact that in this data, the real population of students is finite and fully observed, rather than a subsample of a larger population.
(Note that the choice to treat $q^a_i$ as observed rather than latent does not affect the model in the large $m$ limit, where the identification formula applies.)
We also assume $C_i$ and $\mu_i$ depend on $S_i$ and $q^{a}_i$ only through the product $S_i \lambda_i$, the total number of students who are treated at school $i$.

With these assumptions in place, we consider the following hierarchical Bayesian model, which describes the joint distribution over $\lambda, C, S$ and $\mu$. On the left hand side we annotate each part of the parametric model with the term of the causal model it is describing. We use the parameterization of the Beta distribution in terms of its mean and precision~\citep{Ferrari2004-ag}.
\begin{equation*}
\begin{split}
        \kappa_{A} &\sim \mathrm{Normal}(0, 10) \text{ and } \zeta_{A} \sim \mathrm{Normal}(0, 10)\\
        Q^a_i \sim \pr(q^a)\,\quad\quad\quad \quad \lambda_i &\sim \mathrm{Beta}(mean = \sigma(\kappa_A), precision = \log(1 + \exp(\zeta_A)))\\
        \kappa_S &\sim \mathrm{Normal}(50, 100) \text{ and } \zeta_S \sim \mathrm{HalfCauchy}(100)\\
        S_i \sim \pr(s)\,\,\,\,\quad\quad\quad \quad S_i &\sim \mathrm{Normal}(\kappa_S, \zeta_S)\\
        \alpha_C &\sim \mathrm{Normal}(0, 50) \text{ and } \beta_C \sim \mathrm{Normal}(10, 100) \text{ and } \zeta_C \sim \mathrm{HalfCauchy}(10)\\
        C_i \sim \pr(c \mid q^a, s)\quad \quad C_i &\sim \mathrm{Normal}( \alpha_C s_i \lambda_i\,+\beta_C, \zeta_C)\\
        \alpha_Y &\sim \mathrm{Normal}(0, 100), \beta_Y \sim \mathrm{Normal}(0, 100), \omega_Y \sim \mathrm{Normal}(0, 500), \tau \sim \mathrm{HalfCauchy}(5)\\
        \mu_i \sim \pr(\mu \mid q^a, s, c) \quad \mu_i &\sim \mathrm{Normal}(\alpha_Y c_i + \beta_Y s_i \lambda_i + \omega_Y, \tau)\\
        \hat{\mu}_i &\sim \mathrm{Normal}(\mu_i, \sigma_i).
\end{split}
\end{equation*}

We draw samples from the posterior using the NUTS sampler in NumPyro~\citep{Hoffman2014-ic,Phan2019-om,Bingham2019-aa}.
We use these samples to form a Monte Carlo approximation of \Cref{eqn:eight_schools_interfere_simplified_ate}.
Note that since we have limited data, we draw samples from the model's estimate of $\pr(s)$ and $\pr(q^a, s)$, rather than use the empirical distribution.
When computing the Monte Carlo approximation of the treatment effect, we clip posterior samples of $s$ and $c$ that are physically impossible, setting negative or zero values of $s$ and $c$ to 1, and setting values of $c$ greater than $s$ to $s$. 
\section{Framing Previous Models as HCMs} \label{apx:previous}

\subsection{Fixed-effects} \label{apx:fixed_effect}

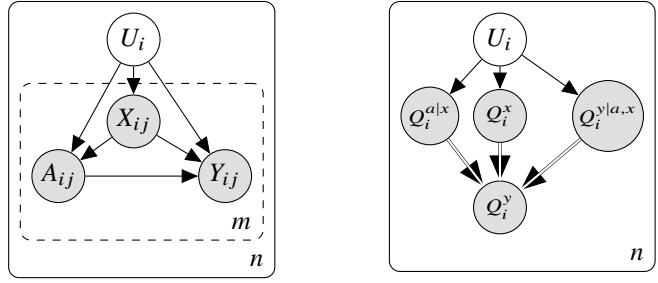
\begin{figure}[t!]
\centering
\begin{subfigure}[t]{0.3\textwidth}
\caption{Hierarchical causal model.} \label{fig:fixed_effect_hcm}
\centering
\begin{tikzpicture}

  \node[obs]                               (y) {$Y_{ij}$};
  \node[obs, left=1.5cm of y] (a) {$A_{ij}$};
  \node[obs, above=.01cm of a, xshift=1cm] (x) {$X_{ij}$};
  \node[latent, above=1.1cm of a, xshift=1cm]  (u) {$U_i$};

  \edge {a,u} {y} ;
  \edge {u} {a} ;
  \edge {u} {x} ;
  \edge {x} {a} ;
  \edge {x} {y} ;

  \plate[dashed] {in} {(a)(y)(x)} {$m$} ;
  \plate {out} {(in)(u)(x)} {$n$} ;

\end{tikzpicture}
\end{subfigure}
\begin{subfigure}[t]{0.3\textwidth}
\caption{Collapsed and augmented.} \label{fig:fixed_effect_augment}
\centering
\begin{tikzpicture}

\node[obs]                               (qyax) {$\scriptstyle Q_i^{y \mid a, x}$};
  \node[obs, left=1.5cm of qyax] (qax) {$\scriptstyle Q_{i}^{a|x}$};
  \node[obs, left=0.6cm of qyax] (qx) {$\scriptstyle Q_{i}^{x}$};
  \node[latent, above=.3cm of qx]  (u) {$U_i$};
  \node[obs, below=.5cm of qx]  (qy) {$\scriptstyle Q_i^{y}$};

\edge {u} {qyax,qax,qx} ;
  \path (qax) edge [very thick, ->]  (qy) ;
  \path (qax) edge [color=white, thick] (qy) ;
  \path (qx) edge [very thick, ->]  (qy) ;
  \path (qx) edge [color=white, thick] (qy) ;
  \path (qyax) edge [very thick, ->]  (qy) ;
  \path (qyax) edge [color=white, thick] (qy) ;

\plate {ayu} {(qax)(qyax)(u)(qy)} {$n$} ;

\end{tikzpicture}
\end{subfigure}
\caption{\textbf{Hierarchical causal model for fixed-effect models and related methods.}} \label{fig:fixed_effect}
\end{figure}

In this section we detail how fixed-effects models can be seen as examples of HCMs.
Fixed-effects models are a staple of econometrics and related fields.
They are perhaps most often applied to panel data, in which observations are made of the same set of people at different timepoints. 
In this context, we can think of each person as a unit, and each timepoint as a subunit.

We can understand fixed-effect models in terms of the HCM in \Cref{fig:fixed_effect}.
The idea of the method is to correct for unobserved confounders at the unit level, $U_i$.
In the context of econometric panel data this unobserved confounder could represent, for example, the latent ability of each individual.
We observe covariates $X_{ij}$, treatment status $A_{ij}$ and outcome $Y_{ij}$ for each unit.

Standard fixed-effects models posit a parametric, linear causal mechanism.
In our framework, this corresponds to a hierarchical structural causal model with,
\begin{equation}
	y_{ij} = \f^y(u_i, \gamma^y_i, a_{ij}, x_{ij}, \epsilon_{ij}^y) = \alpha a_{ij} + \beta^\top x_{ij} + \delta^\top u_i + \epsilon_{ij}^y.
\end{equation}
(Note that although the unit noise $\gamma^y_i$ does not appear in this expression, it can be absorbed into $u_i$ without loss of generality, since $u_i$ is latent.)  
The noise distribution $\pr(\epsilon^y)$ is assumed to have mean zero; we will take it to be Gaussian with standard deviation $\sigma$ for simplicity. 
In the HCGM we have, 
\begin{equation} \label{eqn:fixed_effects_hcgm}
\begin{split}
	Y_{ij} \sim q^{y \mid a, x}_i(y \mid a_{ij}, x_{ij}) = \mathrm{Normal}(\alpha a_{ij} + \beta^\top x_{ij} + z_i, \sigma),
\end{split}
\end{equation}
where $z_i \triangleq \gamma^\top u_i$ is a latent scalar, per-unit offset. 
Estimation methods for fixed-effect models proceed based on \Cref{eqn:fixed_effects_hcgm}, fitting $\alpha$, $\beta$ and $\{z_i\}_{i=1}^n$ to the dataset and thus inferring $\{q^{y \mid a, x}_i\}_{i=1}^n$.

The primary goal of inference in fixed-effect models is to learn $\alpha$, the coefficient on the treatment. In the HCM framework, we can understand $\alpha$ as the average treatment effect on $Y$ of a hard intervention on $A$. In particular, we can compute, using the collapsed and augmented model (\Cref{fig:fixed_effect_augment}),
\begin{equation}
\begin{split}
	\mathbb{E}_\pr[\mathbb{E}_{Q}[Y] \s &\rmdo(q^a = \delta_{a_\star})] = \int \int \mathbb{E}_{q^{y \mid a, x}}[Y \mid A = a_\star, x] q^x(x) \di x\, \pr(q^x, q^{y \mid a,x}) \di q^x \di q^{y \mid a,x}\\
	 & = \int [\alpha a_\star + \beta^\top\mathbb{E}_{q^x}[X] + z] \pr(q^x, z) \di q^x \di z
	= \alpha a_\star + \beta^\top\mathbb{E}_\pr[\mathbb{E}_{Q^x}[X]] + \mathbb{E}_\pr[Z],
\end{split}
\end{equation}
where in the second line we have reparameterized the integral, writing it in terms of the parameter $z$ that determines $q^{y \mid a, x}$ rather than $q^{y \mid a, x}$ itself.
Now we have the average treatment effect,
$\mathbb{E}_\pr[\mathbb{E}_{Q^y}[Y] \s \rmdo(q^a = \delta_1)] -  \mathbb{E}_\pr[\mathbb{E}_{Q^y}[Y] \s \rmdo(q^a = \delta_0)] = \alpha$.                                                                                                                         
In summary, we can understand fixed-effects models as examples of the HCM in \Cref{fig:fixed_effect}, parameterized with a linear model for $Y$. 

There are many extensions and variations of this basic fixed-effect model that also fall within the HCM framework, by similar arguments. 
The coefficients $\alpha$ and $\beta$ may be allowed to vary across units, in which case the average treatment effect in the HCM coincides with the standard target of estimation, $\mathbb{E}_\pr[\alpha]$~\citep{Wooldridge2005-bk}.
Time may be included as a covariate in $X$ to allow for time trends in the model.
Difference-in-difference estimators and synthetic control models can also be derived from the same HCM, by a similar logic~\citep{Abadie2010-nn}.

\subsection{Interference} \label{apx:interference}

\begin{figure}[t!]
\centering
\begin{subfigure}[t]{0.2\textwidth}
\caption{HCM.} \label{fig:interfere_unobserve_hcm}
\centering
\begin{tikzpicture}

  \node[obs]                               (y) {$Y_{ij}$};
  \node[obs, left=0.8cm of y] (a) {$A_{ij}$};
  \node[obs, above=.4cm of a, xshift=0.7cm] (x) {$X_{ij}$};
  \node[latent, below=1.8cm of x]  (z) {$Z_i$};
  
  \edge {a,x} {y} ;
  \edge {x} {a} ;
  \edge {a} {z} ;
  \edge {z} {y} ;

  \plate[dashed] {in} {(a)(y)(x)} {$m$} ;
  \plate {out} {(in)(z)(x)} {$n$} ;

\end{tikzpicture}
\end{subfigure}
\begin{subfigure}[t]{0.2\textwidth}
\caption{Collapsed model.} \label{fig:interfere_unobserve_collapse}
\centering
\begin{tikzpicture}

  \node[obs]                               (qyax) {$\scriptstyle Q^{y|a,x}_{i}$};
  \node[obs, left=0.8cm of y] (qax) {$\scriptstyle Q^{a|x}_{i}$};
  \node[obs, above=.4cm of a, xshift=0.7cm] (qx) {$\scriptstyle Q^x_{i}$};
  \node[latent, below=1.8cm of x]  (z) {$Z_i$};
  
  \edge {qax,qx} {z} ;
  \edge {z} {qyax} ;

  \plate {out} {(z)(qx)(qax)(qyax)} {$n$} ;

\end{tikzpicture}
\end{subfigure}
\begin{subfigure}[t]{0.27\textwidth}
\caption{Augmented model.} \label{fig:interfere_unobserve_augment}
\centering
\begin{tikzpicture}

  \node[obs]                               (qyax) {$\scriptstyle Q^{y|a,x}_{i}$};
  \node[obs, left=2cm of y] (qax) {$\scriptstyle Q^{a|x}_{i}$};
  \node[obs, right=0.6cm of qax] (qa) {$\scriptstyle Q^a_{i}$};
  \node[obs, above=0.6cm of qa] (qx) {$\scriptstyle Q^x_{i}$};
  \node[latent, below=1.3cm of x]  (z) {$Z_i$};
  \node[obs, above=.5cm of qyax]  (qy) {$\scriptstyle Q^y_{i}$};
  
  \edge {qa} {z} ;
  \edge {z} {qyax} ;
  \path (qyax) edge [very thick, ->]  (qy) ;
  \path (qyax) edge [color=white, thick] (qy) ;
  \path (qax) edge [very thick, ->]  (qy) ;
  \path (qax) edge [color=white, thick] (qy) ;
  \path (qx) edge [very thick, ->]  (qy) ;
  \path (qx) edge [color=white, thick] (qy) ;
  \path (qx) edge [very thick, ->]  (qa) ;
  \path (qx) edge [color=white, thick] (qa) ;
  \path (qax) edge [very thick, ->]  (qa) ;
  \path (qax) edge [color=white, thick] (qa) ;

  \plate {out} {(z)(qx)(qax)(qyax)(qy)} {$n$} ;

\end{tikzpicture}
\end{subfigure}
\begin{subfigure}[t]{0.27\textwidth}
\caption{Marginalized model.} \label{fig:interfere_unobserve_marginalize}
\centering
\begin{tikzpicture}

  \node[obs]                               (qyax) {$\scriptstyle Q^{y|a,x}_{i}$};
  \node[obs, left=2cm of y] (qax) {$\scriptstyle Q^{a|x}_{i}$};
  \node[obs, right=0.6cm of qax] (qa) {$\scriptstyle Q^a_{i}$};
  \node[obs, above=0.6cm of qa] (qx) {$\scriptstyle Q^x_{i}$};
  \node[obs, above=.5cm of qyax]  (qy) {$\scriptstyle Q^y_{i}$};
  
  \edge {qa} {qyax} ;
  \path (qyax) edge [very thick, ->]  (qy) ;
  \path (qyax) edge [color=white, thick] (qy) ;
  \path (qax) edge [very thick, ->]  (qy) ;
  \path (qax) edge [color=white, thick] (qy) ;
  \path (qx) edge [very thick, ->]  (qy) ;
  \path (qx) edge [color=white, thick] (qy) ;
  \path (qx) edge [very thick, ->]  (qa) ;
  \path (qx) edge [color=white, thick] (qa) ;
  \path (qax) edge [very thick, ->]  (qa) ;
  \path (qax) edge [color=white, thick] (qa) ;

  \plate {out} {(qx)(qax)(qyax)(qy)} {$n$} ;

\end{tikzpicture}
\end{subfigure}

\caption{\textbf{Unobserved interference model.}} \label{fig:interfere_unobserve}
\end{figure}
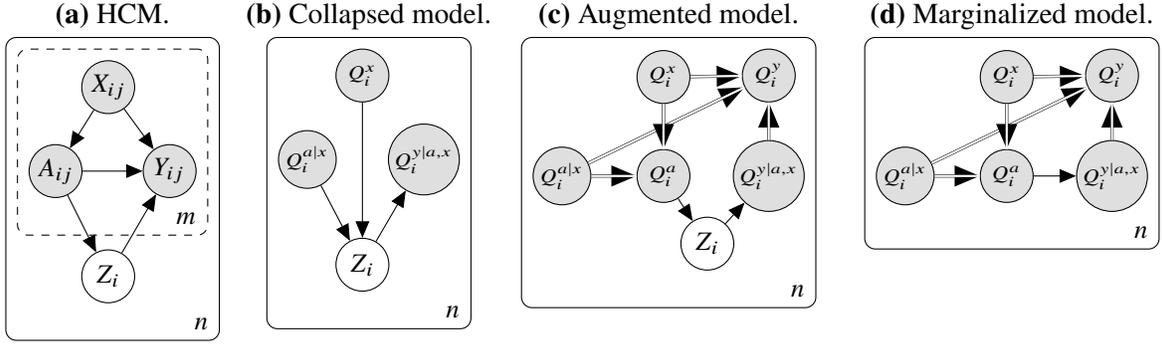 
In this section we describe in more detail how existing models of interference can be understood in terms of HCMs.
Consider the model in \Cref{fig:interfere_unobserve}, which has an unobserved interferer $Z_i$. 
From the collapsed, augmented and marginalized model, we can identify the effect of a soft intervention on the treatment $A$ with a backdoor correction,
\begin{equation} \label{eqn:unobserve_interfere}
\begin{split}
	\mathbb{E}_\pr[&\mathbb{E}_Q[Y] \s \rmdo(A \sim q_\star^{a})] \\
	=&\, \mathbb{E}_\pr[\mathbb{E}_Q[Y] \s \rmdo(q^{a|x} = q_\star^{a})] = \mathbb{E}_\pr[\mathbb{E}_{Q^x}[\mathbb{E}_{Q^{a|x}}[\mathbb{E}_{Q^{y|a,x}}[Y|A,X]]]\s \rmdo(q^{a|x} = q_\star^{a})]\\
	=&\, \int\int\int \int \int y q^{y|a,x}(y\mid a, x) q^x(x) q^a_\star(a) \di a \di x \di y\,  \pr(q^{y|a,x} \mid q^{a|x} = q_\star^{a}, q^x) \pr(q^x) \di q^{y|a,x} \di q^x\\
	=&\, \int\int\int \int \int y q^{y|a,x}(y\mid a, x) q^x(x) q^a_\star(a) \di a \di x \di y\,  \pr(q^{y|a,x} \mid q^{a} = q_\star^{a}) \pr(q^x) \di q^{y|a,x} \di q^x.
\end{split}
\end{equation}
Note we identify the effect by equating it to a hard intervention on $Q^{a|x}$, rather than on $Q^a$, as \Cref{prop:intervene_valid} does not apply in \Cref{fig:interfere_unobserve_marginalize}.
To obtain the final equality above, we use the fact that $Q^{y|a,x}$ must depend on $Q^x$ and $Q^{a|x}$ only through their marginal $Q^a$.
A crucial consequence of unobserved interference is the appearance of the term $\pr(q^{y|a,x} \mid q_\star^{a})$ in the identification formula, which implies that we must predict $q^{y|a,x}$ from the distribution of the treatment, $q^a$.

From this HCM identification formula, we can derive, as special cases, some standard techniques for correcting for interference.
Assume the treatment is binary, so that $q^a$ is Bernoulli, and $q^a$ is characterized entirely by its mean $\mu^a$. Then, consider a linear HSCM mechanism for $Y$,
\begin{equation} \label{eqn:unobserve_interfere_linear}
\begin{split}
	\gamma^y_i \sim \mathrm{Normal}(0, \sigma), \quad\quad \epsilon^y_{ij} \sim \mathrm{Normal}(0, \tau), \quad\quad  y_{ij} = \alpha a_{ij} + \beta^\top_i x_{ij} + \kappa \mu^a_i + \gamma^y_i + \epsilon_{ij}^y. 
\end{split}
\end{equation}
With this parameterization, the mean treatment within each unit $\mu^a$ becomes essentially just another covariate, and one can estimate the coefficients $\alpha$, $\beta$ and $\kappa$ via a fixed-effects regression model.
Now, in the HCGM, $\pr(q^{y|a,x} \mid q^a = q_\star^{a})$ is given by,
\begin{equation} \label{eqn:unobserv_interfere_hcgm}
\begin{split}
\gamma^y &\sim \mathrm{Normal}(0, \sigma)\\
q^{y|a, x}(\cdot \mid a, x) &= \mathrm{Normal}(\alpha a + \beta^\top x + \kappa \mu^a_\star + \gamma^y, \tau).
\end{split}
\end{equation}
Plugging \Cref{eqn:unobserv_interfere_hcgm} into \Cref{eqn:unobserve_interfere} we can derive the effect,
\begin{equation}
\begin{split}
	\mathbb{E}_\pr[\mathbb{E}_Q[Y] \s \rmdo(A \sim \mathrm{Bernoulli}(\mu^a_\star))] - \mathbb{E}_\pr[\mathbb{E}_Q[Y] \s \rmdo(A \sim \mathrm{Bernoulli}(\mu^a_{\star\star}))] = (\alpha + \kappa)(\mu^a_\star - \mu^a_{\star\star}).
\end{split}
\end{equation}
We can understand this effect as a combination of the direct effect of $A$, which has coefficient $\alpha$, and an interference effect, which has coefficient $\kappa$.

So, one way to account for interference is to include as a covariate, in a model for $Y$, an estimate of the mean $\mu^a$ of the treatment distribution within each unit. This approach is widely used in practice in studies of clustered interference, peer effects, etc.~\citep{Duflo2011-ru,Angrist2014-ms}.
Here we saw $\mu^a$ enter as a covariate in a linear model, but it can also enter into nonlinear models~\citep{Lee2022-jd}.

One subtlety is that our identifying equations are based on large $m$ asymptotics, and are agnostic to how $\mu^a$ is estimated.
Many interference methods recommend using the leave-one-out mean, estimated based on all the subunits except the one being predicted. For example, in \Cref{eqn:unobserve_interfere_linear}, one would replace $\mu^a$ with $\hat{\mu}^a_{-j} = \frac{1}{m-1}\sum_{j'=1}^m a_{ij'} \mathbb{I}(j'\neq j)$.
This can be viewed as a sample-split estimate of $\mu^a$.

\subsection{Multi-site instrumental variables} \label{apx:multisite_iv}

The multi-site instrumental variable models studied by~\citet{Reardon2014-fy,Raudenbush2012-es} can be seen as HCMs in which the instrument, treatment, and outcome are all subunit level, and there are both unit level and subunit level confounders between the treatment and outcome.
In the linear models proposed by \citet{Reardon2014-fy,Raudenbush2012-es}, the bias in standard IV methods introduced by the unit confounder is referred to as ``compliance-effect covariance bias''.
While these models are HCMs, our nonparametric identification methods do not apply, as identification here depends on parametric assumptions.

\subsection{Multi-environment learning} \label{apx:multi-enviro}

We next connect multi-environment causal models to HCMs.
In the context of HCMs, different environments correspond to different units.

Several multi-environment methods have been developed to address problems in which the graph of subunit variables is at least partially unknown.
To discover or account for the unknown graph, these methods make assumptions about the causal relationships between subunit treatments $A_{ij}$ and subunit outcomes $Y_{ij}$.
If these assumptions hold, there will be a detectable signature in the data that $A_{ij}$ indeed causes $Y_{ij}$, and not the other way around.
Broadly speaking, there are two typical classes of assumptions: the independent causal mechanism assumption, and the invariance assumption.

\parhead{Independent causal mechanisms}
Under the hypothesis that $A$ causes $Y$, the independent causal mechanism assumption states that the mechanism generating $A$ and the mechanism generating $Y$ are independent across units~\citep[Chapters 2, 4][]{Peters2017-ww,Perry2022-ah,Guo2022-bl}.
In the framing of HCMs, this means that $Q^a$ is independent of $Q^{y|a}$.
So, one way to recover the independent causal mechanism assumption is to assume: There are no unit confounders nor interferers between $A$ and $Y$.

\parhead{Invariance}
Under the hypothesis that $A$ causes $Y$, the strong invariance assumption states that the stochastic mechanism generating $Y$ is fixed across units~\citep{Peters2015-uw,Yin2021-ma}.
In the context of HCMs, this means that $q^{y|a}_i$ is constant across units, i.e. $q^{y|a}_1 = q^{y|a}_2 = \cdots$.
In an HCM, $Q^{y|a}_i$ will in general vary across units, unless there is no unit-level noise $\gamma^Y_i$ or other unit-level causes. 
So, one way to recover the strong invariance assumption is to assume: There are no unit-level causes of $Y$.

\end{document}